\documentclass[twocolumn,trackchanges,twocolappendix]{aastex63}
\graphicspath{{./}{figures/}}

\received{\today}
\revised{\today}
\accepted{\today}
\submitjournal{AJ}

\shorttitle{TESS Data for Asteroseismology (T'DA) Classification pipeline}
\shortauthors{Audenaert et al.}

\usepackage{amsmath}
\usepackage{amssymb}
\usepackage{graphicx}
\usepackage[caption=false]{subfig}
\usepackage{footnote}
\usepackage{multirow}
\usepackage{booktabs}
\usepackage[flushleft]{threeparttable}
\usepackage{xspace}
\newcommand{\TESS}{TESS\xspace} 

\newcommand{\rom}[1]{#1} 

\newcommand{\citsem}[2]{\citeauthor{#2} \citetext{\citeyear{#2}; #1}}

\begin{document}
\title{\TESS Data for Asteroseismology (T'DA) Stellar Variability Classification Pipeline: Set-Up and Application to the \textit{Kepler} Q9 Data}

\correspondingauthor{J.\,Audenaert}
\email{jeroen.audenaert@kuleuven.be}

\author[0000-0002-4371-3460]{J.\,Audenaert}
\affiliation{Institute of Astronomy, KU Leuven,  Celestijnenlaan 200D, 3001, Leuven, Belgium}

\author{J.~S.\,Kuszlewicz}
\affiliation{Landessternwarte, Zentrum f\"{u}r Astronomie der Universit\"{a}t Heidelberg, K\"{o}nigstuhl 12, 69117, Heidelberg, Germany }
\affiliation{Stellar Astrophysics Centre, Department of Physics and Astronomy, Aarhus University, Ny Munkegade 120, DK-8000 Aarhus C, Denmark}

\author[0000-0001-8725-4502]{R.\,Handberg}
\affiliation{Stellar Astrophysics Centre, Department of Physics and Astronomy, Aarhus University, Ny Munkegade 120, DK-8000 Aarhus C, Denmark}

\author[0000-0003-0842-2374]{A.\,Tkachenko}
\affiliation{Institute of Astronomy, KU Leuven,  Celestijnenlaan 200D, 3001, Leuven, Belgium}

\author[0000-0002-5080-4117]{D.\,Armstrong}
\affiliation{Department of Physics, University of Warwick, Gibbet Hill Road, Coventry CV4 7AL, UK}
\affiliation{Centre for Exoplanets and Habitability, University of Warwick, Gibbet Hill Road, Coventry, CV4 7AL, UK}

\author[0000-0003-2400-6960]{M.\,Hon}
\affiliation{Institute for Astronomy, University of Hawai`i,
2680 Woodlawn Drive, Honolulu, HI 96822, USA}
\affiliation{School of Physics, The University of New South Wales, Sydney, NSW 2052, Australia}

\author{R.\,Kgoadi}
\affiliation{College of Science and Engineering, James Cook University, Townsville, Australia, 4811}

\author[0000-0002-0786-7307]{M.\,N.\,Lund}
\affiliation{Stellar Astrophysics Centre, Department of Physics and Astronomy, Aarhus University, Ny Munkegade 120, DK-8000 Aarhus C, Denmark}

\author[0000-0002-0656-032X]{K.~J.\,Bell}
\affiliation{DIRAC Institute, Department of Astronomy, University of Washington, Seattle, WA-98195, USA}
\affiliation{NSF Astronomy and Astrophysics Postdoctoral Fellow}
    
\author[0000-0003-0142-4000]{L.\,Bugnet}
\affiliation{Flatiron Institute, Simons Foundation, 162 Fifth Ave, New York, NY 10010, USA}
\affiliation{AIM, CEA, CNRS, Université Paris-Saclay, Université Paris Diderot, Sorbonne Paris Cité, F-91191 Gif-sur-Yvette, France}

\author[0000-0001-7402-3852]{D.~M.\,Bowman}
\affiliation{Institute of Astronomy, KU Leuven,  Celestijnenlaan 200D, 3001, Leuven, Belgium}

\author[0000-0002-3054-4135]{C.\,Johnston}
\affiliation{Institute of Astronomy, KU Leuven,  Celestijnenlaan 200D, 3001, Leuven, Belgium}
\affiliation{Department of Astrophysics, IMAPP, Radboud University Nijmegen, NL-6500 GL, Nijmegen, the Netherlands}

\author[0000-0002-8854-3776]{R.~A.\,Garc\'{i}a}
\affiliation{AIM, CEA, CNRS, Université Paris-Saclay, Université Paris Diderot, Sorbonne Paris Cité, F-91191 Gif-sur-Yvette, France}

\author{D.\,Stello}
\affiliation{School of Physics, The University of New South Wales, Sydney, NSW 2052, Australia}
\affiliation{Sydney Institute for Astronomy (SIfA), School of Physics, University of Sydney, Sydney, NSW 2006, Australia}
\affiliation{Stellar Astrophysics Centre, Department of Physics and Astronomy, Aarhus University, Ny Munkegade 120, DK-8000 Aarhus C, Denmark}

\author[0000-0002-8159-1599]{L.\,Moln\'ar}
\affiliation{Konkoly Observatory, Research Centre for Astronomy and Earth Sciences, E\"otv\"os Lor\'and Research Network (ELKH), Konkoly Thege Mikl\'os \'ut 15-17, H-1121 Budapest, Hungary}
\affiliation{MTA CSFK Lend\"ulet Near-Field Cosmology Research Group, 1121, Budapest, Konkoly Thege Mikl\'os \'ut 15-17, Hungary}
\affiliation{ELTE E\"otv\"os Lor\'and University, Institute of Physics, 1117, P\'azm\'any P\'eter s\'et\'any 1/A, Budapest, Hungary}
    
\author[0000-0002-5481-3352]{E.\,Plachy}
\affiliation{Konkoly Observatory, Research Centre for Astronomy and Earth Sciences, E\"otv\"os Lor\'and Research Network (ELKH), Konkoly Thege Mikl\'os \'ut 15-17, H-1121 Budapest, Hungary}
\affiliation{MTA CSFK Lend\"ulet Near-Field Cosmology Research Group, 1121, Budapest, Konkoly Thege Mikl\'os \'ut 15-17, Hungary}
\affiliation{ELTE E\"otv\"os Lor\'and University, Institute of Physics, 1117, P\'azm\'any P\'eter s\'et\'any 1/A, Budapest, Hungary}

\author[0000-0002-1988-143X]{D.\,Buzasi}
\affiliation{Department of Chemistry and Physics, Florida Gulf Coast University, 10501 FGCU Blvd. S., Fort Myers, FL 33965 USA}

\author[0000-0003-1822-7126]{C.\,Aerts}
\affiliation{Institute of Astronomy, KU Leuven,  Celestijnenlaan 200D, 3001, Leuven, Belgium}
\affiliation{Department of Astrophysics, IMAPP, Radboud University Nijmegen, NL-6500 GL, Nijmegen, the Netherlands}
\affiliation{Max Planck Institute for Astronomy, Koenigstuhl 17, 69117 Heidelberg, Germany}

\author{and the T'DA collaboration}


\begin{abstract}
The NASA Transiting Exoplanet Survey Satellite (TESS) is observing tens of millions of stars with time spans ranging from $\sim$ 27 days to about 1 year of continuous observations. This vast amount of data contains a wealth of information for variability, exoplanet, and stellar astrophysics studies but requires a number of processing steps before it can be fully utilized. In order to efficiently process all the TESS data and make it available to the wider scientific community, the TESS Data for Asteroseismology working group, as part of the TESS Asteroseismic Science Consortium, has created an automated open-source processing pipeline to produce light curves corrected for systematics from the short- and long-cadence raw photometry data and to classify these according to stellar variability type. We will process all stars down to a TESS magnitude of 15. This paper is the next in a series detailing how the pipeline works. Here, we present our methodology for the automatic variability classification of TESS photometry using an ensemble of supervised learners that are combined into a metaclassifier. We successfully validate our method using a carefully constructed labelled sample of \textit{Kepler} Q9 light curves with a 27.4 days time span mimicking single-sector TESS observations, on which we obtain an overall accuracy of 94.9\%. We demonstrate that our methodology can successfully classify stars outside of our labeled sample by applying it to all $\sim$ 167\,000 stars observed in Q9 of the \textit{Kepler} space mission.

\end{abstract}


\section{Introduction}

Understanding stellar variability is important for many fields of astrophysics. Asteroseismology and stellar astrophysics in general have been revolutionized with the launch of space missions that delivered (and continue delivering) months- to years-long high precision, high cadence, and high-duty cycle brightness measurements for large numbers of stars. Following the MOST \citep{Walker2003}, WIRE \citep{Buzasi2004,Bruntt2006} and CoRoT \citep{Auvergne2009} space missions that were among the pioneers in the field of ``space asteroseismology'' \citep[e.g.][for historical notes]{Aerts2010}, \textit{Kepler} \citep{borucki2009} observed around 160,000 stars in 30-minute (long) and 1-minute (short) cadence intervals for up to four years. After the failure of its second reaction wheel, the \textit{Kepler} mission was turned into the \textit{Kepler} Second Light \citep[K2;][]{howell2014} mission that observed a large number of stars along the ecliptic plane during 20 further campaigns, each of about 80 days duration. The TESS mission \citep{ricker2015} was launched in 2018 and is covering almost the entire sky. With millions of stars observed, it offers many times the number of targets as \textit{Kepler} did, but most will be observed for only a fraction of the duration of that mission. The TESS targets in the Full Frame Images (FFIs) were observed at 30-minute cadence intervals during its first 2 years while a pre-selected list of targets was observed at 2-minute cadence. For the first extended mission the FFI cadence is reduced to 10~minutes, with an additional 20~sec cadence introduced as well. The observing periods in a single cycle range from 27.4\,d up to 352\,d, depending on the position on the sky.

\begin{figure*}
  \centering
  \includegraphics[width=18cm]{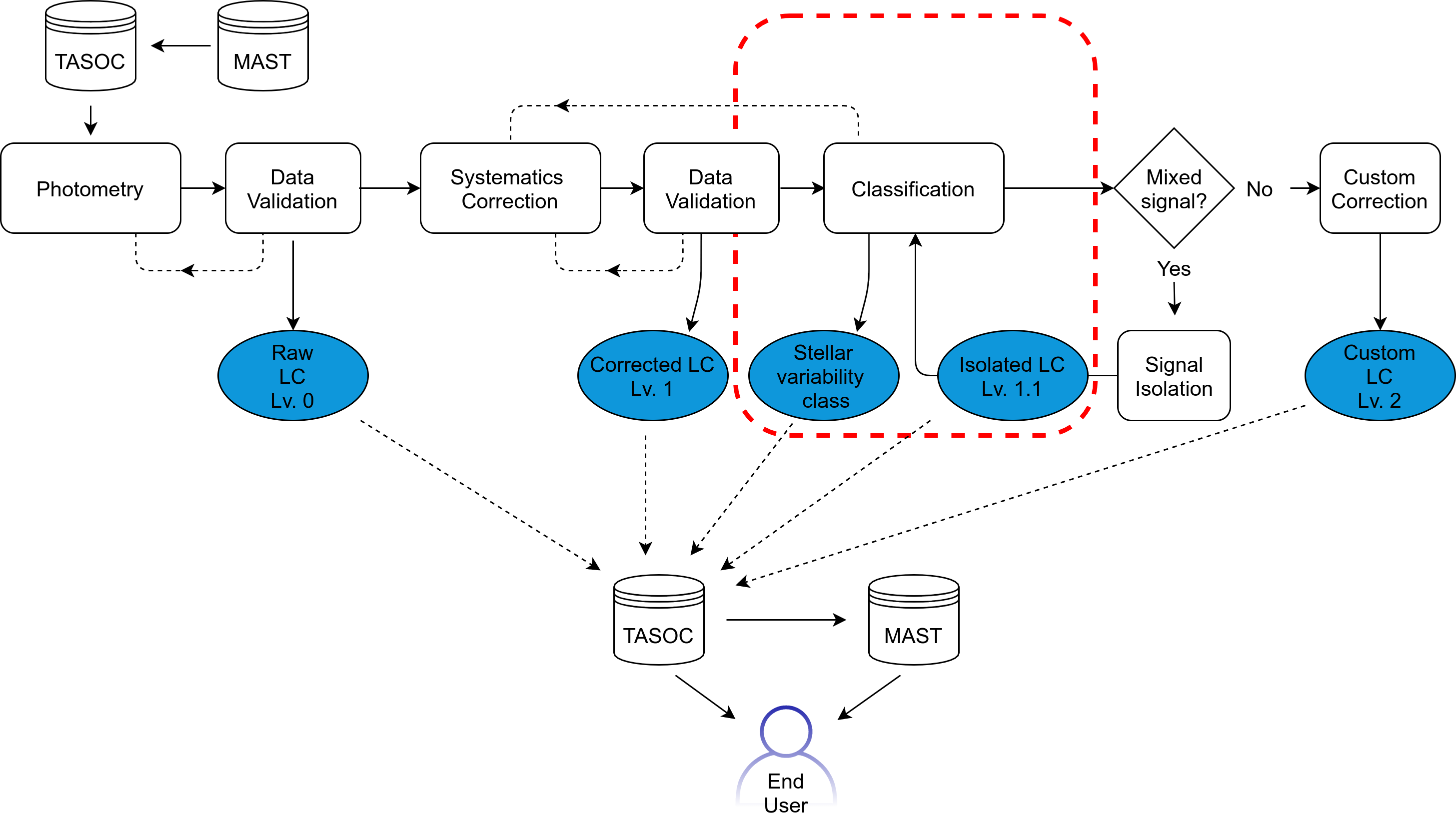}
    \caption{The overall structure of the full T'DA pipeline, with modules given as rectangular boxes, data products as ellipses, and ``TASOC'' and ``MAST'' indicate the databases hosting the data products.  Dashed lines between modules indicate that an iteration might take place.  The part enclosed by the red dashed line indicates the pipeline component described in this paper. The ``photometry'' part of the pipeline is described in \citet{handberg2021}, while the ``correction'' is detailed in Lund et al. (in prep.).}
        \label{Fig:TDA-pipeline}
\end{figure*}

\begin{figure*}
   \centering
   \includegraphics[width=18cm]{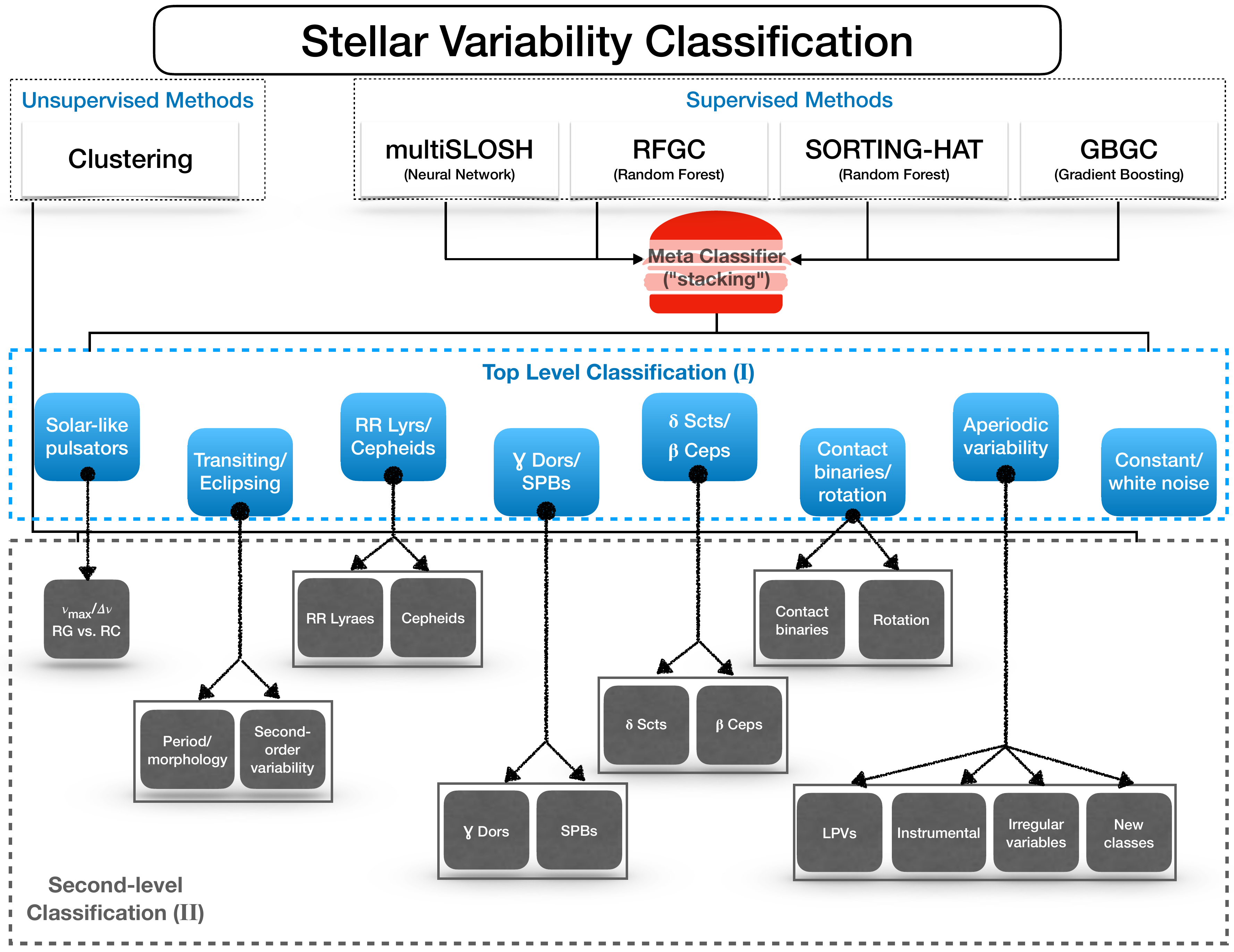}
      \caption{Graphical representation of the TASC classification scheme that encompasses two major stages: ``Level\,\rom{1}'' is the most general, largely light curve based classification, while ``Level\,\rom{2}'' stands for a detailed classification based on external features, such as parallaxes, colours, effective temperatures, etc. Rather than only relying on supervised learning, we also make use of unsupervised learning algorithms in Level 2.}
         \label{Fig:ClassificationScheme}
   \end{figure*}

Coping with the large volume of data obtained by various space-missions, and in particular by the currently operational TESS mission, requires a coordinated effort. To that end, the TESS Data for Asteroseismology (T'DA\footnote{\url{https://tasoc.dk/tda/}}) coordinated activity has been created within the TESS Asteroseismic Consortium (TASC\footnote{\url{https://tasoc.dk/}}). The major task of the T'DA unit is to serve the community with optimal processing of TESS data (both short cadence and full frame images) for all stars in the sky down to a TESS magnitude of 15. This includes raw light curve extraction, correction of the extracted light curves for systematics, and their automated classification into variability classes. Putting it into context, thanks to the observing strategy of the TESS mission and a high-level integration of the raw TESS image data into our pipeline which allows us to handle large amounts of data quickly and efficiently, we will ultimately produce an all-sky variability catalogue containing tens of millions of stars. While being a treasure trove on its own, our variability catalogue also forms a rich legacy for future space- and ground-based missions/surveys. The overall scheme of the T'DA operations is depicted in Fig.~\ref{Fig:TDA-pipeline}, and includes the data processing and classification pipeline itself as well as the ways our data products are made available to the community. The steps of the light curves extraction and their optimal corrections for systematic effects are described in detail in \citet{handberg2021} and Lund et al. (in prep.), respectively.

This paper is the next in a series of the T'DA papers and concerns the automated stellar variability classification. This component within the T'DA pipeline structure is highlighted by the red dashed box in Fig.~\ref{Fig:TDA-pipeline}, while the classification scheme itself is depicted in Fig.~\ref{Fig:ClassificationScheme}. It comprises two major steps: (i) ``top-level classification'' that is based solely on the information encoded in the light curves themselves; and (ii) ``second-level classification'' that involves using extra information, such as Gaia parallaxes, photometric colours, etc. This latter classification step also involves using unsupervised methods for variability classification that help us identify potential misclassifications and to search for new (sub-)groups of variable stars within our predefined general variability classes, and will be the subject of a separate future study. The final result is a variability catalog of the whole sky down to a magnitude of 15, containing all the tens of millions of stars observed by TESS. The creation of this large catalog is only possible thanks to the efforts of the entire T'DA team contributing to the pipeline development.

Automated variability classification based on light curves (and frequency spectra) resulted from large-scale surveys such as the Hipparcos mission\footnote{\url{https://www.cosmos.esa.int/web/hipparcos}}, Optical Gravitational Lensing Experiment (OGLE\footnote{\url{ http://ogle.astrouw.edu.pl/}}), All Sky Automated Survey (ASAS\footnote{\url{http://www.astrouw.edu.pl/asas/}}), Sloan Digital Sky Survey (SDSS\footnote{\url{https://www.sdss.org}}), etc. The classifications varied in scale from general ones, e.g. \citsem{OGLE}{Wyrzykowski2008}, \citsem{ASAS}{Pojmanski2002}, \citsem{SDSS}{Ball2006} and \citsem{Hipparcos}{EyerGrenon1998}, to those focused on specific types of stars, e.g. \citet[][]{Aerts1998} and \citet[][]{Waelkens1998} from Hipparcos. \citet{debosscher2007}, \citet{sarro2009}, and \citet{debosscher2011} presented an automated classification of light curves of variable stars in a supervised manner, employing Gaussian Mixtures and Bayesian Networks to classify OGLE, CoRoT, and \textit{Kepler} Quarter 1 (Q1) data. \cite{richards2011} also used a feature-based approach in combination with a Random Forest to classify variable stars in the OGLE and Hipparcos datasets. More recently, \citet{Kim2016} and \citet{Armstrong:2016br} respectively used a Random Forest, and Self-Organizing Maps (SOM) in combination with a Random Forest, to perform classification of variable stars in the ASAS, MACHO (MAssive Compact Halo Objects), LINEAR (LincoIn Near-Earth Asteroid Research), and K2 (Campgains 0-4) surveys. \cite{naul2018} took a hybrid approach and reverted to automated feature learning by means of an unsupervised autoencoder in order to capture the stellar variability, and then subsequently used the latent layer as input into a Random Forest. \cite{jamal2020} extended this approach by making a comprehensive analysis of neural architectures suited for light curve classification. Unsupervised light curve classification is much less prevalent in the literature with a few application examples being by \citet{EyerBlake2005}, \citet{valenzuela2018} and \citet{modak2018}. Other notable large-scale variability studies include the work by \citet{Gaia2016,GaiaEyer2019} for the Gaia mission.

Here, we present a method for the supervised classification of light curves into broad variability classes as depicted by the blue boxes in Fig.~\ref{Fig:ClassificationScheme} (``top-level classification''). We discuss the feature engineering and the collection of the training set, including a detailed description of each variability class, in Sections~\ref{Sect:Features} and \ref{sec:training set}, respectively. Our individual classifiers are described in Section~\ref{Sect:Methods} while their testing and validation is presented in Section~\ref{Sect:Test_Validation}. The individual classifiers form the basis for the metaclassifier that is tested and validated in Section~\ref{sec:meta} and is ultimately applied to the truncated 27.4-d segment \textit{Kepler} Q9 data to mimick the single sector TESS case (Section~\ref{sec:Kepler Q9 application}). We close the paper with the discussion, conclusions, and an outline of future prospects in Section~\ref{sec:conclusion}.

\section{Classification features}
\label{Sect:Features}

\begin{table*}
\begin{threeparttable}
\caption{Overview of classification features employed by the individual algorithms.}
\label{Table:LightCurveFeatures}      
\centering                          
\begin{tabular}{l | c c c c | l}        
\toprule
Algorithm/Feature & SLOSH & RFGC & SORTING-HAT & GBGC & Notes\\    
\midrule \hline                        
  PDS & x & & & & Power density spectrum. \\      
  $f_i$, $jf_i^{\bf (a)}$ & & x & x & x & Frequencies and their harmonics. \\
  $A_{ij}$ & & & & x & Amplitudes. \\ 
  \hspace*{3mm} $\frac{A_{21}}{A_{11}}$, $\frac{A_{31}}{A_{11}}$ & & x & & & Amplitude ratios. \\
  $\phi_{ij}$ & & & & x & Phases. \\
  \hspace*{3mm} $\phi_{i1}-\phi_{11}, i = 2,3$ & & x & & & Phase differences. \\  
  FliPer (F$_{\rm p}$)$^{\bf (b)}$ & & & & & Mean power in a given frequency range\\
  \hspace*{3mm} F$_{\rm p}$07,7,20,50 & & x & & & 0.7, 7, 20, 50$\mu \rm Hz$ onwards. \\
  SOM\_loc & & x & & & Location on the trained self-organizing maps. \\
  $\phi$\_p2p\_98 & & x & & & Point-to-point difference, 98$_{\rm th}$ percentile,\\
  p2p\_98 & & x & & & $\phi$ refers to the phase-folded light curve.\\
  $\phi$\_p2p\_mean & & x & & & Mean of the point-to-point difference,\\
  p2p\_mean & & x & & & $\phi$ refers to the phase-folded light curve. \\
  $\phi$\_range & & x & & & Range of phase-folded light curve. \\
  $D_k$ & & x & & & Number of zero-crossings in a light curve. \\
  $\psi^2$ & & x & & & Coherency parameter. \\
  $\eta_e^{\bf (d)}$ & & & & x & Variability index. \\
  skewness$^{\bf (c)}$ & & & x & x & Light curve skewness. \\
  MAD$^{\bf (e)}$ & & x & & x & Median absolute deviation. \\
  Rcs$^{\bf (f)}$ & & & & x & Range of the cumulative sum of the fluxes. \\
  $\sigma^2$ & & & & x & Variance.\\
  SW$^{\bf (g)}$ & & & & x & Shapiro-Wilk test for normality.\\
  kurt$^{\bf (h)}$ & & & & x & Kurtosis. \\
  varrat$^{\bf (i)}$ & & & x & & Variance ratio. \\
  SH & & & x & & Number of significant harmonics of $f_1$. \\
  FR & & & x & & Flux ratio.\\
  $h(x)$ & & & x & & Differential entropy. \\
  MSE & & & & & Multiscale entropy \\
\hspace*{3mm} MSE avg,std,max,pow & & & x & & mean, standard deviation, max and power. \\
\bottomrule
\end{tabular}
\begin{tablenotes}
      \small
      \item $^{\bf (a)}$ $i \in [1,6]$ and $j \in [1,10]$; the number of frequencies and harmonics used is algorithm-dependent
      \item $^{\bf (b)}$ $\textnormal{F}_{\textnormal{p},f_i}=\overline{\textnormal{PDS}[f \rightarrow f_{\rm max}]}-\textnormal{P}_n$, where $\textnormal{P}_n$ is the photon noise computed by considering the averaged power at high frequencies \citep{Bugnet2018}.
      \item $^{\bf (c)}$ skewness is defined by ${\rm skew} = \frac{m_3}{m_2^{3/2}},$ where $m_{\rm r} = \frac{1}{n}\sum_{i=1}^{n} (x_i - \overline{x})^r$ is the $r_{th}$ moment about the mean $\overline{x}$
      \item $^{\bf (d)}$ variability index $\eta_e$ is computed as ratio of the mean square of successive differences to the variance of the data points
      \item $^{\bf (e)}$ MAD = ${\rm median}(|X_{0,i} - {\rm median}(X_0)|)$, where $X_0$ stands for the whole time series while the subscript $i$ refers to a single data point in the time series $X_0$
      \item $^{\bf (f)}$ cumulative sum of the fluxes is defined by $S_i = S_{i-1} + \left(x_i - \overline{x} \right), i\in[1,N]$, where $\overline{x}$ is the mean flux
      \item $^{\bf (g)}$ $\text{SW}=\frac{(\sum_{i=1}^n{a_ix_{(i)}})^2}{\sum_{i=1}^n (x_i-\overline{x})^2}$, where $x_{(i)}$ are the ordered fluxes, $\overline{x}$ the mean flux and $a_i$ the generated constants (see \citet{ShapiroWilk1965} for a detailed description)
      \item $^{\bf (h)}$ kurtosis is defined by ${\rm kurt} = \frac{m_4}{m_2^{2}} -3,$ where $m_{\rm r} = \frac{1}{n}\sum_{i=1}^{n} (x_i - \overline{x})^r$ is the $r_{th}$ moment about the mean $\overline{x}$
      \item $^{\bf (i)}$ varrat = $ (\sigma_{init}^2 - \sigma_{sines}^2) / \sigma_{init}^2 $, where  $\sigma_{sines}^2 = \sum_{i=1}^j A_i^2$, $A$ the amplitude and $j$ the number of harmonics
    \end{tablenotes}
\end{threeparttable}
\end{table*}

The optimally extracted and corrected light curves are subject to parameterization; a step that is often referred to as {\it feature engineering}. The T'DA classification pipeline provides the means for an automated feature extraction that is tuned to the needs of the individual classification algorithms (cf. Sect.~\ref{Sect:Methods}). Two main types of features are extracted and used in the process: (i) Fourier-based features and (ii) time-domain features.

\subsection{Fourier-based features}
\label{Subsect:Fourier}

An efficient way of extracting periodic signals from a time series of data is to take its Fourier transform. We employ the Lomb-Scargle periodogram method \citep{lomb1976,scargle1982} to represent input light curves in the Fourier domain and perform classical iterative prewhitening \citep[see e.g.,][]{Roberts1987,Brown1991,Kjeldsen1995,Montgomery1999,degroote2009,Antoci2019} to extract individual frequencies with their corresponding amplitudes and phases. In this process, stellar flux is represented as
\begin{equation}
X(t) = C + \sum_{i=1}^{n} \sum_{j=1}^{m} \left(a_{ij}\sin{\left(2\pi f_ijt\right)} + b_{ij}\cos{\left(2\pi f_ijt\right)}\right),
\end{equation}
with $C$ representing the mean value of flux, $n$ and $m$ are the number of extracted frequencies $f_i$ and their harmonic terms, and $a_{ij}$ and $b_{ij}$ are the Fourier coefficients. The coefficients are converted into time-translation invariant frequency amplitude ($A_{ij}$) and phases ($\phi_{ij}$) \citep[see e.g.,][]{Bracewell1986,Aerts2010}.

\begin{figure}
   \centering
   \includegraphics[width=9cm]{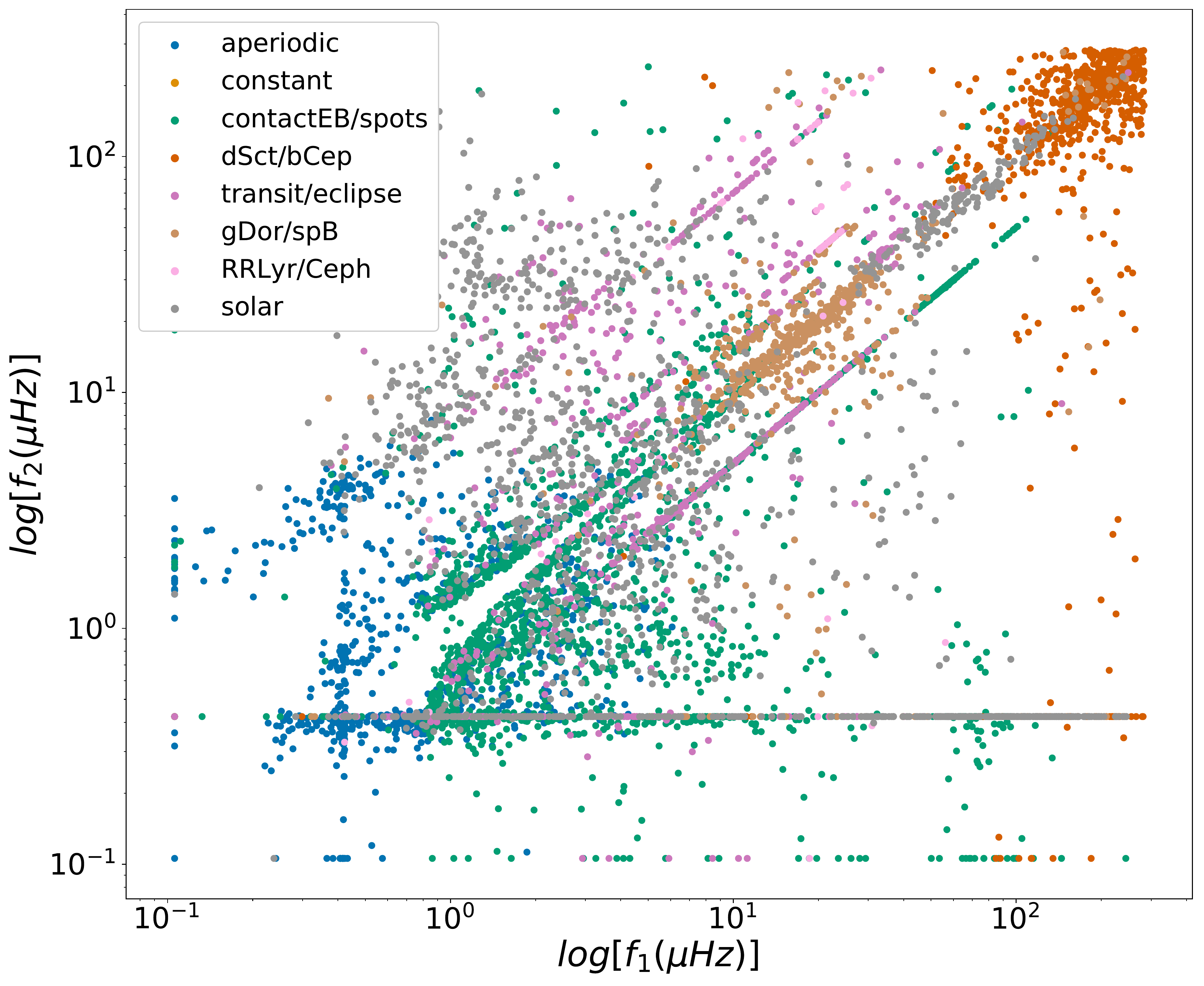}
      \caption{Scatter plot of the full training set for $log(f_1)$ and $log(f_2)$ colored per variability class in our classification scheme as defined in Table~\ref{tab:classlabels}.}
         \label{Fig:scatter_fourier}
\end{figure}

Iterative prewhitening assumes obtaining and subtracting the optimal fit for the frequency $f_i$ and its $m$ harmonic terms from the flux $X(t)$, and repeating the procedure until $n$ frequencies are extracted from the data. The total number of extracted frequencies varies between individual time series and is determined by a significance criterion. This criterion can be based on an amplitude signal-to-noise or on a threshold calculated in the Fourier domain, as in e.g. \cite{papics2012}. Such an approach prevents the extraction of spurious frequencies which are the residual signal from the preceding prewhitening steps. We refer the reader to \citet{VanReeth2015a} and \citet{Antoci2019} for a detailed discussion on the method. The obtained set of frequencies, amplitudes, and phases form a basis for calculation of the Fourier-based classification features whose overview is provided in Table~\ref{Table:LightCurveFeatures}. In Fig.~\ref{Fig:scatter_fourier} we show as an example the ability of Fourier attributes $f_1$ and $f_2$ to separate the different classes. It is clear from this that the dSct/bCep class is the most well separated, followed by the gDor/SPB class. The latter does have a small but non-neglegible overlap with contactEB/spots stars. In general we see a good structure in the distribution, but it is far from perfect. In order to obtain good classifications they are therefore complemented with other Fourier and non-Fourier attributes.

The Fourier-based feature selection assumes periodic signals as a good representation of the light curve. This is however not particularly suitable for stars that exhibit either stochastic variability or no variability within the detection limit of an instrument. For that reason, some of our individual algorithms work with image-like features where the power density spectrum (PDS) is represented as an image. The PDS is the dominant frequency analysis method for stochastically excited oscillations \citep{HekkerCD2017,GarciaBallot2019}. The multiclass solar-like oscillation shape hunter algorithm (multiSLOSH, see Sect.~\ref{Methods:multiSLOSH} for details) therefore performs image recognition on the PDS of variable stars. 

\subsection{Time-domain features}

Other classification features are extracted directly from the time series and are statistical measures of the distribution of data points in the time series. Some of those are well-known, general statistics features (e.g., skewness and variance). Below we provide a short description of features that are less intuitive and hence require a certain level of insight. All time-domain features are listed in Table~\ref{Table:LightCurveFeatures} with the reference to classifiers that use them.

The {\it zero-crossings} parameter is computed from the ``clipped'' time series $Z_{k,i}$ defined as 
\begin{equation}
  Z_{k,i} =
    \begin{cases}
      1 & \text{if X$_{k,i} \geq$ 0}\\
      0 & \text{if X$_{k,i} < $ 0},
    \end{cases}       
\end{equation}
where $X_{k,i}$ stands for the input time series comprising $N$ data points and with a mean of zero, and $k$ for the $k$\textsuperscript{th} order difference (see next paragraph). The number of zero-crossings $D_k$ is then computed directly from the ``clipped'' time-series and is given by
\begin{equation}
    D_k = \sum\limits_{i=2}^N \left(Z_{k,i}-Z_{k,i-1} \right)^2.
\end{equation}
We normalize the number of zero-crossings to the total number of points $N$ in the light curve to account for a possibly different length of the time series for the individual targets. Setting $k = 0$ gives the number of zero-crossings in the original light curve while $k>0$ refers to the number of zero-crossings in the time series of higher-order differences. The $k$th order differences is defined by
\begin{equation}
    X_{k,i} = X_{k-1,i} - X_{k-1,i-1}.
\end{equation}
For example, the 1\textsuperscript{st} order differences $X_{1,i}$ is given by the point-to-point differences in the original time series $X_0$, the 2\textsuperscript{nd} order differences $X_{2,i}$ is given by the point-to-point differences in the time series $X_1$, and so on \citep{Kuszlewicz2020, Kedem1981,Kedem1982,Bae1996}.

The {\it coherency} parameter $\psi^2$ is a measure of the coherence (or stochasticity) of the signal in a time series and is computed from the zero-crossings of the higher-order differences in the time series. It is given by 
\begin{equation}
    \psi^2 = \sum\limits_{k=0}^5 \frac{\left(\Delta_k - \phi_k \right)^2}{\phi_k},
\end{equation}
where $\Delta_k$ gives the rate of change (i.e. increments) of the number of zero-crossings in the time series of higher-order differences and $\phi_k = (0.167, 0.066, 0.038, 0.025, 0.018)$ are the increments computed from simulated time series of white noise \citep{Kuszlewicz2020}.

The {\it flux ratio} is the ratio of the sum of squared residuals of the fluxes either brighter or fainter than the mean flux \citep{Kim2016} and is meant to capture eclipse-like variability. It is defined as
\begin{equation}
    {\rm FR} = \frac{\frac{1}{N} \sum_{i=1}^{N} (x_i - \overline{x})^2}{\frac{1}{M} \sum_{j=1}^{M} (x_j - \overline{x})^2},
\end{equation}
where $\overline{x}$ is the mean flux of the light curve and $x_i$ and $x_j$ the fluxes respectively brighter or fainter than the mean flux. For sinusoidal light curves the ratio is close to unity, while for light curves with eclipses the steep flux gradients cause it to be larger than unity.

The {\it differential entropy} is an extension of the Shannon Entropy \citep{Shannon1948} into the continuous domain. It is a measure of the average uncertainty of a variable, and thus a quantification of its unpredictability. The Shannon entropy $H(x)$ of a discrete random variable $x$ is defined as
\begin{equation}
     H(x) = - \sum_{i=1}^{n} p(x_i) \log p(x_i) = - E[\log p(x_i)],
     \label{eq:shannon}
\end{equation}
where $E$ is the expected value.

We use the differential entropy because, although the light curves are not continuous, they can typically take on a large range of values, causing the number of discrete states to equal the number of samples. This could distort the calculation in the discrete case, so we therefore opted to use the differential entropy. As an alternative we could have opted to use a binned version of the Shannon entropy. The differential entropy $h(x)$ of a continuous random variable $x$ is defined as 
\begin{equation}
    h(x) = - \int \mu(x) \log \Big( \mu(x) \Big) dx,
\end{equation}
where $\mu(x)$ is the density function.

\begin{figure*}
   \centering
   \includegraphics[width=18cm]{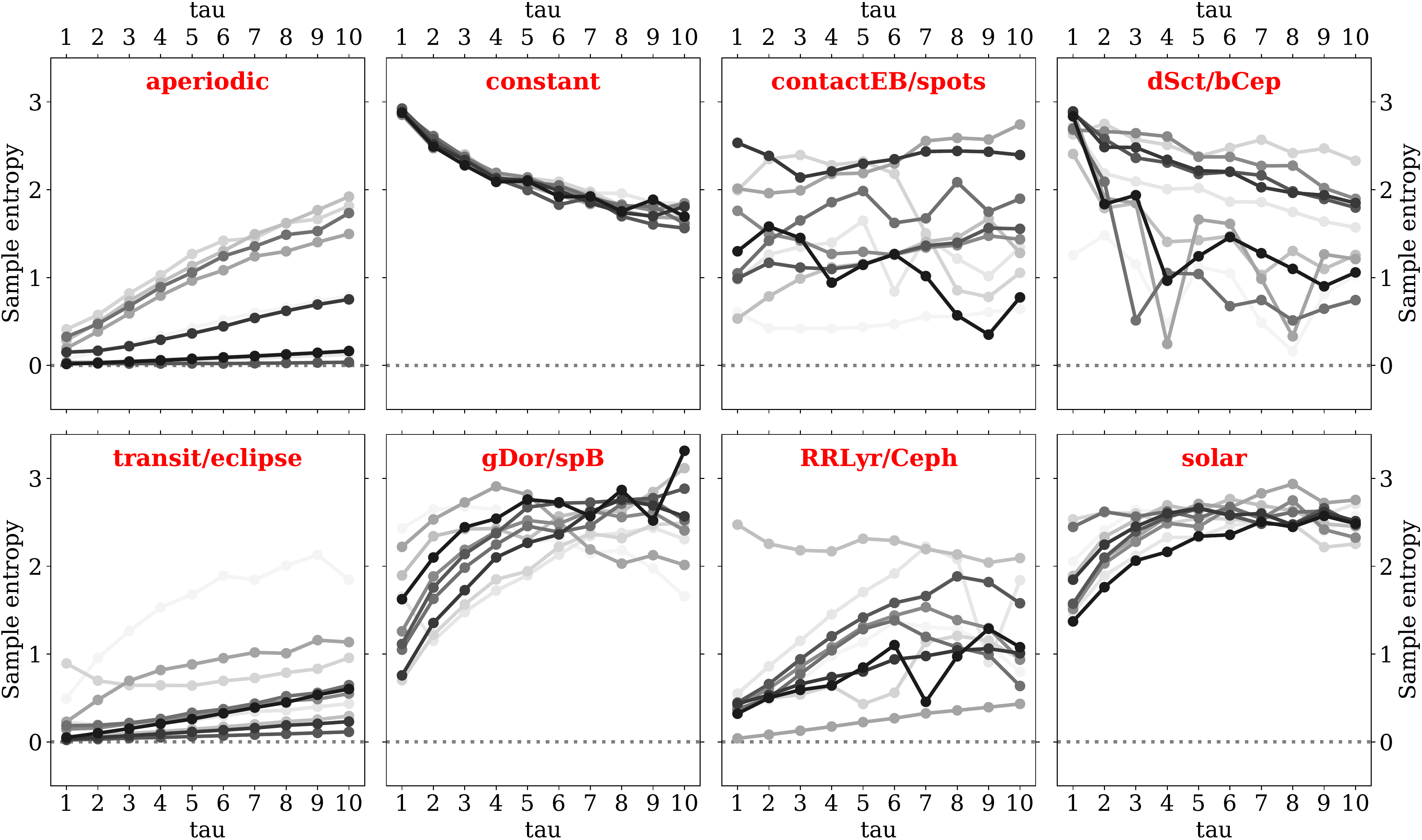}
      \caption{Examples of the Multiscale Entropy (MSE) curves for 10 random samples per variability class in our classification scheme as defined in Table~\ref{tab:classlabels}.}
         \label{Fig:MSE_Curves}
\end{figure*}

The entropy $h(x)$ can be calculated for a light curve or power density spectrum, where in the latter case it essentially becomes the spectral entropy. Although both are strongly correlated, they complement each other in specific areas. The calculations of $h(x)$ are done with the Python-based Non-parametric Entropy Estimation Toolbox (NPEET)\footnote{\url{https://github.com/gregversteeg/NPEET}}, which uses the Kozachenko-Leonenko estimate \citep{Kozachenko1987} to calculate the differential entropy as defined in \cite{Kraskov2004}.

The sample entropy \citep{Richman2000} is a different type of entropy metric that evaluates the complexity of a time series. The Sample entropy $S_E$ of a signal is defined as
\begin{equation}
    S_E(m,N,r) = -\ln \frac{A}{B} = \ln \frac{\sum_{i=1}^{N-m}n_i^m}{\sum_{i=1}^{N-m}n_i^{m+1}}
    \label{eq:SampEn}
\end{equation}
where $m$ is the number of consecutive data points or the embedding dimension, $r$ the tolerance, $N$ the number of data points and $n_i$ the number of vectors close to a basis vector, i.e. $d[u_i^m,u_i^m]\leq r$.

In practice we calculate the sample entropy by first identifying all unique sequences consisting of $m$ consecutive data points, where each data point is written as $x_i + r$, with $r$ a tolerance margin usually set to a factor of 0.15 of the time series standard deviation. We then count how many times a sequence or template vector of length $m$ occurs and subsequently extend the template vector to length $m + 1$ and count how many times that occurs. The calculations are repeated for each of the next $m$ and $m + 1$ template vector to determine the ratio between the total number of $m$ and $m + 1$ component templates, A and B respectively in Eq.~(\ref{eq:SampEn}). The sample entropy is the natural logarithm of this ratio and represents the probability that a sequence matching each other for the first $m$ data points also match for the next $m + 1$ data points.

The {\it Multiscale entropy} \citep[MSE;][]{Costa2005} takes advantage of the fact that stellar variability is active on multiple time scales. Rather than calculating one entropy metric for the full series, we calculate the entropy at each time scale, allowing us to capture the full complexity. More specifically, we first coarse-grain the signal and then calculate the sample entropy for each of these new signals. This allows the MSE to assign minimum values to both deterministic/predictable signals and random/unpredictable signals. Given a time series $x_1, ... x_i, ... x_N$, the coarse-graining is achieved by dividing the time-series into non-overlapping windows of length $\tau$. Each element $x_j$ in this new time series is then calculated as
\begin{equation}
    \label{eq:coarse-graining}
    x_{j}^{\tau} = \frac{1}{\tau} \sum_{i=(j-1)\tau + 1}^{j\tau} x_i, \quad 1 \leq j \leq \frac{N}{\tau}, 
\end{equation}
where $\tau$ is the window length, $N$ the time series length and $j$ the index after coarse-graining.

For $\tau = 1$, the time series $\{x_j^\tau\}$ is simply the original series. For each coarse-grained time series we then calculate the sample entropy given by Eq.~(\ref{eq:SampEn}) and plot it as a factor of the scale. The different types of complexity will then be represented by different types of MSE curves. In general we can say that 1) if for most values of $\tau$ the entropy is higher for one signal than for another, that signal is considered more complex, and 2) that a monotonic decrease of the entropy curve indicates that the signal only contains information on the shortest time scale. This monotonic decrease is exactly what we notice in the case of uncorrelated random signals (i.e. white noise in constant stars), as they only contain information on the shortest time scale, while in other signals information is often present across multiple time scales.

In order to obtain consistent Sample Entropy values it is suggested to have 200 data points per window at the minimum \citep{Busa2016}. Given that the shortest light curves observed by the TESS nominal mission will have a time span of $\sim$27.4 days, consisting of slightly over 1300 data points, we set $\tau_{max} = 10$. This means that for the majority of the coarse-grained time series we have more than 200 data points, where at the smallest window length, i.e. when the scaling factor reaches 10, we have around 130 data points, which is still acceptable in terms of stability. We also did experiments with $\tau_{max}=20$, and those provided good results as well. Fig.~\ref{Fig:MSE_Curves} shows the MSE curves for ten random samples per variability class. The figure illustrates the MSE's separating capacity, in particular for constant stars, solar-like oscillators and gDor/SPB stars. Due to complexity associated with implementation of the full curves, we parametrize MSE through its maximum, mean, standard deviation and power\footnote{$\text{MSE}_{\text{power}}=\frac{1}{\tau} \sum_{i=1}^{\tau} S_E ^2$}, and use these as classification features.

Lastly, the random forest general classification algorithm (RFGC, see Sect.~\ref{Sect:RFGC} for details) employs the location of a star on the self-organising map (SOM; \citealt{Kohonen:1990fd}) as one of the features in its classification scheme. The SOM location is obtained by comparing light curve shapes after folding them on the dominant extracted period, essentially grouping similar shapes into clusters.

\begin{figure*}
   \centering
   \includegraphics[width=18cm]{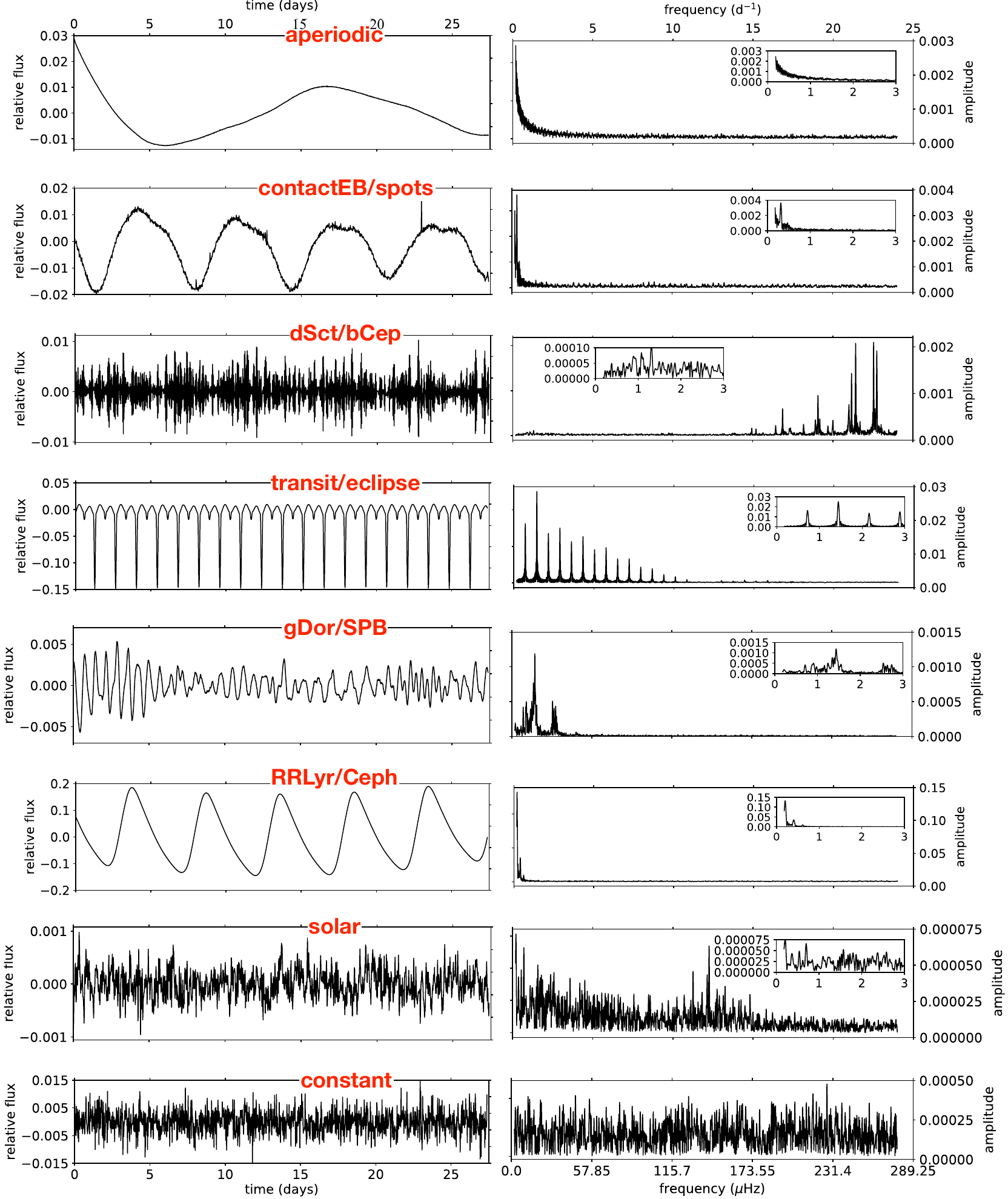}
      \caption{Examples of the light curves (left column) and the respective amplitude spectra (right column) from the training set as defined in Table~\ref{tab:classlabels}. The inset in the amplitude spectrum panel (where provided) shows a zoom-in into the low-frequency domain of 0 to 3\,d$^{-1}$. Note the different scale on the Y-axis.}
         \label{Fig:TrainingSet}
\end{figure*}

\section{Variability classes \& Training set} \label{sec:training set}

\begin{table}
	\centering
	\caption{Description of the training set.}
	\label{tab:classlabels}
	\begin{tabular}{p{3.9cm} p{2.8cm} l}
	\hline
	Class label & Type  & Size \\
	\hline
	aperiodic (Sect.~\ref{sec:aperiodic}) & Aperiodic stars & 830\\
	contactEB/spots (Sect.~\ref{sec:contactrot}) & Contact binaries and rotational variables & 2\,260 \\
	dSct/bCep (Sect.~\ref{sec:dsctbcep}) & $\delta$ Sct and $\beta$ Cep stars & 772\\
	transit/eclipse (Sect.~\ref{sec:eclipse}) & Eclipsing binaries & 974\\
	gDor/SPB (Sect.~\ref{sec:gdorspb}) & $\gamma$ Doradus and SPB stars & 630\\
	RRLyr/Ceph (Sect.~\ref{sec:rrlyrceph}) & RR Lyraes and Cepheids & 62\\
	solar (Sect.~\ref{sec:solarlike}) & Solar-like pulsators & 1\,800\\
	constant (Sect.~\ref{sec:constant}) & Constant stars & 1\,000 \\
	\hline
	\end{tabular}
\end{table}

The scientific needs of the TESS Asteroseismic Consortium drive the selection of the main variability classes (schematically represented in Fig.~\ref{Fig:ClassificationScheme}) and hence our selection of the training set. Below we provide a short description of each of the variability classes listed in Table~\ref{tab:classlabels} alongside the selection criteria that were used to select stars into the respective classes. We made sure to, where possible, maintain a balanced distribution across the different classes, while still incorporating more stars for those classes for which larger known samples exist.
For all but one (constant, see below for details) variability classes, we make use of the latest Kepler data release 25\footnote{\url{https://archive.stsci.edu/missions-and-data/kepler/documents/data-release-notes}} \citep{Thompson2016}, specifically the first 27.4 days of the Q9 PDCSAP data. Our choice of the 27.4~days total time base is dictated by the length of the majority of TESS data -- two full orbits of the satellite around the Earth. The choice of the total length of the light curve and building the training set from white-light space-based \textit{Kepler} photometric data enables a smooth knowledge and methodology transition to the TESS data afterwards. The choice for Q9 was made because it has the least gaps of all Kepler quarters. 

\subsection{Aperiodic variables}
\label{sec:aperiodic}
Aperiodic variability (\underline{aperiodic}) is a class introduced to account for targets whose variability (for one reason or another) appears to be lacking periodicity over time scales shorter than 27.4 days. For example, these can be Mira long-period variables whose variability remains unresolved on the time scale of 27.4 days as only a small fraction of the variability cycle is being captured. Similarly, a fraction of rotational variables may also appear as aperiodic stars due to their rotation periods being much longer than the length of the data set.

Our selection of aperiodic variables is based on the catalog of long-period variables compiled by \citet{Yu2020}. The selection consists of 830 objects with \textit{Kepler} Q9 data and having periods longer than 13.7 days so that less than two variability cycles are covered on the time scale of 27.4 days. An example of the light curve and amplitude spectrum of a \textit{Kepler} aperiodic variable is shown in Fig.~\ref{Fig:TrainingSet} (first row).

\subsection{Contact binaries \& rotational variables}
\label{sec:contactrot}

Contact binaries and rotational variables (\underline{contactEB/spots}) is a combined class of $i$) contact binary systems, and $ii$) objects whose light curves show signatures characteristic of surface inhomogeneities modulated by stellar rotation over time. Contact binaries are short-period gravitationally bound systems of two stars that both fill their Roche-lobes, and are therefore in contact at the Lagrangian point L$_1$. An example of a rotational variable is that of chemically peculiar B, A,  F-spectral type stars that show anomalies in their surface chemical composition often associated with a non-uniform distribution of chemical elements. These surface inhomogeneities of either enhanced or depleted abundances of certain chemical elements are often termed ``spots'' as they appear to a distant observer as darker/brighter regions with respect to the bulk of the star due to significantly modified local opacities \citep{Preston1974}.

The term ``surface spots'' in application to B, A, F-spectral type stars with radiative envelopes should not be confused with surface spots observed in cooler stars that have extended convective envelopes, e.g. in the Sun. In the latter case, these are regions of reduced surface temperature associated with the contribution of a magnetic field to the total pressure, reducing the gas pressure. Solar-type spots are typically short-lived and vary in their appearance on the time-scales ranging from a day to a few months \citep{McQuillan2014,garcia2014,santos2019}. Depending on the level of stellar magnetic activity, such ``temperature spots'' can be covering up to a few percent of the stellar surface, hence notably modulating the light curves of the respective stars \citep[e.g.][]{Namekata2019}. Because many spots with varying temperature gradients and surface areas can be formed at the same time, light curves of cool active stars are typically much more complex than those of B, A, F-spectral type chemically peculiar stars whose ``chemical abundance spots'' are long-lived (i.e. timescales ranging from years to decades; \citealp{Mathys2020}).

Our selection of rotational variables of cool stars is based on the catalog by \citet{McQuillan2014}. The catalog contains rotation periods measured for over 30\,000 \textit{Kepler} main-sequence stars and selected to have KIC($T_{\rm eff}$)$<6\,500$~K. In order to make sure at least two rotation cycles are covered with the 27.4 days data, we restricted our selection to systems whose rotation periods are shorter than 13.7 days. A total of 907 objects all having \textit{Kepler} Q9 light curves were selected this way. The training set was enriched with rotational variables of hotter stars, i.e. with stars having KIC($T_{\rm eff}$)$\geq 6\,500$~K. For this, we used the catalogs by \citet{Nielsen2013} and \citet{Hummerich2018} which were cross-matched with the lists of dSct/bCep and gDor/SPB variables (see below) to check for and remove possible duplicates. Furthermore, we excluded stars that do not have \textit{Kepler} Q9 data and/or whose rotational modulation signal is nowhere near the dominant signal in the light curve/amplitude spectrum. A total of 656 objects passed the above selection criteria and were added to the list of 907 cool rotational variables. An example of the light curve and amplitude spectrum of a rotational variable is shown in Fig.~\ref{Fig:TrainingSet} (second row).

By analogy with the transit/eclipse class (see below), we queried the \textit{Kepler} Eclipsing Binary Catalog for stars that have \textit{Kepler} Q9 data and whose light curve morphology parameter is larger than 0.6 \citep[high probability contact systems according to][]{Matijevic2012}, given that their light curve morphologies look similar to rotational variables. All 1054 systems selected that way were subject to visual inspection to remove misclassified stars of semi-detached type, resulting in the final selection of 697 contact binaries. Altogether, the contactEB/spots class comprises 2\,260 objects, of which 70\% are rotational variables.

\subsection{\texorpdfstring{$\delta$}{delta} Scuti \& \texorpdfstring{$\beta$}{beta} Cephei stars}
\label{sec:dsctbcep}
$\delta$\,Sct and $\beta$\,Cep (\underline{dSct/bCep}) stars are two classes of variables pulsating in radial and low-order non-radial pressure (p) and gravity (g) modes which are mostly excited by means of the $\kappa$ mechanism acting on the zone of partial ionization of helium ($\delta$\,Sct stars) and of iron-group elements \citep[$\beta$\,Cep stars,][]{Aerts2010}. The instability regions of the $\delta$\,Sct and $\beta$\,Cep stars (partially) overlap in the HR\,diagram with those of $\gamma$\,Dor and SPB stars (see below), respectively, giving rise to hybrid pulsators that exhibit both low-order p modes and high-order g modes simultaneously. $\beta$\,Cep stars have masses between 8 and 25 M$_{\odot}$ and the periods of their pulsations range from about 2 to 8 hours, although none were observed by the \textit{Kepler} mission \citep{Bowman2020}. Less massive $\delta$\,Sct stars cover the mass range from 1.5 to 2.5 M$_{\odot}$ and have periods from some 15 minutes to about 8 hours \citep{Aerts2010}, hence a significant overlap with $\beta$\,Cep stars in terms of pulsation periods. 

It is difficult to distinguish between $\delta$\,Sct and $\beta$\,Cep stars solely based on their light curve information. Hence, we introduce a joint class of coherent p-mode ($\delta$\,Sct/$\beta$\,Cep) pulsators in our classification scheme. Our selection of the training set for this class is based on the $\delta$\,Sct catalog compiled by \citet{Bowman2016}. All 983 objects from that catalog were cross-matched with the catalogs we used to select g-mode pulsators (see below) to search for and remove possible duplicates. Light curves of the remainder of stars were subject to a visual inspection in order to exclude objects with pronounced signatures of rotational modulation as well as stars whose dominant pulsation signal was found to be in the g-mode regime (those hybrid pulsators were included in the class of g-mode pulsators; see below). Ultimately, we selected 772 objects into the class of p-mode pulsators, among those are stars showing p-modes only and hybrid pulsators whose dominant signal is in the p-mode frequency domain. An example of the light curve and amplitude spectrum of a \textit{Kepler} $\delta$\,Sct p-mode pulsator is shown in Fig.~\ref{Fig:TrainingSet} (third row).

\subsection{Eclipsing binaries and transit events}
\label{sec:eclipse}
Eclipsing/Transiting (\underline{transit/eclipse)} systems are a class of objects that show extrinsic variability in the form of periodic transits/eclipses. The latter occur due to a partial or total obscuration of the stellar disk by the companion that can be of either stellar (eclipses) or a planetary (transit) mass. We do not make a distinction between transits and eclipses, neither do we intend to distinguish between binary/multiple stellar systems with different Roche geometries (e.g., detached or semi-detached configurations, etc.). Instead, we introduce a general class of eclipsing/transiting objects in our classification scheme which is also likely to contain members whose stellar components are intrinsically variable stars. Many of eclipsing and transiting systems have been discovered in the \textit{Kepler} space-photometry in recent years, with a variety of orbital and stellar/planetary configurations. The most up-to-date overview of the detections in the \textit{Kepler} field can be obtained from the \textit{Kepler} Eclipsing Binary Catalog\footnote{\url{http://keplerebs.villanova.edu}} and from the NASA Exoplanet Archive\footnote{\url{https://exoplanetarchive.ipac.caltech.edu}}. 

Our selection of the training set for the transit/eclipse class is based on the latest release of the \textit{Kepler} Eclipsing Binary Catalog \citep{Prsa2011,Slawson2011,Kirk2016,Abdul-Masih2016}. We started by selecting all systems with the morphology parameter smaller than 0.6 which allows us to filter out contact binaries while keeping the majority of detached and semi-detached systems \citep{Matijevic2012}. The light curves of all those 1679 objects were subject to a visual inspection in order to remove i) systems whose eclipses are hidden in the noise or any other astrophysical signal and are not traceable in the time domain without aggressive cleaning of the light curve; and 2) (long-period) systems that do not show a single eclipse event in the first 27.4 days segment of their \textit{Kepler} Q9 light curve. Our final training set for the class comprises 974 objects; an example of the light curve and amplitude spectrum of a \textit{Kepler} eclipsing binary is shown in Fig.~\ref{Fig:TrainingSet} (fourth row).

\subsection{\texorpdfstring{$\gamma$}{gamma} Doradus \& Slowly Pulsating B stars}
\label{sec:gdorspb}
$\gamma$\,Dor and Slowly Pulsating B (SPB) stars (\underline{gDor/SPB}) are members of a class of high non-radial order g-mode pulsators whose oscillations are excited by means of the flux blocking mechanism at the base of their convective envelope \citep[$\gamma$\,Dor stars;][]{Guzik2000} and by means of the $\kappa$ mechanism operating on the zone of partial ionization of iron-group elements \citep[SPB stars;][]{Aerts2010}. Although $\gamma$\,Dor and SPB stars occupy different locations in the HR\,diagram representing F- (mass range between some 1.2 and 2.0~M$_{\odot}$) and B- (with masses from some 3 to 9~M$_{\odot}$) type stars, respectively, their light curves are remarkably similar. The light curves of $\gamma$\,Dor and SPB stars are shaped by an ensemble of g-mode pulsations whose periods range from $\sim$0.2 to $\sim$3 days.

Our selection of g-mode pulsators is based on several intermediate- to large-scale studies of F- and B-type stars in the \textit{Kepler} field. The sample of lower mass $\gamma$\,Dor stars was adopted from \citet{tkachenko2013,VanReeth2015a,VanReeth2015b,VanReeth2016,Li2020}, making sure to cross-match between the catalogs to exclude possible duplicates. In addition, the catalog of $\delta$\,Sct stars compiled by \citet{Bowman2016} was used to complement pure g-mode pulsators with stars that show both g- and p-modes simultaneously, the so-called hybrid pulsators. We selected only those hybrid pulsators from \citet{Bowman2016} whose dominant variability was found in the g-mode frequency domain. Finally, the training set of g-mode pulsators was enlarged with SPB stars from \citet{Papics2017} and \citet{Pedersen2020}, providing us with a total of 694 stars, of which 630 objects have \textit{Kepler} Q9 data. Because of the similar observational properties of their light curves, we do not distinguish $\gamma$\,Dor stars from their higher-mass SPB counterparts and combine them into a joint class of g-mode pulsators in our classification scheme. A typical light curve and amplitude spectrum of a g-mode pulsator is shown in Fig.~\ref{Fig:TrainingSet} (fifth row).

\subsection{RR Lyrae and Cepheid stars}
\label{sec:rrlyrceph}
Classical pulsators (\underline{RRLyr/Ceph} class) are low- to intermediate-mass evolved stars whose intrinsic pulsation variability is driven by the opacity ($\kappa$) mechanism acting on the partial ionisation zone of helium. The majority of these stars pulsate in a single dominant radial mode and have characteristic non-sinusoidal light curves. However, a small fraction of these objects show two or even three radial modes with comparable amplitudes. Variability of RR Lyrae stars occurs at periods shorter than 1 day, while Cepheids cover a much larger period range, from half a day to several months. 

About 50 RR Lyrae stars were identified in the \textit{Kepler} field during the mission \citep{Szabo2018}. In Q9, 42 of those were observed: 34 fundamental-mode and 8 first-overtone pulsators. No double-mode RR Lyrae stars have been targeted in the field, and only two Cepheids have been confirmed: a classical Cepheid, V1154\,Cyg, and a medium-period, type II Cepheid, DF\,Cyg \citep{Szabo2011,Derekas2017,Kiss2017,Vega2017,Plachy2018,Manick2019a}. From the list of 44 RR Lyrae stars and Cepheids, we excluded one object whose 27.4~days segment of the \textit{Kepler} light curve and Fourier transform do not display any significant signal. To increase the training sample, we collected 19 further Cepheids from K2 observations, and created artificial light curves for them. We extrapolated the Fourier decomposition of the light curves to the Q9 time stamps and added appropriately scaled white noise to the data. Together, the 19 simulated Cepheid-type light curves and 43 \textit{Kepler} Q9 RR Lyrae/Cepheid light curves provide us with a total of 62 objects in the final training set for the class. We do not differentiate between RR\,Lyrae stars and Cepheids in our classification scheme, but consider them as being members of the joint class of classical radial pulsators. Fig.~\ref{Fig:TrainingSet} (sixth row) shows an example of a \textit{Kepler} light curve of a RR\,Lyrae star along with its amplitude spectrum.

\subsection{Solar-like pulsators}
\label{sec:solarlike}
Solar-like pulsators (\underline{solar} class) are intrinsically variable stars showing oscillations driven by turbulent convective motions near their surfaces. Any star with an outer convective zone is expected to show such stochastically excited oscillations. Indeed, following the detection of solar-like oscillations in a number of main-sequence and evolved stars from ground-based data, space-based photometry with the Hubble Space Telescope (HST), WIRE, MOST, SMEI, and in particular CoRoT and \textit{Kepler}, revealed a treasure of pulsational variability in stars with outer convective regions and enabled extraordinary probes of their interiors and improvement of the respective models (see \citealp{HekkerCD2017} for a review). Stochastically driven solar-like oscillations are well characterized with two global asteroseismic quantities, namely the frequency of maximum power $\nu_{\rm max}$ and the large frequency separation $\Delta \nu$, which were shown by \citet{Kjeldsen_Bedding1995} to scale with mass, radius, and effective temperature of the star. We do not provide an estimate of the global asteroseismic parameters of solar-like pulsators in our classification scheme, hence no differentiation is made between different evolutionary stages of stars.

Our selection of a sample of solar-like pulsators for the training set is based on the latest release of the APOKASC Catalog \citep{Pinsonneault2018}. A total of 1\,800 objects were selected in a random way but making sure each of the targets had a Q9 \textit{Kepler} light curve and oscillations detected with the CAN pipeline \citep{Kallinger2019}. That being said, our selection of solar-like pulsators for the training set is biased towards red-giant stars with very few main-sequence stars. The majority of those will have a high signal-to-noise ratio detection. An example of the light curve and amplitude spectrum of a solar-like pulsator is shown in Fig.~\ref{Fig:TrainingSet} (seventh row).

\subsection{Constant stars}
\label{sec:constant}
Constant stars (\underline{constant}) are a class of objects that do not show any statistically significant variability on the time scale of 27.4 days. We made a random selection of 1\,000 objects from the TESS Input Catalog\footnote{\url{https://tess.mit.edu/science/tess-input-catalogue/}} \citep{Stassun2019}  and simulated their light curves with pure white noise on the 27.4 days \textit{Kepler} time stamps. The noise level was calculated by adding shot, read, zodiacal and a TESS instrumental baseline noise of 60ppm/$\sqrt{\text{hour}}$ in quadrature, using the magnitude, effective temperature and galactic coordinates of each object. An example of the light curve and amplitude spectrum is shown in Fig.~\ref{Fig:TrainingSet} (last row).

\section{Methods -- Individual Classifiers}\label{Sect:Methods}
We first train four individual classifiers each using different feature sets and learning algorithms. In the next step we then combine these different classifiers using stacked generalization by means of a metaclassifier. The benefit of using this stacked ensemble of classifiers is that we can leverage the individual strengths and weaknesses of each classifier to come to the optimal combination of classifiers and obtain a better predictive performance compared to using just one single classifier.

We constructed the classification framework in a modular way, meaning that the different classifiers can use the same functionality without requiring the use of duplication. We have done this by creating a general \verb"BaseClassifier" class that implements all common functionalities between the different classifiers. The different classifiers then inherit all methods and properties and can define new specific functionalities themselves. This modular set-up makes our framework very flexible and easily allows for additional classifiers to be added later on. As in the other modules of the TASOC pipeline, we make use of Message Passing Interface (MPI) to parallelize our computations. During runtime, all features are also cached in a local SQLite database. In the following subsections we discuss each individual classifier.

\subsection{Multiclass \textbf{S}olar-\textbf{L}ike \textbf{O}scillation \textbf{S}hape \textbf{H}unter (multiSLOSH)}\label{Methods:multiSLOSH}

The multiSLOSH classifier uses image recognition via deep learning to visually determine the presence of the desired signal on a 2D plot of the power density of a star. This is the multiclass generalization of the method described by \citet{Hon_2018}, where now we classify other types of variability at once instead of only solar-like oscillations. To summarize, a 128$\times$128 binary image of a star's power density spectrum in log-log space is used as input into a 2D deep learning network. The log-log representation of the power density spectrum is used because stars with different types of variability distinctly show different frequency-power profiles in log-log power density spectra. For example, in the case of a solar-like oscillator, one can see the convective granulation background and the Gaussian-like power excess containing the oscillation modes.

While the original method has shown to be effective in classifying red giants observed in long-cadence during any amount of time as obtained by TESS \citep{Hon_2018}, SLOSH can be very easily generalized towards stars only observed in short-cadence, for example, main sequence, dwarf or subgiant stars. This can be done by modifying the training set that the networks use to learn.
To allow for the detection of signals in main sequence or subgiant stars, the plotting range in the 2D image has to be modified. The range in frequency, $f_{\mathrm{range}}$, and power density, $P_{\mathrm{range}}$, for the different evolutionary states are defined by the following:

\begin{equation}\label{SLOSH_Ranges}
\begin{gathered}
 f_{\mathrm{range}} (\mu \mathrm{Hz})= 
  \begin{cases} 
   [3, 283] & \text{for LC } \\
   [40, 4160]       & \text{for SC }
  \end{cases}
  \\
  P_{\mathrm{range}} (\text{ppm}^2 \mu \text{Hz}^{-1}) = 
  \begin{cases} 
   3\times[10^1,10^7] & \text{for LC } \\
    1\times[10^{-1},10^5]      & \text{for SC }
  \end{cases}
\end{gathered}
\end{equation}
where respectively LC and SC stand for \textit{Kepler} long- and short-cadence data. These ranges are defined in $\mu \text{Hz}$ (where one cycle per day (d$^{-1}$) amounts to 11.57~$\mu \text{Hz}$) given that this frequency unit is commonly used in the solar-like community.

The original deep learning implementation from \citet{Hon_2018} saved generated plots to image files to be read in later. In this work, we implemented a new method to directly create 128$\times$128 binary array representations of the power density spectra without using a plotting library or input/output to disk. We define 128 even bins in log-log space between the bounds indicated in Eq.~\ref{SLOSH_Ranges} that represent image pixels. The default pixel values are one, except for bins that the plotted power spectrum passes through, which take the value of zero.
Compared to the original approach, the image arrays that we now generate are computed faster, maintain higher data fidelity, and are better suited for parallel processing.

\subsection{\textbf{R}andom \textbf{F}orest \textbf{G}eneral
\textbf{C}lassification (RFGC)}\label{Sect:RFGC}
The RFGC uses a hybrid self-organising-map (SOM; \citealt{Kohonen:1990fd,Brett:2004cr}) and Random Forest \citep{Breiman:fb} classifier, as previously demonstrated on data from the \emph{K2} satellite \citep{Armstrong:2015bn,Armstrong:2016br}. A full methodological description is provided in \cite{Armstrong:2016br}. While the underlying methodology is the same, the features used here have been updated to better account for the new datasets and variability classes considered. 

Light curves are initially phase folded, using 64 equal width bins, on the dominant frequency as extracted in Section~\ref{Subsect:Fourier}. We also test each light curve using half the dominant frequency, and if the resulting phase-folded light curve shows significantly reduced dispersion, the half-frequency is used. This test ensures the correct value is picked for the orbital frequency of an eclipsing binary, where the presence of primary and secondary eclipses often results in the dominant frequency being double the true binary orbital frequency. 

The training set of phase-folded light curves is then used to train a SOM with shape (1,400) using 300 training iterations and a learning rate of 0.1. Training a SOM involves creating a set of template `pixels' which steadily approach similarity to underlying shapes in the input data. In the end the pixels contain representations of various common and uncommon shapes seen in the training set. The index of the closest matching pixel to a test input is then a powerful feature for parameterizing the phase-folded light curve shape.

The actual classification is performed by a Random Forest, implemented through scikit-learn \citep{scikit-learn}. The 22 features used are listed in Table~\ref{Table:LightCurveFeatures}, including the SOM location described above. We set the parameters of the Random Forest by optimising the out-of-bag score. This led to a Random Forest with 1000 component decision trees, considering a maximum of three features at each node split, with a minimum of two samples required to split an internal node and a maximum tree depth of 15. We use the Gini impurity to measure the quality of a split and in this way select the best splits at the decision tree nodes \citep{Breiman1984}.

\subsection{\textbf{S}upervised rand\textbf{O}m fo\textbf{R}est variabili\textbf{T}y class\textbf{I}fier using high-resolution p\textbf{H}otometry \textbf{A}ttributes in \textbf{T}ESS data (SORTING-HAT)}\label{Sect:sorting-hat}
The SORTING-HAT is a Random Forest classifier with an architecture similar to RFGC.  It does not use a SOM, but relies on a set of 13 carefully constructed features in the entropy, Fourier and time domain, as described in Table~\ref{Table:LightCurveFeatures}. The use of entropy metrics allows it to differentiate light curves based on their unpredictability and complexity.

The set of hyperparameters is the same as in RFGC, but was independently confirmed by optimising the weighted $F_1$ score\footnote{$F_1 = 2*\frac{\text{Precision} \times \text{Recall}}{\text{Precision} + \text{Recall}}$} in an initial version of the classifier, through a general randomized grid search followed by a narrow but complete grid search. This led to a Random Forest with 1000 decision trees, a maximum tree depth of 15, a minimum of two samples required to split an internal node and the usage of the Gini impurity measure. 

\subsection{\textbf{G}radient \textbf{B}oosting \textbf{G}eneral \textbf{C}lassification (GBGC)}\label{Sect:MethodsGBGC}

Similar to the RFGC discussed in section \ref{Sect:RFGC}, GBGC is a tree-based ensemble method whose trees were constructed with Gradient Boosted Machines \citep{Freidman2001}. In contrast to RFGC, the GBGC is an adaptive method of constructing a model where the classifier aims to correct previous trees in the ensemble by assigning higher weights to the incorrectly predicted samples. The efficiency and generalisation abilities of the GBGC classifier were established using a sample of labelled light curves from the OGLE catalog of variable stars in the LMC \citep{Udalski2008,Udalski2015} and from the \textit{Kepler/K2} missions by \citet{Kgoadi2019}. Eight hyper-parameters were adjusted to improve the performance of the classifier. In addition to the number of trees in the ensemble (\verb"n_estimators") and the optimal depth of the trees (\verb"max_depth"), the fraction of samples in the training set (\verb"subsample") and features (\verb"colsample_bytree") during training were also tuned. To ensure convergence was reached in a timely manner, the learning rate of the gradient descent (\verb"learning_rate") was tuned once the \verb"n_estimators" were determined. Adjustment of the hyperparameters was done to prevent over-fitting and to reduce running time complexities. Optimal hyperparameters were established using a grid search with 10-fold cross validation. This resulted in 500 estimators, a maximum tree depth of 6, a training sample ratio of 0.8, a feature sample ratio of 0.7, and a learning rate of 0.1.

The finalized GBGC classifier was trained on the set of features indicated in Table~\ref{Table:LightCurveFeatures}. These were selected using Recursive Feature Elimination with Cross-Validation supplemented by Correlation Based Feature Selection as introduced in \citet{Kgoadi2019}. This is a two step feature selection process where recursive feature elimination with cross-validation \citep{Granitto2006} is applied to select features that best describe light curves and can be mapped to the star classes. To reduce redundancy, the Pearson correlation coefficients were used to remove correlated features from the selected subset through the Correlation Based Feature Selection process \citep{Hall1999}, in which a correlation threshold of 0.65 was applied to remove features. In order to accommodate the class imbalance in our training set, feature selection was done with stratified cross-validation. The GBGC classifier was constructed using \verb"XGBoost" \citep{xgboost2016} as the base estimator of the model.  

\section{Testing and Validation of the Individual Classifiers}
\label{Sect:Test_Validation}

The individual classifiers are tested and validated in two different ways to ensure that they are not overfitting the training data. For a given training set, we hold out 20\% of the data from the start for testing both individual classifiers and the metaclassifier (see Section~\ref{sec:meta}). We partition the remaining 80\% of data into five folds or splits of equal size, making sure to include a balanced proportion of all variability classes in each fold. We train on four of these folds and validate on the fifth fold to report one iteration of the performance for all individual classifiers. We repeat this process four more times, but using different folds to train and validate; we are thus cross-validating the individual classifiers over the training set.

\begin{table}
	\centering
	\caption{Accuracy of each classifier on the training data using 5-fold cross validation and on the holdout set (see Fig.~\ref{Fig:ClassificationScheme} for a graphical explanation of the classifier training and testing procedure). For the Training set (5-fold CV) case we report the mean of the accuracy over each of the five different tested folds. The uncertainty here is equal to standard deviation.}
	\label{tab:accuracies}
	\begin{tabular}{l c c}
	\hline
	\multirow{2}{*}{Classifier} & \multicolumn{2}{c}{Accuracy}  \\    \cline{2-3}
     & Training set (5-fold CV) & Holdout set \\  \hline
	\hline
	multiSLOSH & $92.39\pm0.89$\% & $91.48$\% \\
	RFGC & $93.41\pm0.27$\%  & $92.56$\%\\
	SORTING-HAT & $93.79\pm0.26$\% & $93.70$\%	 \\
	GBGC & $93.79\pm0.26$\% & $91.36$\% \\
	Meta &  & 94.90\%\\
	\hline
	\end{tabular}
\end{table}

Cross-validation is the first approach we use to validate the performance of each classifier. The variance of each classifier over the different folds should be relatively low if they are not overfitting the training set. In Table \ref{tab:accuracies} we report the mean of the accuracy over the five cross-validation folds and report the uncertainty as the standard deviation. The mean scatter of $\sim0.5$\% over the cross-validation is due to the small size of some of the classes and the initial training set ($0.5$\% corresponds to $\sim8$ stars). 

All of the classifiers perform well on the training set, with SORTING-HAT performing best. As we shall see in Section~\ref{sec:meta}, we are not concerned with a single classifier performing better than all the others but more so with the classifiers being uncorrelated with one another. It is important that the individual classifiers have different strengths and each perform best on different parts of the training set if we are to leverage this information in a meta-classification stage.

The second way we validate the individual classifiers is on the 20\% initial hold out set that was not used in the previous training and cross-validation step. Whilst validating the classifiers over cross-validation folds gives a good grasp of how well the classifiers generalize to unseen data, testing the classifiers on a holdout set gives an idea of pure performance and accuracy. We report the accuracy of each classifier in the third column of Table \ref{tab:accuracies}.

Overall the holdout set accuracy of the set of classifiers is comparable to their mean accuracy over the cross-validation folds, as for most classifiers the holdout set accuracy lies almost within one standard deviation of the mean cross-validation accuracy. For RFGC alone, we notice that the holdout set accuracy is about 0.6 percent lower than the lower uncertainty bound. In absolute numbers, however, this is still very small and only represents $\pm$10 out of the 1666 stars in the holdout set. Given that the accuracies on both sets are so similar, we can safely assume that the individual classifiers are fitting the data well and are not overfitting significantly. 

We use SHAP (SHapley Additive exPlanations; \citealt{NIPS2017_7062, lundberg2019explainable}), a unified approach that connects game theory with local explanations to explain the output of a machine learning model, to compute the feature importance scores. The feature importance plots for SORTING-HAT, RFGC and GBGC are shown in Appendix \ref{Appendix:features} including a more detailed explanation. For RFGC we find that the zero-crossing parameter and point-to-point differences are the most important features, while for SORTING-HAT it is clear that the multiscale entropy (MSE) together with the first fundamental frequency and skewness are the most important attributes in the classification process. Lastly, for GBGC the variability index is by far the most important. We do not plot the feature importance scores for multiSLOSH given that it is a neural network classifier that does not rely on a set of predefined features, but rather learns a set of weights that define the importance of each region in the power density spectrum image.

\begin{figure}
   \centering
   \includegraphics[width=8cm]{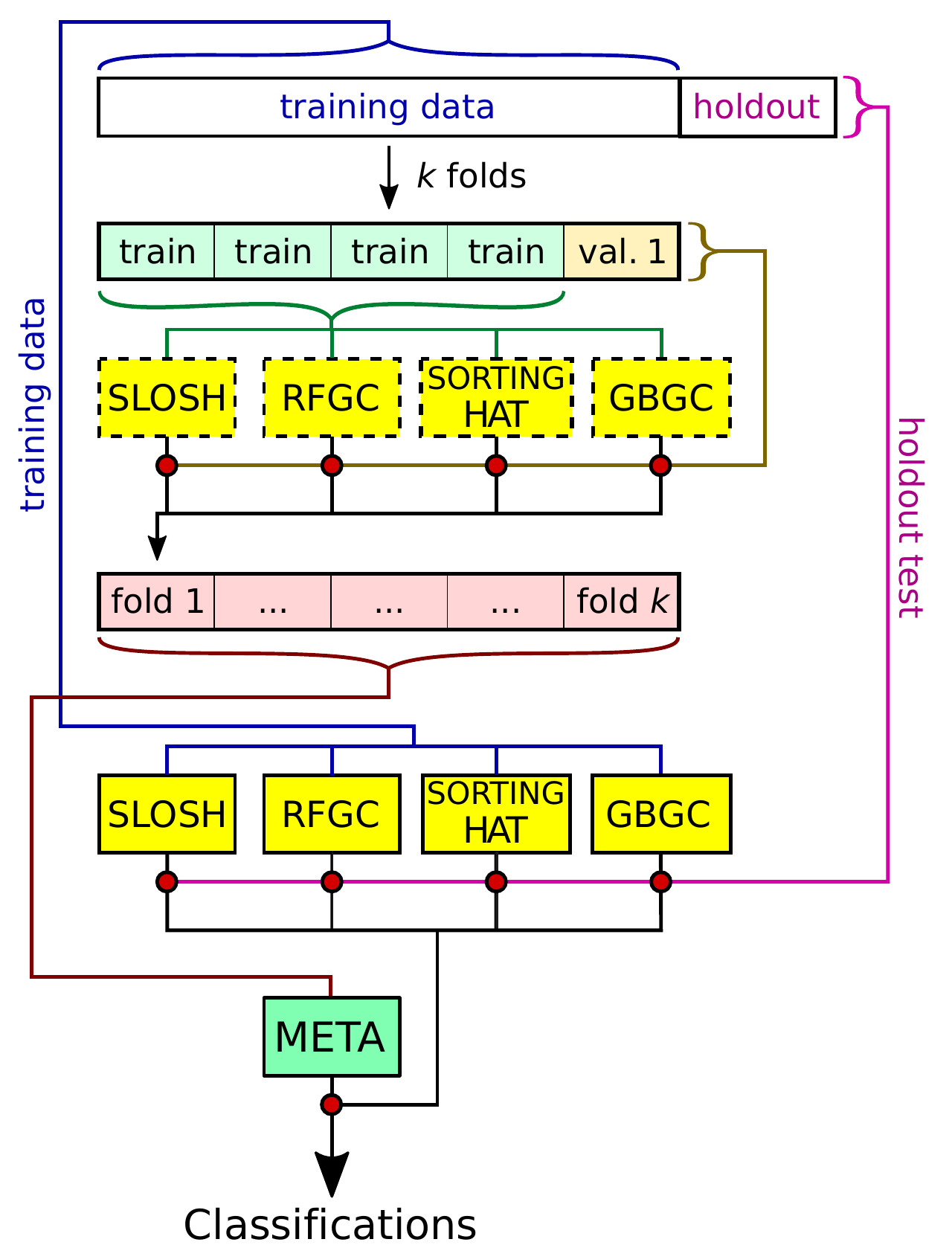}
      \caption{Graphical representation of the classifier training and testing procedure. 80\% of the data set is split into $k$ stratified folds for cross-validation, where $k=5$. Class probabilities for data in each fold are predicted by the supervised individual classifiers trained on the other $k-1$ folds. The training class probabilities from each individual classifier are used to train the metaclassifier. The individual classifiers used to characterize the unseen data are trained on all of the training data. The success of the overall classification is tested by classifying the holdout data with the individual classifiers, and then by using their predictions as input into the metaclassifier.
      }
         \label{Fig:metatraining}
   \end{figure}
   
\begin{figure*}
   \centering
   \includegraphics[width=12cm]{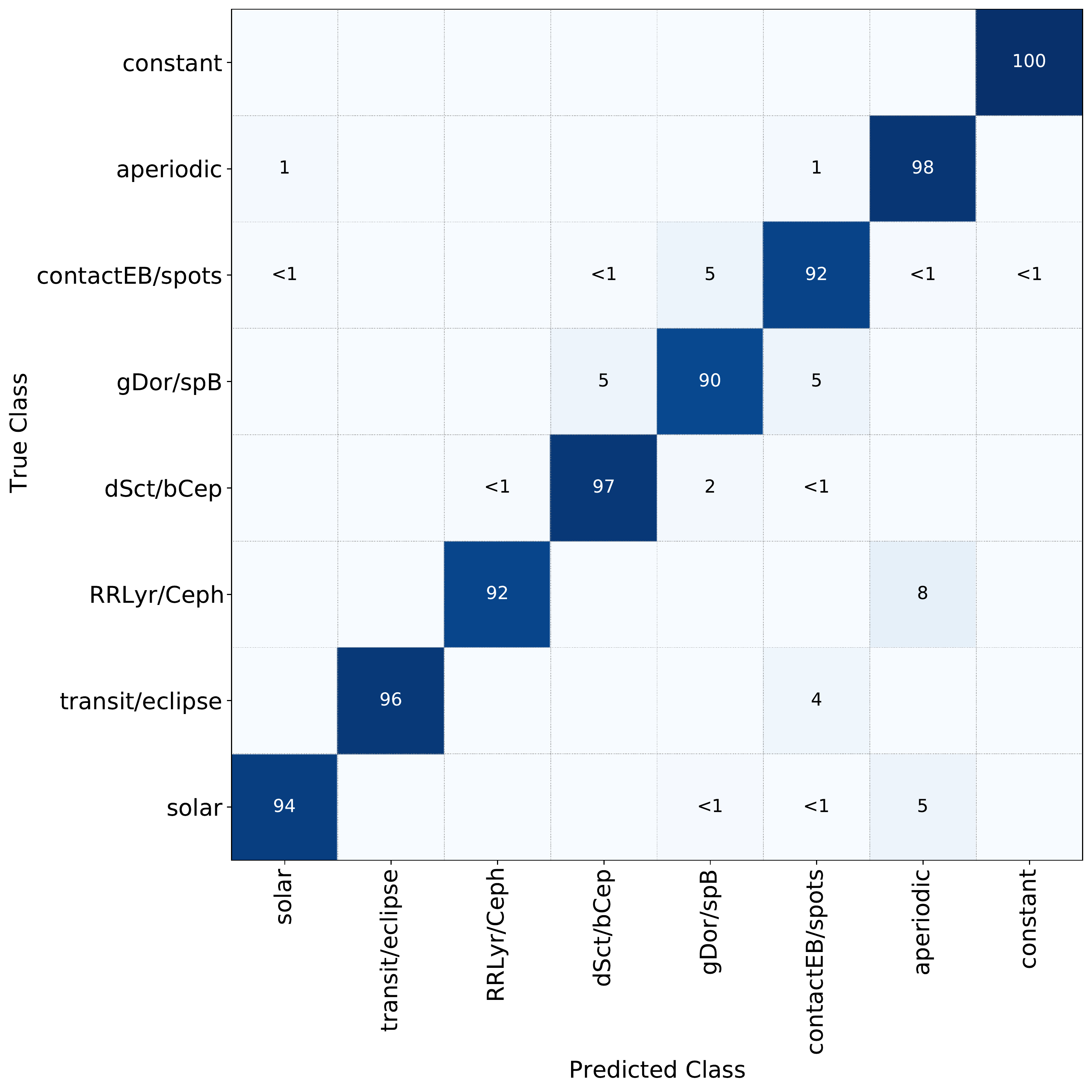}
      \caption{Normalized confusion matrix of the metaclassifier for the holdout set in percentages. Each element shows the fraction of stars that were predicted as positive for a particular class (column) over the total number of stars that truly belong to that class (row). The diagonal shows the fraction of stars that the classifier correctly predicted as positive for that class (i.e. the recall rate = $\frac{TP}{TP+FN}$, where $TP$ is the number of True Positives and $FN$ the number of False Negatives).}
         \label{Fig:confusion-meta}
\end{figure*}
   
\section{The Metaclassifier}\label{sec:meta}

Each of the individual classifiers described in Section~\ref{Sect:Methods} predicts the class probability scores for each light curve. We combine the predictions from this ensemble of classifiers using stacked generalization \citep{Wolpert1992}, in which we turn to a metaclassifier that takes the probabilities outputted by the individual classifiers as its features to produce overall class probabilities for each light curve. This metaclassifier accounts for the relative strengths of the individual strong classifiers in the ensemble (see \citealt{schapire1990} for a description of strong versus weak).
   
\subsection{Training the Metaclassifier}

The stacked nature of our overall classification scheme could lead to overfitting and poor generalization to the unseen TESS data if the classifier is not trained carefully. Our approach to training the metaclassifier and individual classifiers together closely follows Algorithm~19.7 in \citet{Aggarwal2014}. We represent our application of this training algorithm graphically in Figure~\ref{Fig:metatraining}.\footnote{Inspired by the illustration at \url{http://rasbt.github.io/mlxtend/user_guide/classifier/StackingCVClassifier/}} 

As explained in Section \ref{Sect:Test_Validation}, for a given training set, we take 20\% of the data from the start as a holdout set to test the trained ensemble of individual classifiers. We split the remaining data set into $k$ folds and produce class probabilities for each by cross validation. We predict the class probabilities for each fold using the classifiers that are trained on the other $k-1$ folds. We assume that the performance of the classifiers trained across $k$ folds approximates the performance of the models trained on all of the training data. The cross-validated class probabilities from each of the individual classifiers on the training data are the inputs used to train the metaclassifier. The performance of the metaclassifier is finally tested on the holdout data by using the holdout set class probabilities predicted by the individual classifiers trained on the training data (indicated in blue on Fig.~\ref{Fig:metatraining}) as input.

The algorithm we use for the metaclassifier is a Random Forest with a similar architecture to RFGC (see Section~\ref{Sect:RFGC}), but with the number of estimators and maximum tree depth constrained to respectively 100 and 7, to avoid overfitting. This is chosen over a simpler scheme such as majority/soft voting because we want to leverage the potential correlations between classes. The meta classifier, like the individual classifiers, predicts the class probabilities per star. We note that the class probabilities predicted by the metaclassifier are well calibrated, but not perfect. It might thus be better to interpret them as a ranked confidence score rather than in a purely probabilistic fashion. If the metaclassifier assigns a confidence of 0.8 to 100 predictions, we should not expect that exactly 80 of those are correct. However, if we have a star with a confidence of 0.3 and a star with a confidence of 0.7, we can safely assume that the second one has a much higher probability of belonging to the class than the first one.

\subsection{Metaclassifier Testing and Validation}
\label{Subsect:meta-testingval}

The metaclassifier obtains an accuracy of $94.90$\%, a substantial improvement over any single individual classifier. However, given that there are class imbalances in our training set the overall accuracy only provides a limited amount of information. We therefore also look at the confusion matrix as this gives a more detailed view on the metaclassifier's performance per class. The confusion matrix for the metaclassifier is shown in Fig.~\ref{Fig:confusion-meta}. The classification rates (or recall scores) per class range from $90$\% for the $\gamma$~Dor/SPB class to a near perfect score for the constant class. A detailed look reveals that the lower score for $\gamma$~Dor/SPB  is mostly caused by confusion with the $\delta$~Sct/$\beta$~Cep and contactEB/spots class. Our visual analysis shows that the former can be explained by the presence of hybrid pulsators in the training sample, while the latter is caused by $\gamma$~Dor/SPBs containing either some rotational signal or low frequencies that resemble those of contactEB/spots. We also notice some confusion between the aperiodic and contactEB/spots classes. This is mostly caused by the fact that both classes can mimic each other on the short time scale of 27.4 days. Lastly, there is a fraction of solar-like oscillators being predicted as aperiodic variables, where we find that they all have low $\nu_{\rm max}$ values, hence their light curve and power spectrum properties are similar to those of aperiodic stars. The high percentage for RRLyr/Ceph misclassifications is due to the small class size and in absolute numbers only concerns one star.

\begin{figure}
   \centering
   \includegraphics[width=\columnwidth]{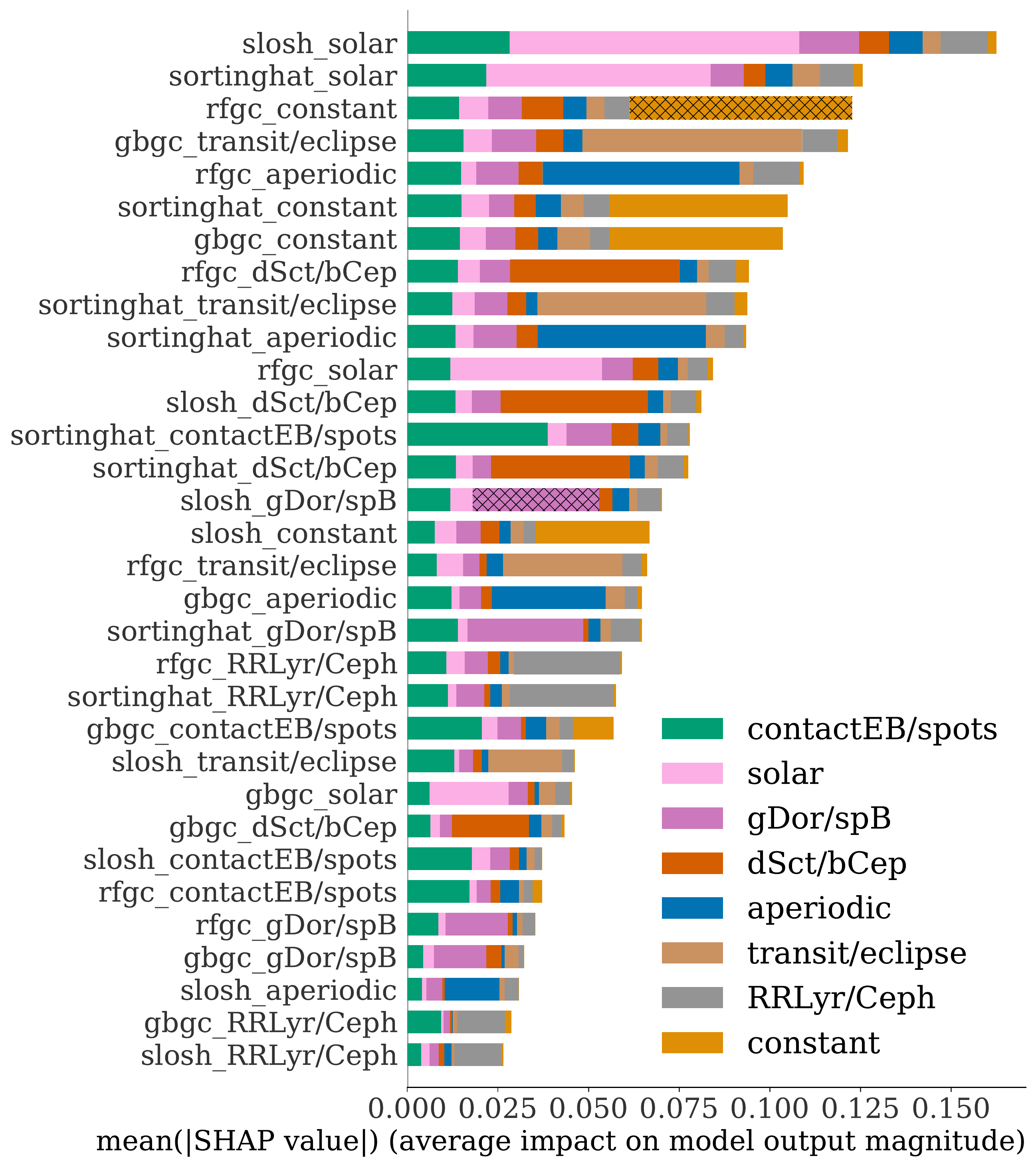}
      \caption{Metaclassifier feature importance from SHAP. The hatched regions indicate the most important feature per class.}
         \label{Fig:meta_shap}
\end{figure}

In Fig~\ref{Fig:meta_shap} we show the feature importance plot for the metaclassifier, which allows us to analyze the contribution of each individual classifier towards the final prediction. The hatched regions indicate the most important feature for a specific class here.
This reveals that the multiSLOSH\_solar probability is the most important feature in the classification of solar-like oscillators, followed by SORTING-HAT. This could be expected given that SLOSH was initially designed to classify this type of star and in the case of SORTING-HAT the entropy features allow it to capture the stochastic nature of the signal. The same order holds for the gDor/SPB class. GBGC is most the important classifer in classifying transit/eclipse signals while it is RFGC for the aperiodic, RRLyr/Ceph, constant and dSct/bCep classes. SORTING-HAT is the primary classifier for contactEB/spots stars. It is interesting, however, that in the contactEB/spots class, SORTING-HAT is followed by the multiSLOSH probability of being a solar-like star. By plotting the SHAP values of every feature for every star, specifically for the solar class, we analyze the impact of each feature on the model output (i.e. the probability of being classified as solar). This reveals that a high multiSLOSH\_solar probability lowers the predicted probability of being a contactEB/spots star and vice-versa. The feature importance plot in Fig.~\ref{Fig:meta_shap} clearly shows that the metaclassifier's strength is in combining the different individual classifier results.

\begin{figure}
   \centering
   \includegraphics[width=\columnwidth]{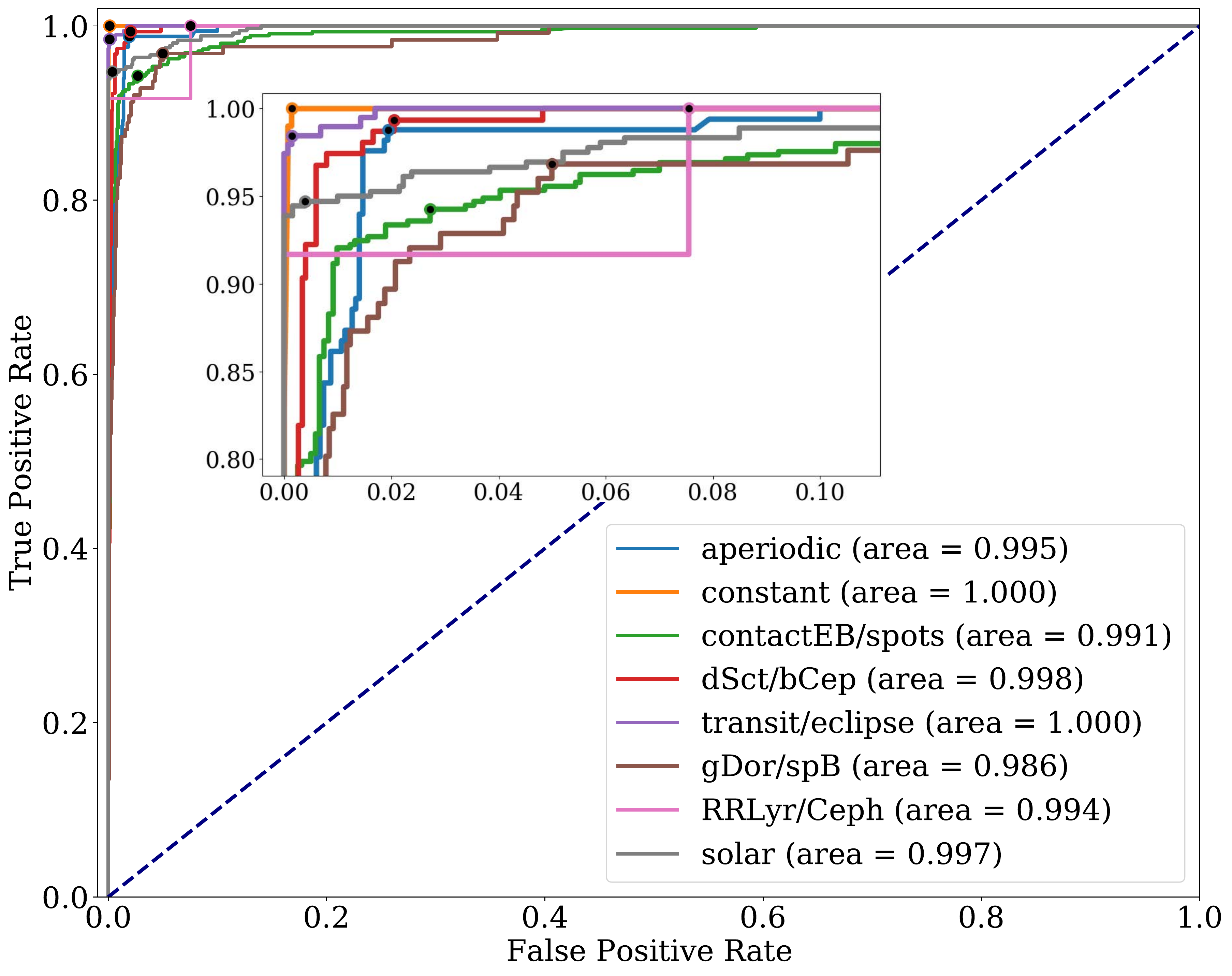}
      \caption{Receiver-Operator Characteristic (ROC) curves of the metaclassifier for each class. The Area Under the ROC (AUROC) curve is indicated next to the classes in the legend and the dots represent the TPR and FPR for the chosen probability thresholds. The dashed line indicates the `random chance' curve.}
         \label{Fig:roc-meta}
\end{figure}

We also assess performance by looking at the receiver-operator characteristic (ROC) curves. The ROC curves illustrate the diagnostic ability of a classifier by plotting the True Positive Rate (TPR) against the False Positive Rate (FPR) for different classification thresholds. This allows us to assess the performance of the classifier at each threshold. We calculate the ROC curves for each class using a one-vs-rest methodology \citep{fawcett2006}. The ROC curves per class are shown in Fig.~\ref{Fig:roc-meta}. Ideally, the ROC curve should be as close to the top left corner (0,1) as possible, because for this threshold on the curve the classifier is making a high number of correct classifications with a small amount of false positives.

Final class labels are commonly assigned to the class with the highest probability, which is equivalent to using a probability threshold of $1/C$, where $C$ is the number of classes. In case more than one of the predicted class probabilities of a star exceeds its respective threshold, the star is assigned to the class with the highest probability. When dealing with class imbalance, however, this $1/C$ approach often does not lead to the optimal results \citep{provost2000}. We therefore opt to fine-tune the classification threshold by choosing for each class the threshold that maximizes the TPR and minimizes the FPR, which is the point on the ROC curve that is closest to the top left corner. This point can be determined by finding the threshold that maximizes Youden's J statistic \citep{youden1950}, which is the difference between the TPR and FPR. Given that we have one ROC curve per class, this implies that we also have a different threshold for each, reflecting the classifier's differing ability in identifying the class members of each variability class. The obtained thresholds per class are given in Table \ref{tab:thresholds}.

As an aggregate performance measure across all probability thresholds used in the ROC curve, we can measure the Area Under the ROC curve (AUROC). The AUROC represents the probability that the classifier assigns a higher probability to a random positive example than to a random negative example. It is thus a measure of how well the classifier predicts the correct class. Given that we are working with a one-vs-rest methodology, it means that the respective ROC class is the positive class and that the other classes belong to the negative class. The confusion between the gDor/SPB and contactEB/spots class causes their AUROC values to be slightly lower compared to the other classes.

\begin{table}
	\centering
	\caption{Classification thresholds per class.}
	\label{tab:thresholds}
	\begin{tabular}{p{3cm} l}
	\hline
	Class label & Probability threshold \\
	\hline
	aperiodic & 0.211 \\
	constant & 0.593 \\
	contactEB/spots & 0.262	 \\
	dSct/bCep & 0.049 \\
	transit/eclipse &  0.157 \\
	gDor/SPB & 0.135 \\
	RRLyr/Ceph & 0.101 \\
	solar & 0.296	\\
	\hline
	\end{tabular}
\end{table}

\begin{table}
    \centering
    \caption{\textit{Kepler} Q9 classification summary: number of stars per class for each thresholding method.}
    \label{tab:keplerq9predictions}
    \begin{tabular}{l c c}
      \hline
      \multirow{2}{*}{Class label} & \multicolumn{2}{c}{\# stars per threshold type}  \\    \cline{2-3}
     & 1/C  & Youden's J \\  \hline
	aperiodic & 3\,711 & 3\,711\\
	constant & 5\,061 &  0 \\
	contactEB/spots  & 140\,566	& 139\,059 \\
	dSct/bCep & 1\,758 & 1\,758\\
	transit/eclipse &  1\,563 & 1\,563\\
	gDor/SPB & 2\,263 & 2\,263\\
	RRLyr/Ceph & 96 & 96\\
	solar & 12\,225	& 12\,185\\
	\textit{unknown}  &  & 6\,608 \\
      \hline
    \end{tabular}
\end{table}

\begin{figure*}
   \centering
   \includegraphics[width=18cm]{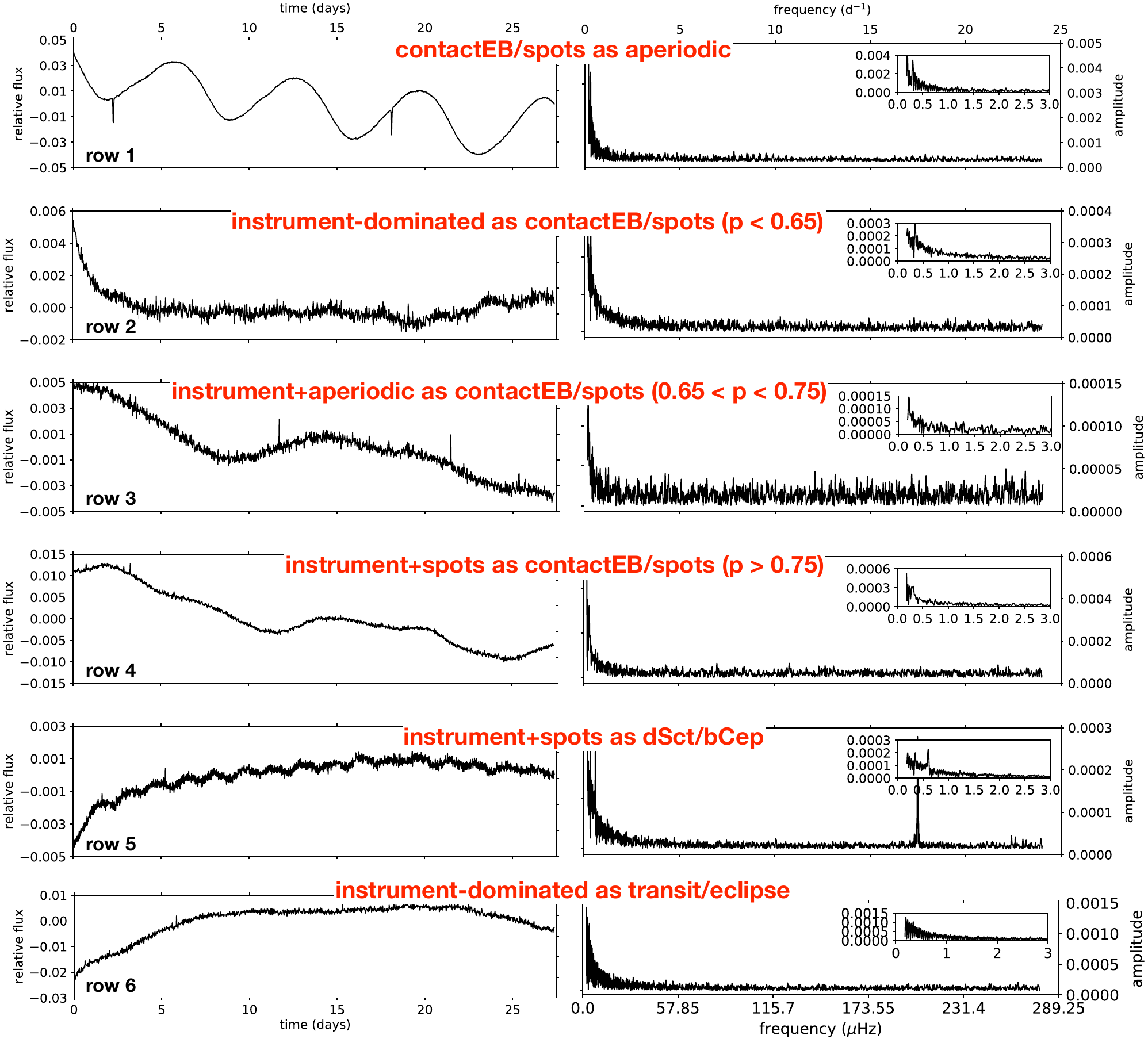}
      \caption{Examples of the misclassified \textit{Kepler} Q9 light curves. Left and right columns show the light curves and the amplitude spectra, respectively. Note the different scale on the Y-axis of the plots.}
      \label{Fig:Q9Misclassifications}
\end{figure*}

\begin{figure*}
   \ContinuedFloat
   \centering
   \includegraphics[width=18cm]{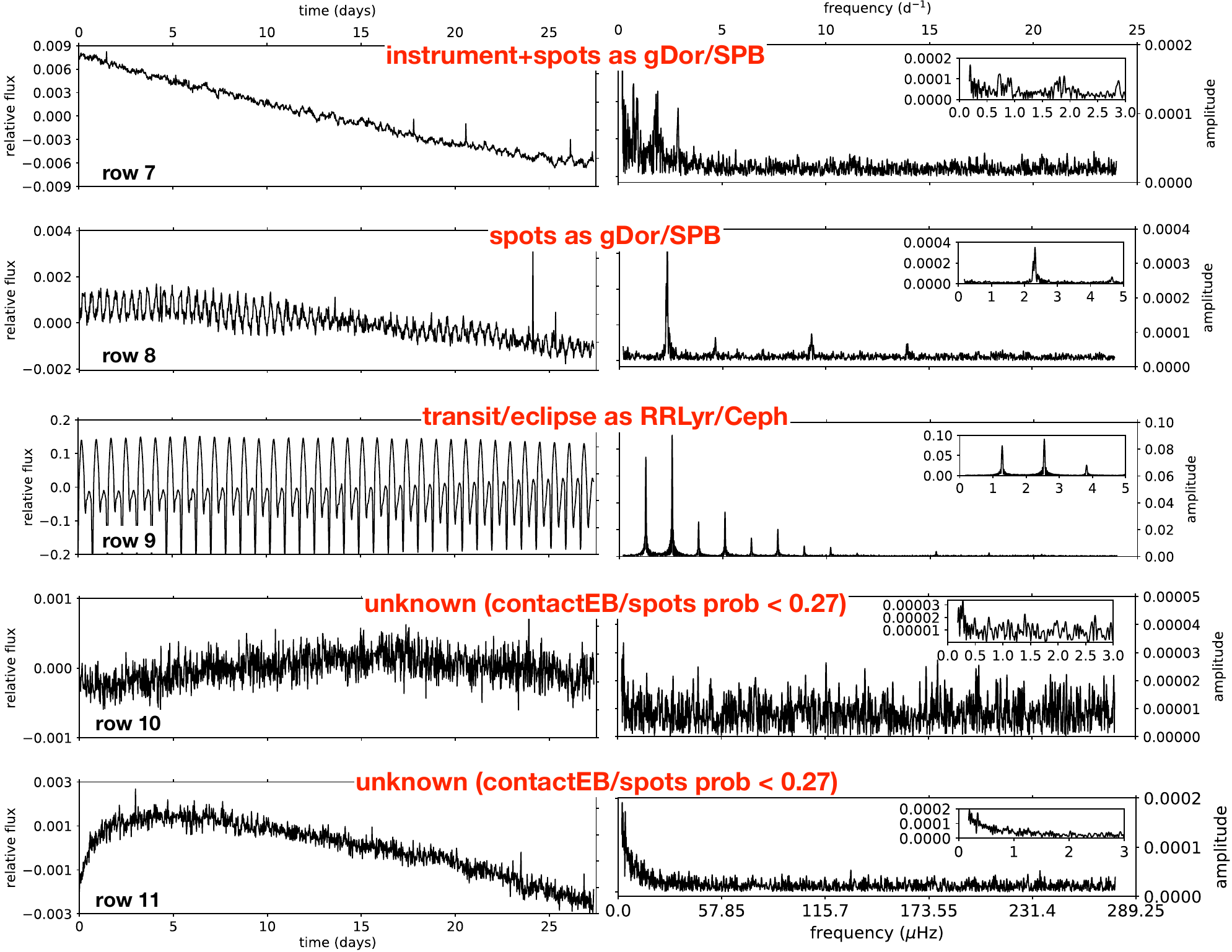}
      \caption{Continued}
      \label{Fig:Q9Misclassifications}
\end{figure*}

\section{Validation on full \textit{Kepler} Q9 data set}\label{sec:Kepler Q9 application}
We validate our classification scheme by applying it to all 167\,243 stars observed in \textit{Kepler} Q9, but with their light curves cut to the first 27.4 days. We start with the default methodology as described in the previous sections, then test the effect of linear detrending, and ultimately assess the advantage of introducing an instrumental class. For each of those additional scenarios, we assess both the results on the holdout set and on the \textit{Kepler} Q9 data set. The assessment is achieved by the summary statistics of both data sets and by visually inspecting random sub-samples of 1000 light curves in each class for the \textit{Kepler} Q9 classification results.\footnote{Before taking the random sample we first removed the stars that were included in the training set.} 

\subsection{The default scenario}
\label{Sec:Q9 base scenario}

The results obtained by applying our framework to \textit{Kepler} Q9 data are summarized in Table~\ref{tab:keplerq9predictions}. The left column lists the numbers for the label assignments being made according to the highest probability, while the right column gives those according to the optimized probability thresholds. Overall, we see that all predicted classes, apart from contactEB/spots, have classification rates similar to those in the confusion matrix in Fig.~\ref{Fig:confusion-meta}. The high number of stars in the contactEB/spots class can be explained by the fact that light curves that are not assigned to any of the other classes, for example with a dominant instrumental signal in the low-frequency domain, end up in this bin. A careful visual inspection of the light curves and amplitude spectra of 1000 random sub-samples in each class strengthens the above conclusion; below we present a concise summary of our visual analysis.

The \underline{aperiodic} variables are identified robustly by our methodology with an overall low number of misclassifications. After inspecting 1000 light curves randomly selected from the respective class, we confirm that some 97\% of the light curves indeed exhibit aperiodic type variability as demonstrated in the first row in Fig.~\ref{Fig:TrainingSet}. The most common misclassifications (about 3\% in total) belong to the 
contactEB/spots class and are the light curves resembling rotational modulation and/or binary ellipsoidal signals, in many cases with the coverage of a single rotation/orbital period. The median probability value for misclassifications is found to be $p(x)\approx 0.50$. The worst-case scenario misclassification light curve is shown in Fig.~\ref{Fig:Q9Misclassifications} (first row) where likely a close eclipsing binary got (mis)classified as an aperiodic variable.

\begin{figure}
   \centering
   \includegraphics[width=\columnwidth]{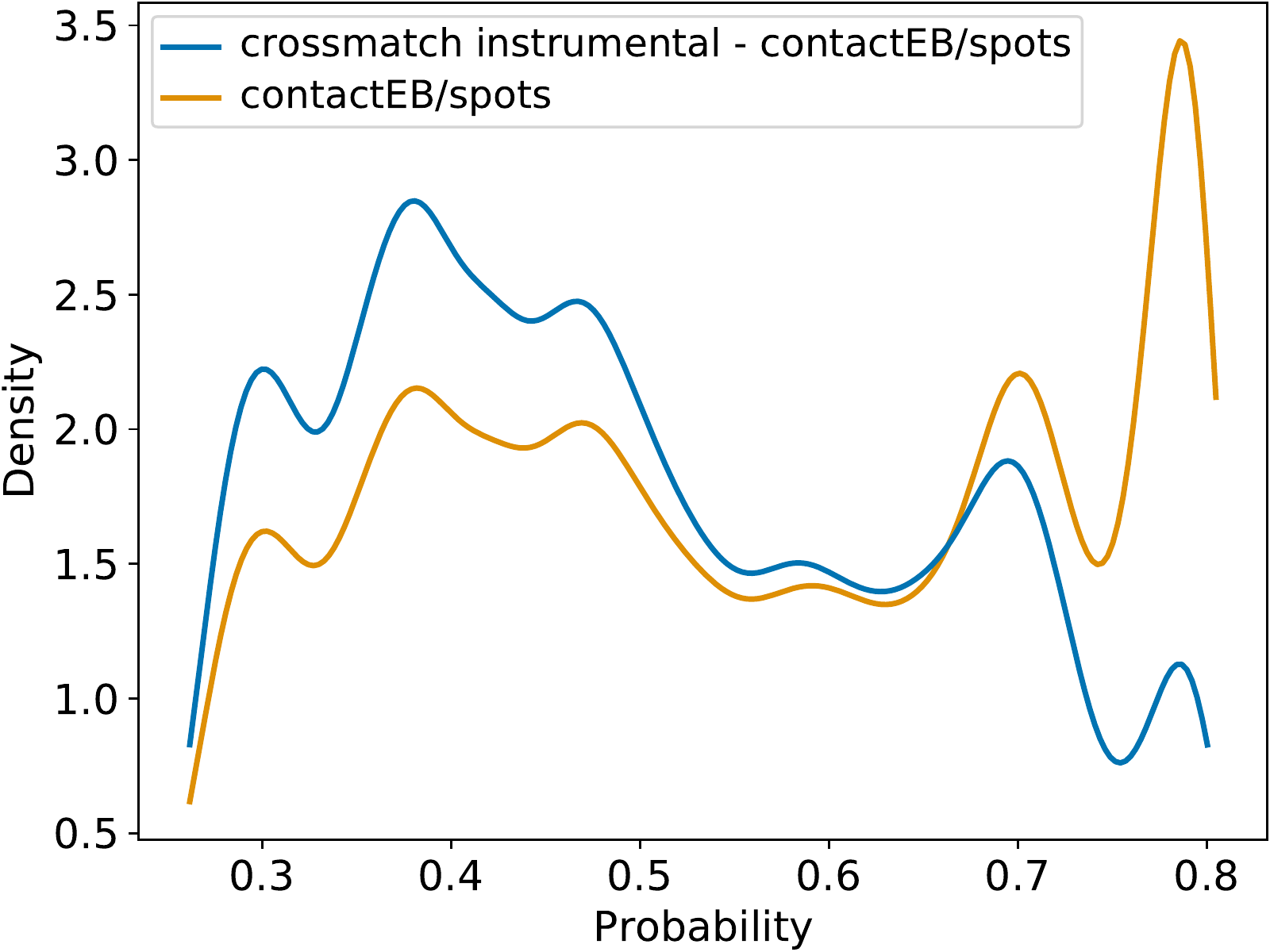}
      \caption{Kernel density estimate (KDE) plot comparing $p_{\text{contactEB/spots}}^{\text{default scenario}}(x)$ of the instrumental class cross-matched to the original (i.e. from the default set-up) contactEB/spots class, against the complete original contactEB/spots class, with class assignments based on Youden's J statistic (see Sect.~\ref{Subsect:meta-testingval} for a description of Youden's J). Stars from the training set have been subtracted from both sets.}
         \label{Fig:kde-contactinstr}
\end{figure}

The \underline{contactEB/spots} class suffers the most from misclassifications and partially resembles properties of miscellaneous classes often employed by other light curve classification methods \citep[e.g.,][]{debosscher2011}. Fig.~\ref{Fig:kde-contactinstr} (orange line) shows the probability density function for the contactEB/spots class. We immediately notice an excess of objects in the low probability regime ($p(x)\lesssim 0.55$) as well as a double-peak feature at high probabilities ($p(x)\gtrsim 0.65$). Owing to this distribution, we divide the contactEB/spots class into three probability bins and visually inspect 500 randomly selected light curves in each of them: $i$) $p(x) < 0.65$, $ii$) $0.65 < p(x) < 0.75$, and $iii$) $p(x) > 0.75$. We find that the lowest probability bin ($p(x) < 0.65$) contains some 97\% misclassifications. Among those the dominant fraction (about 90\%) are light curves that exhibit some sort of instrumental signal (see the second row in Fig.~\ref{Fig:Q9Misclassifications}). This can be either truly instrumental in origin or due to inferior data processing. The intermediate-probability bin that is associated with the first peak in the kernel-density plot ($0.65 < p(x) < 0.75$, Fig.~\ref{Fig:kde-contactinstr}) is also found to be rich in misclassifications (overall about 92\%). Yet, the major difference with the low-probability regime is that the fraction of light curves that exhibit pure instrumental signal is significantly lower, at around 55\%. In the rest of the light curves, the instrumental and the true astrophysical signals are found to co-exist, as illustrated in the third row in Fig.~\ref{Fig:Q9Misclassifications}. In this particular example, a weak astrophysical aperiodic signal (on the time scale of 27.4 days) co-exists with low-frequency signal due to inferior data processing. Lastly, the highest probability bin associated with the tallest peak in the probability density function ($p(x) > 0.75$, Fig.~\ref{Fig:kde-contactinstr}) contains 16\% misclassifications that are pure instrumental in origin. About 34\% show both instrumental and astrophysical signals. We note that the latter are not necessarily misclassifications, it is just that we visually identify the instrumental signal as being the dominant one in the respective light curves (the fourth row in Fig.~\ref{Fig:Q9Misclassifications}). Finally, we note that pure astrophysical misclassifications are dominated by aperiodic variables and is at the level of some 17\%. Many of those seemingly aperiodic signals might in fact be rotational variables with periods longer than 13.7 days and therefore spanning less than half of the rotation cycle. Hence they are visually classified as aperiodic
stars. That said, we recommend a probability threshold of $p(x)\gtrsim 0.75$ for the high-confidence selection of contactEB/spots variables for this class (an example is shown in the second row in Fig.~\ref{Fig:TrainingSet}). This threshold deviates from the one calculated using Youden's J statistic because our training set is largely free from any instrumental signal, while this type of signal is present in the full \textit{Kepler} Q9 data (we will elaborate on this point in sect.~\ref{subsect:instr}), resulting in a suboptimal contactEB/spots threshold. The results with probabilities $p(x)\lesssim 0.75$ should be taken with caution and one should keep in mind that the number of genuine astrophysical signals in this particular class drops substantially towards low probability values.

The \underline{dSct/bCep} variables are identified with high confidence by our methodology, where the overall fraction of misclassifications amounts to some 3.5\%. The vast majority of misclassifications are due to spurious frequencies found in the high-frequency domain (the fifth row in Fig.~\ref{Fig:Q9Misclassifications}). We consider those frequencies as spurious because exactly the same frequencies are found in multiple objects indicating their non-astrophysical origin. At the same time, we did not find any indication of these particular frequencies being instrumental in nature as those are not listed as such the latest $Kepler$ Data Release Notes. The median probability value for the misclassified light curves is $p(x)\approx 0.45$. A considerable fraction of the identified $\delta$~Sct stars also exhibit low-frequency variability (either due to g-mode oscillations or rotational modulation), yet the high-frequency component is significant in all the detections and is the dominant one in the majority of them.

The \underline{transit/eclipse} class is among the cleanest identified with our method, containing about 10\% misclassifications overall. All misclassifications look alike and are due to imperfections in the data processing mimicking a flux drop in the light curve, most often at the beginning/end of the dataset. A typical example of the transit/eclipse misclassification is shown in the sixth row in Fig.~\ref{Fig:Q9Misclassifications}. We also note that the type of light curve shown in the fifth row in Fig.~\ref{Fig:Q9Misclassifications} has high chances of being misclassified as a transit/eclipse variable, in absence of the high-frequency peak. We find the median probability value for the misclassifications to be $p(x)\approx 0.45$.

The \underline{gDor/SPB} class suffers from about 30\% misclassifications, either from astrophysical signal of different origin or from low-frequency signal due to imperfections of data processing. The median probability value for misclassifications appears to be $p(x)\approx 0.52$ and the most common astrophysical misclassifications are stars that belong to the contactEB/spots class. We show a typical example of a misclassified light curve in the seventh and eighth rows in Fig.~\ref{Fig:Q9Misclassifications}. The Fourier transform of the light curve reveals a rich variability spectrum at low frequencies, possibly with a harmonic structure. Owing to the frequency range of gravity-mode oscillations observed in $\gamma$~Dor/SPB stars and to the short time span of light curves that are being classified (see the fifth row in Fig.~\ref{Fig:TrainingSet}), contactEB/spots class members are indeed the primary candidates for an astrophysical misclassification in the gDor/SPB class. We also note that the particular example shown in the seventh row in Fig.~\ref{Fig:Q9Misclassifications} might not be the actual misclassification but is identified by us visually as such because of a short duration of the respective light curve.

The \underline{RRLyr/Ceph} class of classical pulsators is found to be small (see Table~\ref{tab:keplerq9predictions}) which is expected given the location of the \textit{Kepler} field. Misclassifications amount to about 60\%, are mostly astrophysical in origin, and are dominated by contactEB/spots or transit/eclipse class members. All three types of objects (including RRLyr/Ceph) have their dominant signals in the low-frequency domain and will often show a harmonic structure. However, owing to the characteristic shapes of their light curves (as shown in the sixth row of Fig.~\ref{Fig:TrainingSet}), RRLyr/Ceph stars are usually readily distinguished from other classes of variable stars. Indeed, only a small fraction of binaries and/or rotational variables with light curves that closely resemble those of the RRLyr/Ceph class are expected to exist and hence get misclassified to the class of classical pulsators. The high relative number of binaries and rotational variables that we find in this class is very likely the result of the contactEB/spots and transit/eclipse classes being at least two orders of magnitude larger than the RRLyr/Ceph class itself, hence increasing the chance of misclassification. An example of the light curve and amplitude spectrum of an eclipsing binary misclassified as RRLyr/Ceph is shown in the ninth row in Fig.~\ref{Fig:Q9Misclassifications}.

The \underline{solar} class is the second largest. The number of misclassifications is found to be small in this class (about 4\%) and is mostly instrumental in origin. The median probability value for misclassifications is found to be $p(x)\approx 0.55$. 

The \underline{unknown} class contains objects that do not satisfy the Youden's J statistics-based thresholds per class as listed in Table~\ref{tab:thresholds} (see Sect.~\ref{Subsect:meta-testingval} for a description). By comparing the class sizes before and after applying the thresholds (see Table~\ref{tab:keplerq9predictions}) we notice that the unknown class comprises the entire class of constant stars as well as a small fraction of the lowest probability objects from the contactEB/spots class. The constant class gets marked as unkown due to the fact that the calculated probability threshold of $0.593$ is very high. This happens because the classifier achieves a near perfect classification rate for the constant class (see Fig.~\ref{Fig:confusion-meta}) and is thus very confident in classifying stars as such. Therefore, during testing, stars are either predicted not belong to the constant class at all (i.e. $p(x)\approx 0$) or they are confidently classified as constant. In the latter case, the lowest probability (which is still high in absolute terms) of a star that is being classified as constant is set as the threshold (due to the mathematical calculation). In reality, however, it appears that there are no stars as distinctly constant as in our training set. This makes sense given that this class is simulated in the training set while existence of constant stars is not assured. 99\% of objects found in the unknown class do not show any significant astrophysical signal, with two typical examples being shown in the two bottom rows (10 and 11) in Fig.~\ref{Fig:Q9Misclassifications}.

\subsection{Additional classification set-ups}
\label{subsect:instr}

We also test the effect of automatically removing a linear trend from all \textit{Kepler} 27.4 days light curves prior to computing the Fourier- and time-domain features. The results on the holdout set and \textit{Kepler} Q9 show no performance gain over the default set-up, hence we stick to using the original light curves. We do not test detrending with higher degree polynomials because this can have undesirable effects on the classification as $i$) the original light curves may be significantly distorted, and $ii$) the signal of long-period variables may be largely filtered out during the process.

One of the key findings from running our default set-up is that the contactEB/spots class is largely overpopulated, with a clear tendency for a large number of misclassifications towards the low probability values by light curves containing some sort of an instrumental signal. We note that we use the term ``instrumental signal'' to mark a signal that is either truly instrumental in origin or is the result of sub-optimal detrending/correction of the data. To overcome the above-mentioned drawback, we opt to introduce an instrumental class with properties resembling those of light curves affected by the instrumental trends and/or sub-optimal data processing. We use a sub-sample of the contactEB/spots class light curves whose probability values were found to be of $p(x)\lesssim 0.65$ to manually select a training set for the instrumental class based on the visual inspection of the light curves. To preserve the balance with other variability classes in the training set a total of about 1100 light curves are selected.

The most notable differences after introducing the instrumental class are i) a considerable reduction of the size of the contactEB/spots variability class by about a factor 3.5, and ii) a much smaller size of the unknown class. This happens because the originally low-probability (lower than the respective thresholds reported in Table~\ref{tab:thresholds}) objects in the various classes are now classified with high confidence as members of the newly introduced instrumental class. Hence there are considerably less candidates to feed the unknown class in the latter.

Furthermore, we cross-match the newly obtained instrumental class with the original contactEB/spots class and find about 70\% of overlap. The probability density plot for the cross-matched sample is shown in Fig.~\ref{Fig:kde-contactinstr} (blue line) where the distribution is evidently skewed towards low probabilities. Therefore, we conclude that introducing an instrumental class does not necessarily improve the overall performance of the method, instead a considerable fraction of light curves that receive low confidence values in their respective classes are moved to the class of instrumental variables.

While not having clear advantages, the disadvantages of introducing an instrumental class are that it is not only instrument-dependent but that it is also extremely sensitive to the way data from a given instrument are being processed. Therefore, an instrumental class proves impractical as it has to be re-designed each time the data from a given instrument are being reprocessed and/or the methodology is being applied to data from a different instrument. A much more practical solution is the one outlined and employed in Sect.~\ref{Sec:Q9 base scenario}, i.e. a recommended probability-based threshold to separate high-confidence detections of genuine contactEB/spots variables from their low-confidence counterparts that are most likely not astrophysical in origin.

\subsection{Effects of photon noise}
\label{Subsect:photon-noise}

The level of recognizable astrophysical signal in a light curve is tightly related to the amount of photon noise present in it, which depends on the stellar magnitude. Given that we did not exert any particular influence on the distribution of the magnitudes in the training set, we test the sensitivity of the metaclassifier with respect to increasing photon noise. As for the relation between magnitude and noise level, we base ourselves on the 10th percentile RMS CDPP (Combined Differential Photometric Precision) measurements presented in the TESS Data Release Notes: Sector 5, DR7. We multiply the values by $\sqrt{2}$ to account for the 30-min sampling.

For each class in the holdout set, we select the 20 stars with the highest probability and remove all stars with \textit{Kepler} magnitude $< 15$. We leave out the constant stars given that they are simulated white noise already. We then add noise to the sampled light curves in steps of 0.5 magnitude, with the maximum number of steps restricted to 8, which is equivalent to a magnitude 4 increase. The added noise is Gaussian with mean zero and standard deviation equal to 30-min CDPP value for the desired magnitude. The new equivalent magnitude is constrained to be brighter than 15.5. If this is reached before the maximum number of allowed steps, no further noise additions are done for the star. We choose this constraint because the T'DA Photometry pipeline \citep{handberg2021} will process TESS stars down to magnitude 15.

Once we have calculated the new noisier light curves, we classify those in each step with the metaclassifier and analyze how the overall predictions change with added photon noise. Fig.~\ref{Fig:sankeyplot} shows how stars move between the different classes when more noise is added to their light curves. The relatively brightest stars are shown on the left and relatively faintest on the right. The colors of the streams indicate the true variability class, and the bars indicate the predicted class. The height of the bars corresponds to the number of stars in that bin. The number of stars decreases from left to right because stars are eliminated once their new equivalent magnitude exceeds the 15.5 threshold. In Fig.~\ref{Fig:violinplot} we show how the magnitudes evolve over the different steps. We start with 115 stars on the left and end up with 36 in the rightmost bin.

We can see from Fig.~\ref{Fig:sankeyplot} that when the noise level increases ($i$) the majority of solar-like stars get classified as constant ($ii$) a signficant fraction of contactEB/spots star get classified as constant and ($iii$) most aperiodic stars end up being classified as contactEB/spots. The reason for ($i$) is physical in origin and results from the fact that the added noise is much larger than the oscillations in the original light curve, causing the new light curves to be dominated by white noise and get classified as such. The solar class is also the most varied one in terms of time scales and so the location of the oscillations dictates to which bin a star moves into when adding noise, causing some of them to be classified as other variability types as well. In a similar manner, we can see that the contactEB/spots stars that get classified as constant ($ii$) are actually cool and spotted stars in which the noise becomes larger than their oscillations. Only the hot and chemically peculiar stars with high amplitude variability that is stable on longer time scales survive. The reason for ($iii$) can be attributed to a training set bias and occurs because aperiodic and contactEB/spots stars can mimic each other on 27.4 day time scales, and because in our training set the contactEB/spots class tends to be more noisy than its aperiodic counterpart, causing these stars to be classified as such.

When we connect these findings to the magnitude distribution of our full training set (Fig.~\ref{Fig:kde-mags}), we conclude that one should be careful when interpreting the predicted probabilities of stars that do not lie within the magnitude range of the training set. More specifically, a decreasing magnitude for stars that are part of the solar, contactEB/spots or aperiodic class corresponds to an increasing uncertainty over their probabilities. Hence, when interpreting the results, it is important to, in addition to the assigned probabilities, also look at the magnitude of the target. If the magnitude is much fainter than those of the training samples and it belongs to one of these three classes, caution should be paid when interpreting the results. For other classes, such as transit/eclipse, this effect is not present because the amplitude of the signal is often much larger than the added noise. It is thus important to note that a bright star, even in the case of ($i$), ($ii$) and ($iii$), does not always mean that there is a very clear signal while a faint stars does not necessarily mean we have an indistinguishable signal. It is the amplitude of the signal relative to the noise that matters.

\section{Discussion and Conclusions}
\label{sec:conclusion}

\subsection{Summary and Discussion}
\label{Subsect:summary}

The TESS Data for Asteroseismology pipeline is designed for a largely automated processing and high-level interpretation of TESS space-based photometric data. As depicted in Fig.~\ref{Fig:TDA-pipeline}, the first two modules of the pipeline are designed for the extraction of light curves from the TESS Full Frame Images \citep{handberg2021} and for their subsequent optimal correction for systematic effects (Lund et al. in prep.). In this work, we have designed a third module of the T'DA pipeline that performs an automated classification of the corrected light curves according to their type of variability.

We combine four individual classifiers into a metaclassifier using stacked generalization. Out of the four individiual classifiers,  RFGC \citep[][Sect.~\ref{Sect:RFGC}]{Armstrong:2016br} and SLOSH \citep[][Sect.~\ref{Methods:multiSLOSH}]{Hon_2018} have been previously published, while SORTING-HAT (Sect.~\ref{Sect:sorting-hat}) and GBGC (Sect.~\ref{Sect:MethodsGBGC}) were additionally developed to enhance the T'DA pipeline classification module appreciably. We show that by stacking the predictions of this set of different individual classifiers we obtain a substantial improvement over any single one, because the metaclassifier is able to learn their relative strengths. We are able accurately classify light curves according to their general variability type, without relying on any extra information other than the light curves.

Although inspired by the amount of TESS data currently collected, our ultimate goal is to design an automated pipeline for the end-to-end processing of high-cadence and duty-cycle space-based photometric data, irrespective of whether these come from the retired CoRoT and \textit{Kepler}/K2 missions, currently operational TESS mission, or future space-missions such as PLATO \citep{Rauer2014}. Hence, in this work, we make use of the \textit{Kepler} mission legacy, both in terms of the available high precision, cadence, and duty-cycle data and the published catalogs of variable stars, to train, validate and test our classifiers. The training set is carefully built from the existing catalogs with a subsequent vetting of light curves in all eight variability classes used in our classification scheme. All individual classifiers as well as the metaclassifier are trained on 80\% of the compiled training set, while the remaining 20\% are kept as a hold-out set to test and validate the method. We obtain an overall accuracy of $94.9\%$ on the holdout set with some small differences between the different classes. 

We further apply the designed classification scheme to the \textit{Kepler} Q9 data set that has been truncated into 27.4 days light curves. In addition to testing our default classification set-up, we also test the effect of linear detrending and the introduction of an extra instrumental class to isolate light curves dominated by the instrumental signal. We show that although the latter allows for a significantly lower number of misclassifications of the sub-optimally processed light curves in some of the classes, it has the disadvantage that the instrumental class has to be re-designed each time the method is applied to the re-processed data from the same space-mission and/or data obtained by another mission.

Given that we currently use 27.4 days light curves, one of the expected and detected (astro)physical limitations of our method are apparent misclassifications of objects whose variability on a 27.4 days time scale does not necessarily resemble their true origin. A common example is non-resolved rotational variability in cool stars that gives rise to an overdensity of low frequencies in the Fourier transform of the light curve causing a confusion with the class of g-mode pulsators and/or aperiodic variables. Another example is the flux drop in a light curve due to sub-optimal data processing which mimics a single transit/eclipse event in the time-domain and gives rise to a misclassification as a transiting/eclipsing object. Other than that, we find that the predicted classes have classification scores similar to those in the confusion matrix based on the hold-out validation set (see Fig.~\ref{Fig:confusion-meta}).

Generalizing our framework to TESS will still require adjustments because we are currently training our classifiers on \textit{Kepler} data. Not only does \textit{Kepler} have a different underlying distribution compared to TESS, possibly requiring domain adaptation techniques, TESS also has a worse photometric precision (and hence more noise), more blending, and more systematics that we cannot characterize very well yet. That said, there is no one-to-one correlation between the results obtained based on \textit{Kepler} data in this work and the expectations for TESS data. In order words, we cannot simply extrapolate the results of the performance of our classifiers to TESS data prior to exploring domain adaptation, performing initial classification of TESS dataset, and ultimately (re)training based on the actual TESS data. We note, however, that the performance will not necessarily drop when transitioning to TESS data, it can also be as high or higher than in this work.

We make both the methodology and the results of its application to the \textit{Kepler} Q9 27.4 days data using the default set-up publicly available to the community. Our training set, individual classifiers, and the metaclassifier can be accessed through the dedicated GitHub repository\footnote{\url{https://github.com/tasoc/starclass}} as well as through the TESS Asteroseismic Consortium (TASOC) Wiki pages\footnote{\url{https://tasoc.dk/tda/}}. The predicted class probabilities and class labels for the \textit{Kepler} Q9 27.4 days are released in electronic format; a snippet of the class probabilities table is shown in the Appendix (Table.~\ref{tab:classprob_table}).

\subsection{Future prospects}

With the machinery built, our immediate future prospects include:
\begin{itemize}
    \item Classification of all \textit{Kepler} stars based on $i$) 1-year data to mimic TESS Continuous Viewing Zone (CVZ) operations and enable direct comparison with the results presented in this work; $ii$) 2-year data to mimic PLATO Long Pointing Field (LPF) operations enabling an important set of tests for the PLATO Consortium; and $iii$) 4-year data to provide a full \textit{Kepler} classification catalogue and quantitatively assess performance of our method on ultra-high precision data by cross-matching with the existing \textit{Kepler} catalogues. At this step, we will consider using extra information, such as photometric colours, Gaia parallaxes, etc., in order to break the existing degeneracies within and between the individual variability classes. This particular step covers our intended ``second-level classification'' (as depicted in Fig.~\ref{Fig:ClassificationScheme}) where we aim to distinguish between different evolutionary states of solar-like pulsators (dwarfs vs. sub-giants, RGB stars vs. red-clump stars), between sub-groups of g- ($\gamma$~Dor vs. SPB variables) and p-mode ($\delta$~Sct vs. $\beta$~Cep stars) pulsators, etc.
    \item Inclusion of a learning algorithm capable of identifying transient phenomena, such as stellar flares, Be star outbursts, etc. For this, we will consider existing algorithms such as STELLA\footnote{\url{https://archive.stsci.edu/hlsp/stella}} \citep{Feinstein2020} which will be adapted to the needs of our metaclassifier similarly to the RFGC and multiSLOSH methods.
    \item Inclusion of an unsupervised learning algorithm to help identify misclassifications and search for overdensities in the feature space within the identified supervised classification module variability classes. This particular step is depicted in Fig.~\ref{Fig:ClassificationScheme} as the ``unsupervised methods'' box and will strengthen our classification scheme by allowing for the detection of additional variability (sub)classes.
    \item Inclusion of statistical features for an improved classification of aperiodic autocorrelated signals. For this, we will consider tests such as the Durbin-Watson statistic for serial autocorrelation and the Kullback-Leibler divergence to measure the disparity against white noise.
    \item Transition to TESS data that are processed with the corresponding T'DA pipeline light curve extraction and systematics correction modules. At this step, we also envision an iteration between all three modules of the T'DA pipeline, in particular to inform the light curve correction algorithms on the variability time-scales that should be preserved rather than removed for specific classes of objects. In terms of the corresponding data releases, we plan them jointly with the light curves themselves on a per sector basis and will make our results publicly available through the MAST and TASOC databases. The accompanying TESS classification papers for the nominal missions are also foreseen and will be based on the full year of TESS data, i.e., per TESS observational hemisphere.
    \item Integration of the variability catalog into the TASOC database search-interface\footnote{\url{https://tasoc.dk}}. This interface will allow for a quick and convenient search of variable stars according to user-defined selection criteria. As concrete examples, one will be able to opt for an all-sky search of $\delta$~Sct variables that have been identified with a user-defined confidence with our classifiers, or stars classified with probability in multiple classes (e.g. both $\delta$~Sct and eclipsing binary).
    \item Inclusion of the full variability catalog on both TASOC and MAST as an new high-level data product.
\end{itemize}

\begin{acknowledgments}
The research leading to these results has received funding from the European Research Council (ERC) under the European Union's Horizon 2020 research and innovation programme (grant agreement N$^\circ$670519: MAMSIE), from the KU~Leuven Research Council (grant C16/18/005: PARADISE), from the Research Foundation Flanders (FWO) under grant agreement G0H5416N (ERC Runner Up Project), as well as from the BELgian federal Science Policy Office (BELSPO) through PRODEX grant PLATO. D.J.A acknowledges support from the STFC via an Ernest Rutherford Fellowship (ST/R00384X/1). Funding for the Stellar Astrophysics  Centre is provided by The Danish  National Research Foundation (Grant agreement no.: DNRF106). RH and  MNL acknowledges the ESA PRODEX  programme. This research was supported by the National   Aeronautics and Space Administration (80NSSC18K1585 and 80NSSC19K0379) awarded through the TESS Guest Investigator Program. K.J.B. is supported by the National Science Foundation under Award AST-1903828. J.S.K and K.J.B. were supported by  funding from the European Research Council under the European Community’s Seventh Framework Programme (FP7/2007-2013) / ERC grant agreement no. 338251 (StellarAges). DMB gratefully acknowledges funding from a senior postdoctoral fellowship from the Research Foundation Flanders (FWO) with grant agreement No. 1286521N. The research leading to these results has received funding from the Research Foundation Flanders (FWO) under grant agreement G0A2917N (BlackGEM). R.A.G. acknowledges the support from the GOLF and PLATO CNES grants. L.M. and E.M. were supported by the Premium Postdoctoral Research Program of the Hungarian Academy of Sciences. The research leading to these results has been supported by the Hungarian National Research, Development and Innovation Office (NKFIH) grant KH\_18 130405 and the Lend\"ulet LP2014-17 and LP2018-7/2020 grants of the Hungarian Academy of Sciences. D.B. acknowledges support from the NASA TESS Guest Investigator Program under award 80NSSC19K0385.

This paper includes data collected by the TESS mission, which are publicly available from the Mikulski Archive for Space Telescopes (MAST) and described in Jenkins et al. (2016). Funding for the TESS mission is provided by NASA’s Science Mission directorate. This research has made use of NASA’s Astrophysics Data System, as well as the NASA/IPAC Extragalactic Database (NED) which is operated by the Jet Propulsion Laboratory, California Institute of Technology, under contract with the National Aeronautics and Space Administration.
Funding for the TESS Asteroseismic Science Operations Centre is provided by the Danish National Research Foundation (Grant agreement no.: DNRF106), ESA PRODEX (PEA 4000119301) and Stellar Astrophysics Centre (SAC) at Aarhus University. We thank the TESS team and staff and TASC/TASOC for their support of the present work.

This paper includes data collected by the Kepler mission. Funding for the Kepler and K2 mission was provided by NASA's Science Mission Directorate. The authors acknowledge the efforts of the Kepler Mission team in obtaining the light curve data and data validation products used in this publication. These data were generated by the Kepler Mission science pipeline through the efforts of the Kepler Science Operations Center and Science Office. The Kepler light curves are archived at the Mikulski Archive for Space Telescopes.

The numerical results presented in this work were obtained at the Centre for Scientific Computing, Aarhus\footnote{\url{https://phys.au.dk/forskning/cscaa/}}. This research  made  use  of  Astropy, a community-developed core Python package for Astronomy \citep{Astropy:2013,Astropy:2018}.
\end{acknowledgments}

\software{Scikit-learn \citep{scikit-learn}, Numpy \citep{numpy2020}, Astropy \citep{Astropy:2013,Astropy:2018}, Scipy \citep{scipy2020}, Pandas \citep{pandaspaper2010,pandassoftware2020}, Lightkurve \citep{lightkurve2018}, XGBoost \citep{xgboost2016}, Tensorflow \citep{tensorflow2015-whitepaper} 
}

\bibliographystyle{aa}
\bibliography{Class}

\begin{thebibliography}{120}
\expandafter\ifx\csname natexlab\endcsname\relax\def\natexlab#1{#1}\fi

\bibitem[{Abadi {et~al.}(2015)Abadi, Agarwal, Barham, Brevdo, Chen, Citro,
  Corrado, Davis, Dean, Devin, Ghemawat, Goodfellow, Harp, Irving, Isard, Jia,
  Jozefowicz, Kaiser, Kudlur, Levenberg, Man\'{e}, Monga, Moore, Murray, Olah,
  Schuster, Shlens, Steiner, Sutskever, Talwar, Tucker, Vanhoucke, Vasudevan,
  Vi\'{e}gas, Vinyals, Warden, Wattenberg, Wicke, Yu, \&
  Zheng}]{tensorflow2015-whitepaper}
Abadi, M., Agarwal, A., Barham, P., {et~al.} 2015, {TensorFlow}: Large-Scale
  Machine Learning on Heterogeneous Systems, software available from
  tensorflow.org

\bibitem[{{Abdul-Masih} {et~al.}(2016){Abdul-Masih}, {Pr{\v{s}}a}, {Conroy},
  {Bloemen}, {Boyajian}, {Doyle}, {Johnston}, {Kostov}, {Latham},
  {Matijevi{\v{c}}}, {Shporer}, \& {Southworth}}]{Abdul-Masih2016}
{Abdul-Masih}, M., {Pr{\v{s}}a}, A., {Conroy}, K., {et~al.} 2016, \aj, 151, 101

\bibitem[{{Aerts} {et~al.}(2010){Aerts}, {Christensen-Dalsgaard}, \&
  {Kurtz}}]{Aerts2010}
{Aerts}, C., {Christensen-Dalsgaard}, J., \& {Kurtz}, D.~W. 2010,
  {Asteroseismology}

\bibitem[{{Aerts} {et~al.}(1998){Aerts}, {Eyer}, \& {Kestens}}]{Aerts1998}
{Aerts}, C., {Eyer}, L., \& {Kestens}, E. 1998, \aap, 337, 790

\bibitem[{Aggarwal(2014)}]{Aggarwal2014}
Aggarwal, C.~C. 2014, Data Classification: Algorithms and Applications, 1st
  edn. (Chapman \& Hall/CRC)

\bibitem[{{Antoci} {et~al.}(2019){Antoci}, {Cunha}, {Bowman}, {Murphy},
  {Kurtz}, {Bedding}, {Borre}, {Christophe}, {Daszy{\'n}ska-Daszkiewicz},
  {Fox-Machado}, {Garc{\'\i}a Hern{\'a}ndez}, {Ghasemi}, {Handberg}, {Hansen},
  {Hasanzadeh}, {Houdek}, {Johnston}, {Justesen}, {Kahraman Alicavus},
  {Kotysz}, {Latham}, {Matthews}, {M{\o}nster}, {Niemczura}, {Paunzen},
  {S{\'a}nchez Arias}, {Pigulski}, {Pepper}, {Richey-Yowell}, {Safari},
  {Seager}, {Smalley}, {Shutt}, {S{\'o}dor}, {Su{\'a}rez}, {Tkachenko}, {Wu},
  {Zwintz}, {Barcel{\'o} Forteza}, {Brunsden}, {Bogn{\'a}r}, {Buzasi},
  {Chowdhury}, {De Cat}, {Evans}, {Guo}, {Guzik}, {Jevtic}, {Lampens}, {Lares
  Martiz}, {Lovekin}, {Li}, {Mirouh}, {Mkrtichian}, {Monteiro}, {Nemec},
  {Ouazzani}, {Pascual-Granado}, {Reese}, {Rieutord}, {Rodon}, {Skarka},
  {Sowicka}, {Stateva}, {Szab{\'o}}, \& {Weiss}}]{Antoci2019}
{Antoci}, V., {Cunha}, M.~S., {Bowman}, D.~M., {et~al.} 2019, \mnras, 490, 4040

\bibitem[{Armstrong {et~al.}(2016)Armstrong, Kirk, Lam, McCormac, Osborn,
  Spake, Walker, Brown, Kristiansen, Pollacco, West, \&
  Wheatley}]{Armstrong:2016br}
Armstrong, D.~J., Kirk, J., Lam, K. W.~F., {et~al.} 2016, Monthly Notices of
  the Royal Astronomical Society, 456, 2260

\bibitem[{Armstrong {et~al.}(2015)Armstrong, Kirk, Lam, McCormac, Walker,
  Brown, Osborn, Pollacco, \& Spake}]{Armstrong:2015bn}
Armstrong, D.~J., Kirk, J., Lam, K. W.~F., {et~al.} 2015, Astronomy and
  Astrophysics, 579, A19

\bibitem[{{Astropy Collaboration} {et~al.}(2018){Astropy Collaboration},
  {Price-Whelan}, {Sip{\H{o}}cz}, {G{\"u}nther}, {Lim}, {Crawford}, {Conseil},
  {Shupe}, {Craig}, {Dencheva}, {Ginsburg}, {VanderPlas}, {Bradley},
  {P{\'e}rez-Su{\'a}rez}, {de Val-Borro}, {Aldcroft}, {Cruz}, {Robitaille},
  {Tollerud}, {Ardelean}, {Babej}, {Bach}, {Bachetti}, {Bakanov}, {Bamford},
  {Barentsen}, {Barmby}, {Baumbach}, {Berry}, {Biscani}, {Boquien}, {Bostroem},
  {Bouma}, {Brammer}, {Bray}, {Breytenbach}, {Buddelmeijer}, {Burke},
  {Calderone}, {Cano Rodr{\'\i}guez}, {Cara}, {Cardoso}, {Cheedella}, {Copin},
  {Corrales}, {Crichton}, {D'Avella}, {Deil}, {Depagne}, {Dietrich}, {Donath},
  {Droettboom}, {Earl}, {Erben}, {Fabbro}, {Ferreira}, {Finethy}, {Fox},
  {Garrison}, {Gibbons}, {Goldstein}, {Gommers}, {Greco}, {Greenfield},
  {Groener}, {Grollier}, {Hagen}, {Hirst}, {Homeier}, {Horton}, {Hosseinzadeh},
  {Hu}, {Hunkeler}, {Ivezi{\'c}}, {Jain}, {Jenness}, {Kanarek}, {Kendrew},
  {Kern}, {Kerzendorf}, {Khvalko}, {King}, {Kirkby}, {Kulkarni}, {Kumar},
  {Lee}, {Lenz}, {Littlefair}, {Ma}, {Macleod}, {Mastropietro}, {McCully},
  {Montagnac}, {Morris}, {Mueller}, {Mumford}, {Muna}, {Murphy}, {Nelson},
  {Nguyen}, {Ninan}, {N{\"o}the}, {Ogaz}, {Oh}, {Parejko}, {Parley}, {Pascual},
  {Patil}, {Patil}, {Plunkett}, {Prochaska}, {Rastogi}, {Reddy Janga},
  {Sabater}, {Sakurikar}, {Seifert}, {Sherbert}, {Sherwood-Taylor}, {Shih},
  {Sick}, {Silbiger}, {Singanamalla}, {Singer}, {Sladen}, {Sooley},
  {Sornarajah}, {Streicher}, {Teuben}, {Thomas}, {Tremblay}, {Turner},
  {Terr{\'o}n}, {van Kerkwijk}, {de la Vega}, {Watkins}, {Weaver}, {Whitmore},
  {Woillez}, {Zabalza}, \& {Astropy Contributors}}]{Astropy:2018}
{Astropy Collaboration}, {Price-Whelan}, A.~M., {Sip{\H{o}}cz}, B.~M., {et~al.}
  2018, \aj, 156, 123

\bibitem[{{Astropy Collaboration} {et~al.}(2013){Astropy Collaboration},
  {Robitaille}, {Tollerud}, {Greenfield}, {Droettboom}, {Bray}, {Aldcroft},
  {Davis}, {Ginsburg}, {Price-Whelan}, {Kerzendorf}, {Conley}, {Crighton},
  {Barbary}, {Muna}, {Ferguson}, {Grollier}, {Parikh}, {Nair}, {Unther},
  {Deil}, {Woillez}, {Conseil}, {Kramer}, {Turner}, {Singer}, {Fox}, {Weaver},
  {Zabalza}, {Edwards}, {Azalee Bostroem}, {Burke}, {Casey}, {Crawford},
  {Dencheva}, {Ely}, {Jenness}, {Labrie}, {Lim}, {Pierfederici}, {Pontzen},
  {Ptak}, {Refsdal}, {Servillat}, \& {Streicher}}]{Astropy:2013}
{Astropy Collaboration}, {Robitaille}, T.~P., {Tollerud}, E.~J., {et~al.} 2013,
  \aap, 558, A33

\bibitem[{{Auvergne} {et~al.}(2009){Auvergne}, {Bodin}, {Boisnard}, {Buey},
  {Chaintreuil}, {Epstein}, {Jouret}, {Lam-Trong}, {Levacher}, {Magnan},
  {Perez}, {Plasson}, {Plesseria}, {Peter}, {Steller}, {Tiph{\`e}ne}, {Baglin},
  {Agogu{\'e}}, {Appourchaux}, {Barbet}, {Beaufort}, {Bellenger}, {Berlin},
  {Bernardi}, {Blouin}, {Boumier}, {Bonneau}, {Briet}, {Butler}, {Cautain},
  {Chiavassa}, {Costes}, {Cuvilho}, {Cunha-Parro}, {de Oliveira Fialho},
  {Decaudin}, {Defise}, {Djalal}, {Docclo}, {Drummond}, {Dupuis}, {Exil},
  {Faur{\'e}}, {Gaboriaud}, {Gamet}, {Gavalda}, {Grolleau}, {Gueguen},
  {Guivarc'h}, {Guterman}, {Hasiba}, {Huntzinger}, {Hustaix}, {Imbert},
  {Jeanville}, {Johlander}, {Jorda}, {Journoud}, {Karioty}, {Kerjean},
  {Lafond}, {Lapeyrere}, {Landiech}, {Larqu{\'e}}, {Laudet}, {Le Merrer},
  {Leporati}, {Leruyet}, {Levieuge}, {Llebaria}, {Martin}, {Mazy}, {Mesnager},
  {Michel}, {Moalic}, {Monjoin}, {Naudet}, {Neukirchner}, {Nguyen-Kim},
  {Ollivier}, {Orcesi}, {Ottacher}, {Oulali}, {Parisot}, {Perruchot},
  {Piacentino}, {Pinheiro da Silva}, {Platzer}, {Pontet}, {Pradines},
  {Quentin}, {Rohbeck}, {Rolland}, {Rollenhagen}, {Romagnan}, {Russ}, {Samadi},
  {Schmidt}, {Schwartz}, {Sebbag}, {Smit}, {Sunter}, {Tello}, {Toulouse},
  {Ulmer}, {Vandermarcq}, {Vergnault}, {Wallner}, {Waultier}, \&
  {Zanatta}}]{Auvergne2009}
{Auvergne}, M., {Bodin}, P., {Boisnard}, L., {et~al.} 2009, \aap, 506, 411

\bibitem[{Bae {et~al.}(1996)Bae, Ryu, Chang, Song, \& Kim}]{Bae1996}
Bae, J., Ryu, Y., Chang, T., Song, I., \& Kim, H.~M. 1996, Signal Processing,
  52, 75

\bibitem[{{Ball} {et~al.}(2006){Ball}, {Brunner}, {Myers}, \&
  {Tcheng}}]{Ball2006}
{Ball}, N.~M., {Brunner}, R.~J., {Myers}, A.~D., \& {Tcheng}, D. 2006, \apj,
  650, 497

\bibitem[{{Borucki} {et~al.}(2010){Borucki}, {Koch}, {Basri}, {Batalha},
  {Brown}, {Caldwell}, {Caldwell}, {Christensen-Dalsgaard}, {Cochran},
  {DeVore}, {Dunham}, {Dupree}, {Gautier}, {Geary}, {Gilliland}, {Gould},
  {Howell}, {Jenkins}, {Kondo}, {Latham}, {Marcy}, {Meibom}, {Kjeldsen},
  {Lissauer}, {Monet}, {Morrison}, {Sasselov}, {Tarter}, {Boss}, {Brownlee},
  {Owen}, {Buzasi}, {Charbonneau}, {Doyle}, {Fortney}, {Ford}, {Holman},
  {Seager}, {Steffen}, {Welsh}, {Rowe}, {Anderson}, {Buchhave}, {Ciardi},
  {Walkowicz}, {Sherry}, {Horch}, {Isaacson}, {Everett}, {Fischer}, {Torres},
  {Johnson}, {Endl}, {MacQueen}, {Bryson}, {Dotson}, {Haas}, {Kolodziejczak},
  {Van Cleve}, {Chandrasekaran}, {Twicken}, {Quintana}, {Clarke}, {Allen},
  {Li}, {Wu}, {Tenenbaum}, {Verner}, {Bruhweiler}, {Barnes}, \&
  {Prsa}}]{borucki2009}
{Borucki}, W.~J., {Koch}, D., {Basri}, G., {et~al.} 2010, Science, 327, 977

\bibitem[{{Bowman}(2020)}]{Bowman2020}
{Bowman}, D.~M. 2020, Frontiers in Astronomy and Space Sciences, 7, 70

\bibitem[{{Bowman} {et~al.}(2016){Bowman}, {Kurtz}, {Breger}, {Murphy}, \&
  {Holdsworth}}]{Bowman2016}
{Bowman}, D.~M., {Kurtz}, D.~W., {Breger}, M., {Murphy}, S.~J., \&
  {Holdsworth}, D.~L. 2016, \mnras, 460, 1970

\bibitem[{{Bracewell}(1986)}]{Bracewell1986}
{Bracewell}, R.~N. 1986, {The Fourier Transform and its applications}

\bibitem[{Breiman(2001)}]{Breiman:fb}
Breiman, L. 2001, Machine Learning, 45, 5

\bibitem[{Breiman {et~al.}(1984)Breiman, Friedman, Olshen, \&
  Stone}]{Breiman1984}
Breiman, L., Friedman, J.~H., Olshen, R.~A., \& Stone, C.~J. 1984,
  Classification and regression trees, The Wadsworth statistics/probability
  series (Belmont: Wadsworth)

\bibitem[{Brett {et~al.}(2004)Brett, West, \& Wheatley}]{Brett:2004cr}
Brett, D.~R., West, R.~G., \& Wheatley, P.~J. 2004, Monthly Notices of the
  Royal Astronomical Society, 353, 369

\bibitem[{{Brown} {et~al.}(1991){Brown}, {Gilliland}, {Noyes}, \&
  {Ramsey}}]{Brown1991}
{Brown}, T.~M., {Gilliland}, R.~L., {Noyes}, R.~W., \& {Ramsey}, L.~W. 1991,
  \apj, 368, 599

\bibitem[{{Bruntt} \& {Buzasi}(2006)}]{Bruntt2006}
{Bruntt}, H. \& {Buzasi}, D.~L. 2006, \memsai, 77, 278

\bibitem[{{Bugnet} {et~al.}(2018){Bugnet}, {Garc{\'\i}a}, {Davies}, {Mathur},
  {Corsaro}, {Hall}, \& {Rendle}}]{Bugnet2018}
{Bugnet}, L., {Garc{\'\i}a}, R.~A., {Davies}, G.~R., {et~al.} 2018, \aap, 620,
  A38

\bibitem[{Busa \& van Emmerik(2016)}]{Busa2016}
Busa, M.~A. \& van Emmerik, R. 2016, Journal of Sport and Health Science, 5, 44

\bibitem[{{Buzasi}(2004)}]{Buzasi2004}
{Buzasi}, D.~L. 2004, in ESA Special Publication, Vol. 538, Stellar Structure
  and Habitable Planet Finding, ed. F.~{Favata}, S.~{Aigrain}, \& A.~{Wilson},
  205--213

\bibitem[{Chen \& Guestrin(2016)}]{xgboost2016}
Chen, T. \& Guestrin, C. 2016, in Proceedings of the 22nd ACM SIGKDD
  International Conference on Knowledge Discovery and Data Mining, KDD '16 (New
  York, NY, USA: ACM), 785--794

\bibitem[{Costa {et~al.}(2005)Costa, Goldberger, \& Peng}]{Costa2005}
Costa, M., Goldberger, A.~L., \& Peng, C.-K. 2005, Phys. Rev. E, 71, 021906

\bibitem[{{Debosscher} {et~al.}(2011){Debosscher}, {Blomme}, {Aerts}, \& {De
  Ridder}}]{debosscher2011}
{Debosscher}, J., {Blomme}, J., {Aerts}, C., \& {De Ridder}, J. 2011, \aap,
  529, A89

\bibitem[{{Debosscher} {et~al.}(2007){Debosscher}, {Sarro}, {Aerts}, {Cuypers},
  {Vandenbussche}, {Garrido}, \& {Solano}}]{debosscher2007}
{Debosscher}, J., {Sarro}, L.~M., {Aerts}, C., {et~al.} 2007, \aap, 475, 1159

\bibitem[{{Degroote} {et~al.}(2009){Degroote}, {Aerts}, {Ollivier}, {Miglio},
  {Debosscher}, {Cuypers}, {Briquet}, {Montalb{\'a}n}, {Thoul}, {Noels}, {De
  Cat}, {Balaguer-N{\'u}{\~n}ez}, {Maceroni}, {Ribas}, {Auvergne}, {Baglin},
  {Deleuil}, {Weiss}, {Jorda}, {Baudin}, \& {Samadi}}]{degroote2009}
{Degroote}, P., {Aerts}, C., {Ollivier}, M., {et~al.} 2009, \aap, 506, 471

\bibitem[{{Derekas} {et~al.}(2017){Derekas}, {Plachy}, {Moln{\'a}r},
  {S{\'o}dor}, {Benk{\'{o}}}, {Szabados}, {Bogn{\'a}r}, {Cs{\'a}k},
  {Szab{\'o}}, {Szab{\'o}}, \& {P{\'a}l}}]{Derekas2017}
{Derekas}, A., {Plachy}, E., {Moln{\'a}r}, L., {et~al.} 2017, \mnras, 464, 1553

\bibitem[{{Eyer} \& {Blake}(2005)}]{EyerBlake2005}
{Eyer}, L. \& {Blake}, C. 2005, \mnras, 358, 30

\bibitem[{{Eyer} \& {Grenon}(1998)}]{EyerGrenon1998}
{Eyer}, L. \& {Grenon}, M. 1998, in New Eyes to See Inside the Sun and Stars,
  ed. F.-L. {Deubner}, J.~{Christensen-Dalsgaard}, \& D.~{Kurtz}, Vol. 185, 291

\bibitem[{Fawcett(2006)}]{fawcett2006}
Fawcett, T. 2006, Pattern Recognition Letters, 27, 861, rOC Analysis in Pattern
  Recognition

\bibitem[{{Feinstein} {et~al.}(2020){Feinstein}, {Montet}, {Ansdell}, {Nord},
  {Bean}, {G{\"u}nther}, {Gully-Santiago}, \& {Schlieder}}]{Feinstein2020}
{Feinstein}, A.~D., {Montet}, B.~T., {Ansdell}, M., {et~al.} 2020, \aj, 160,
  219

\bibitem[{Friedman(2001)}]{Freidman2001}
Friedman, J.~H. 2001, The Annals of Statistics, 29, 1189

\bibitem[{{Gaia Collaboration} {et~al.}(2019){Gaia Collaboration}, {Eyer},
  {Rimoldini}, {Audard}, {Anderson}, {Nienartowicz}, {Glass}, {Marchal},
  {Grenon}, {Mowlavi}, {Holl}, {Clementini}, {Aerts}, {Mazeh}, {Evans},
  {Szabados}, {Brown}, {Vallenari}, {Prusti}, {de Bruijne}, {Babusiaux},
  {Bailer-Jones}, {Biermann}, {Jansen}, {Jordi}, {Klioner}, {Lammers},
  {Lindegren}, {Luri}, {Mignard}, {Panem}, {Pourbaix}, {Randich}, {Sartoretti},
  {Siddiqui}, {Soubiran}, {van Leeuwen}, {Walton}, {Arenou}, {Bastian},
  {Cropper}, {Drimmel}, {Katz}, {Lattanzi}, {Bakker}, {Cacciari},
  {Casta{\~n}eda}, {Chaoul}, {Cheek}, {De Angeli}, {Fabricius}, {Guerra},
  {Masana}, {Messineo}, {Panuzzo}, {Portell}, {Riello}, {Seabroke}, {Tanga},
  {Th{\'e}venin}, {Gracia-Abril}, {Comoretto}, {Garcia-Reinaldos}, {Teyssier},
  {Altmann}, {Andrae}, {Bellas-Velidis}, {Benson}, {Berthier}, {Blomme},
  {Burgess}, {Busso}, {Carry}, {Cellino}, {Clotet}, {Creevey}, {Davidson}, {De
  Ridder}, {Delchambre}, {Dell'Oro}, {Ducourant},
  {Fern{\'a}ndez-Hern{\'a}ndez}, {Fouesneau}, {Fr{\'e}mat}, {Galluccio},
  {Garc{\'\i}a-Torres}, {Gonz{\'a}lez-N{\'u}{\~n}ez}, {Gonz{\'a}lez-Vidal},
  {Gosset}, {Guy}, {Halbwachs}, {Hambly}, {Harrison}, {Hern{\'a}ndez},
  {Hestroffer}, {Hodgkin}, {Hutton}, {Jasniewicz}, {Jean-Antoine-Piccolo},
  {Jordan}, {Korn}, {Krone-Martins}, {Lanzafame}, {Lebzelter}, {L{\"o}ffler},
  {Manteiga}, {Marrese}, {Mart{\'\i}n-Fleitas}, {Moitinho}, {Mora}, {Muinonen},
  {Osinde}, {Pancino}, {Pauwels}, {Petit}, {Recio-Blanco}, {Richards}, {Robin},
  {Sarro}, {Siopis}, {Smith}, {Sozzetti}, {S{\"u}veges}, {Torra}, {van Reeven},
  {Abbas}, {Abreu Aramburu}, {Accart}, {Altavilla}, {{\'A}lvarez}, {Alvarez},
  {Alves}, {Andrei}, {Anglada Varela}, {Antiche}, {Antoja}, {Arcay},
  {Astraatmadja}, {Bach}, {Baker}, {Balaguer-N{\'u}{\~n}ez}, {Balm}, {Barache},
  {Barata}, {Barbato}, {Barblan}, {Barklem}, {Barrado}, {Barros}, {Barstow},
  {Bartholom{\'e} Mu{\~n}oz}, {Bassilana}, {Becciani}, {Bellazzini},
  {Berihuete}, {Bertone}, {Bianchi}, {Bienaym{\'e}}, {Blanco-Cuaresma}, {Boch},
  {Boeche}, {Bombrun}, {Borrachero}, {Bossini}, {Bouquillon}, {Bourda},
  {Bragaglia}, {Bramante}, {Breddels}, {Bressan}, {Brouillet},
  {Br{\"u}semeister}, {Brugaletta}, {Bucciarelli}, {Burlacu}, {Busonero},
  {Butkevich}, {Buzzi}, {Caffau}, {Cancelliere}, {Cannizzaro}, {Cantat-Gaudin},
  {Carballo}, {Carlucci}, {Carrasco}, {Casamiquela}, {Castellani},
  {Castro-Ginard}, {Charlot}, {Chemin}, {Chiavassa}, {Cocozza}, {Costigan},
  {Cowell}, {Crifo}, {Crosta}, {Crowley}, {Cuypers}, {Dafonte}, {Damerdji},
  {Dapergolas}, {David}, {David}, {de Laverny}, {De Luise}, {De March}, {de
  Martino}, {de Souza}, {de Torres}, {Debosscher}, {del Pozo}, {Delbo},
  {Delgado}, {Delgado}, {Diakite}, {Diener}, {Distefano}, {Dolding},
  {Drazinos}, {Dur{\'a}n}, {Edvardsson}, {Enke}, {Eriksson}, {Esquej}, {Eynard
  Bontemps}, {Fabre}, {Fabrizio}, {Faigler}, {Falc{\~a}o}, {Farr{\`a}s Casas},
  {Federici}, {Fedorets}, {Fernique}, {Figueras}, {Filippi}, {Findeisen},
  {Fonti}, {Fraile}, {Fraser}, {Fr{\'e}zouls}, {Gai}, {Galleti}, {Garabato},
  {Garc{\'\i}a-Sedano}, {Garofalo}, {Garralda}, {Gavel}, {Gavras}, {Gerssen},
  {Geyer}, {Giacobbe}, {Gilmore}, {Girona}, {Giuffrida}, {Gomes}, {Granvik},
  {Gueguen}, {Guerrier}, {Guiraud}, {Guti{\'e}rrez-S{\'a}nchez}, {Haigron},
  {Hatzidimitriou}, {Hauser}, {Haywood}, {Heiter}, {Helmi}, {Heu}, {Hilger},
  {Hobbs}, {Hofmann}, {Holland}, {Huckle}, {Hypki}, {Icardi}, {Jan{\ss}en},
  {Jevardat de Fombelle}, {Jonker}, {Juh{\'a}sz}, {Julbe}, {Karampelas},
  {Kewley}, {Klar}, {Kochoska}, {Kohley}, {Kolenberg}, {Kontizas}, {Kontizas},
  {Koposov}, {Kordopatis}, {Kostrzewa-Rutkowska}, {Koubsky}, {Lambert},
  {Lanza}, {Lasne}, {Lavigne}, {Le Fustec}, {Le Poncin-Lafitte}, {Lebreton},
  {Leccia}, {Leclerc}, {Lecoeur-Taibi}, {Lenhardt}, {Leroux}, {Liao}, {Licata},
  {Lindstr{\o}m}, {Lister}, {Livanou}, {Lobel}, {L{\'o}pez}, {Lorenz},
  {Managau}, {Mann}, {Mantelet}, {Marchant}, {Marconi}, {Marinoni},
  {Marschalk{\'o}}, {Marshall}, {Martino}, {Marton}, {Mary}, {Massari},
  {Matijevi{\v{c}}}, {McMillan}, {Messina}, {Michalik}, {Millar}, {Molina},
  {Molinaro}, {Moln{\'a}r}, {Montegriffo}, {Mor}, {Morbidelli}, {Morel},
  {Morgenthaler}, {Morris}, {Mulone}, {Muraveva}, {Musella}, {Nelemans},
  {Nicastro}, {Noval}, {O'Mullane}, {Ord{\'e}novic}, {Ord{\'o}{\~n}ez-Blanco},
  {Osborne}, {Pagani}, {Pagano}, {Pailler}, {Palacin}, {Palaversa}, {Panahi},
  {Pawlak}, {Piersimoni}, {Pineau}, {Plachy}, {Plum}, {Poggio}, {Poujoulet},
  {Pr{\v{s}}a}, {Pulone}, {Racero}, {Ragaini}, {Rambaux}, {Ramos-Lerate},
  {Regibo}, {Reyl{\'e}}, {Riclet}, {Ripepi}, {Riva}, {Rivard}, {Rixon},
  {Roegiers}, {Roelens}, {Romero-G{\'o}mez}, {Rowell}, {Royer}, {Ruiz-Dern},
  {Sadowski}, {Sagrist{\`a} Sell{\'e}s}, {Sahlmann}, {Salgado}, {Salguero},
  {Sanna}, {Santana-Ros}, {Sarasso}, {Savietto}, {Schultheis}, {Sciacca},
  {Segol}, {Segovia}, {S{\'e}gransan}, {Shih}, {Siltala}, {Silva}, {Smart},
  {Smith}, {Solano}, {Solitro}, {Sordo}, {Soria Nieto}, {Souchay}, {Spagna},
  {Spoto}, {Stampa}, {Steele}, {Steidelm{\"u}ller}, {Stephenson}, {Stoev},
  {Suess}, {Surdej}, {Szegedi-Elek}, {Tapiador}, {Taris}, {Tauran}, {Taylor},
  {Teixeira}, {Terrett}, {Teyssandier}, {Thuillot}, {Titarenko}, {Torra
  Clotet}, {Turon}, {Ulla}, {Utrilla}, {Uzzi}, {Vaillant}, {Valentini},
  {Valette}, {van Elteren}, {Van Hemelryck}, {van Leeuwen}, {Vaschetto},
  {Vecchiato}, {Veljanoski}, {Viala}, {Vicente}, {Vogt}, {von Essen}, {Voss},
  {Votruba}, {Voutsinas}, {Walmsley}, {Weiler}, {Wertz}, {Wevers},
  {Wyrzykowski}, {Yoldas}, {{\v{Z}}erjal}, {Ziaeepour}, {Zorec}, {Zschocke},
  {Zucker}, {Zurbach}, \& {Zwitter}}]{GaiaEyer2019}
{Gaia Collaboration}, {Eyer}, L., {Rimoldini}, L., {et~al.} 2019, \aap, 623,
  A110

\bibitem[{{Gaia Collaboration} {et~al.}(2016){Gaia Collaboration}, {Prusti},
  {de Bruijne}, {Brown}, {Vallenari}, {Babusiaux}, {Bailer-Jones}, {Bastian},
  {Biermann}, {Evans}, {Eyer}, {Jansen}, {Jordi}, {Klioner}, {Lammers},
  {Lindegren}, {Luri}, {Mignard}, {Milligan}, {Panem}, {Poinsignon},
  {Pourbaix}, {Randich}, {Sarri}, {Sartoretti}, {Siddiqui}, {Soubiran},
  {Valette}, {van Leeuwen}, {Walton}, {Aerts}, {Arenou}, {Cropper}, {Drimmel},
  {H{\o}g}, {Katz}, {Lattanzi}, {O'Mullane}, {Grebel}, {Holland}, {Huc},
  {Passot}, {Bramante}, {Cacciari}, {Casta{\~n}eda}, {Chaoul}, {Cheek}, {De
  Angeli}, {Fabricius}, {Guerra}, {Hern{\'a}ndez}, {Jean-Antoine-Piccolo},
  {Masana}, {Messineo}, {Mowlavi}, {Nienartowicz}, {Ord{\'o}{\~n}ez-Blanco},
  {Panuzzo}, {Portell}, {Richards}, {Riello}, {Seabroke}, {Tanga},
  {Th{\'e}venin}, {Torra}, {Els}, {Gracia-Abril}, {Comoretto},
  {Garcia-Reinaldos}, {Lock}, {Mercier}, {Altmann}, {Andrae}, {Astraatmadja},
  {Bellas-Velidis}, {Benson}, {Berthier}, {Blomme}, {Busso}, {Carry},
  {Cellino}, {Clementini}, {Cowell}, {Creevey}, {Cuypers}, {Davidson}, {De
  Ridder}, {de Torres}, {Delchambre}, {Dell'Oro}, {Ducourant}, {Fr{\'e}mat},
  {Garc{\'\i}a-Torres}, {Gosset}, {Halbwachs}, {Hambly}, {Harrison}, {Hauser},
  {Hestroffer}, {Hodgkin}, {Huckle}, {Hutton}, {Jasniewicz}, {Jordan},
  {Kontizas}, {Korn}, {Lanzafame}, {Manteiga}, {Moitinho}, {Muinonen},
  {Osinde}, {Pancino}, {Pauwels}, {Petit}, {Recio-Blanco}, {Robin}, {Sarro},
  {Siopis}, {Smith}, {Smith}, {Sozzetti}, {Thuillot}, {van Reeven}, {Viala},
  {Abbas}, {Abreu Aramburu}, {Accart}, {Aguado}, {Allan}, {Allasia},
  {Altavilla}, {{\'A}lvarez}, {Alves}, {Anderson}, {Andrei}, {Anglada Varela},
  {Antiche}, {Antoja}, {Ant{\'o}n}, {Arcay}, {Atzei}, {Ayache}, {Bach},
  {Baker}, {Balaguer-N{\'u}{\~n}ez}, {Barache}, {Barata}, {Barbier}, {Barblan},
  {Baroni}, {Barrado y Navascu{\'e}s}, {Barros}, {Barstow}, {Becciani},
  {Bellazzini}, {Bellei}, {Bello Garc{\'\i}a}, {Belokurov}, {Bendjoya},
  {Berihuete}, {Bianchi}, {Bienaym{\'e}}, {Billebaud}, {Blagorodnova},
  {Blanco-Cuaresma}, {Boch}, {Bombrun}, {Borrachero}, {Bouquillon}, {Bourda},
  {Bouy}, {Bragaglia}, {Breddels}, {Brouillet}, {Br{\"u}semeister},
  {Bucciarelli}, {Budnik}, {Burgess}, {Burgon}, {Burlacu}, {Busonero}, {Buzzi},
  {Caffau}, {Cambras}, {Campbell}, {Cancelliere}, {Cantat-Gaudin}, {Carlucci},
  {Carrasco}, {Castellani}, {Charlot}, {Charnas}, {Charvet}, {Chassat},
  {Chiavassa}, {Clotet}, {Cocozza}, {Collins}, {Collins}, {Costigan}, {Crifo},
  {Cross}, {Crosta}, {Crowley}, {Dafonte}, {Damerdji}, {Dapergolas}, {David},
  {David}, {De Cat}, {de Felice}, {de Laverny}, {De Luise}, {De March}, {de
  Martino}, {de Souza}, {Debosscher}, {del Pozo}, {Delbo}, {Delgado},
  {Delgado}, {di Marco}, {Di Matteo}, {Diakite}, {Distefano}, {Dolding}, {Dos
  Anjos}, {Drazinos}, {Dur{\'a}n}, {Dzigan}, {Ecale}, {Edvardsson}, {Enke},
  {Erdmann}, {Escolar}, {Espina}, {Evans}, {Eynard Bontemps}, {Fabre},
  {Fabrizio}, {Faigler}, {Falc{\~a}o}, {Farr{\`a}s Casas}, {Faye}, {Federici},
  {Fedorets}, {Fern{\'a}ndez-Hern{\'a}ndez}, {Fernique}, {Fienga}, {Figueras},
  {Filippi}, {Findeisen}, {Fonti}, {Fouesneau}, {Fraile}, {Fraser}, {Fuchs},
  {Furnell}, {Gai}, {Galleti}, {Galluccio}, {Garabato}, {Garc{\'\i}a-Sedano},
  {Gar{\'e}}, {Garofalo}, {Garralda}, {Gavras}, {Gerssen}, {Geyer}, {Gilmore},
  {Girona}, {Giuffrida}, {Gomes}, {Gonz{\'a}lez-Marcos},
  {Gonz{\'a}lez-N{\'u}{\~n}ez}, {Gonz{\'a}lez-Vidal}, {Granvik}, {Guerrier},
  {Guillout}, {Guiraud}, {G{\'u}rpide}, {Guti{\'e}rrez-S{\'a}nchez}, {Guy},
  {Haigron}, {Hatzidimitriou}, {Haywood}, {Heiter}, {Helmi}, {Hobbs},
  {Hofmann}, {Holl}, {Holland}, {Hunt}, {Hypki}, {Icardi}, {Irwin}, {Jevardat
  de Fombelle}, {Jofr{\'e}}, {Jonker}, {Jorissen}, {Julbe}, {Karampelas},
  {Kochoska}, {Kohley}, {Kolenberg}, {Kontizas}, {Koposov}, {Kordopatis},
  {Koubsky}, {Kowalczyk}, {Krone-Martins}, {Kudryashova}, {Kull}, {Bachchan},
  {Lacoste-Seris}, {Lanza}, {Lavigne}, {Le Poncin-Lafitte}, {Lebreton},
  {Lebzelter}, {Leccia}, {Leclerc}, {Lecoeur-Taibi}, {Lemaitre}, {Lenhardt},
  {Leroux}, {Liao}, {Licata}, {Lindstr{\o}m}, {Lister}, {Livanou}, {Lobel},
  {L{\"o}ffler}, {L{\'o}pez}, {Lopez-Lozano}, {Lorenz}, {Loureiro},
  {MacDonald}, {Magalh{\~a}es Fernandes}, {Managau}, {Mann}, {Mantelet},
  {Marchal}, {Marchant}, {Marconi}, {Marie}, {Marinoni}, {Marrese},
  {Marschalk{\'o}}, {Marshall}, {Mart{\'\i}n-Fleitas}, {Martino}, {Mary},
  {Matijevi{\v{c}}}, {Mazeh}, {McMillan}, {Messina}, {Mestre}, {Michalik},
  {Millar}, {Miranda}, {Molina}, {Molinaro}, {Molinaro}, {Moln{\'a}r},
  {Moniez}, {Montegriffo}, {Monteiro}, {Mor}, {Mora}, {Morbidelli}, {Morel},
  {Morgenthaler}, {Morley}, {Morris}, {Mulone}, {Muraveva}, {Musella},
  {Narbonne}, {Nelemans}, {Nicastro}, {Noval}, {Ord{\'e}novic},
  {Ordieres-Mer{\'e}}, {Osborne}, {Pagani}, {Pagano}, {Pailler}, {Palacin},
  {Palaversa}, {Parsons}, {Paulsen}, {Pecoraro}, {Pedrosa}, {Pentik{\"a}inen},
  {Pereira}, {Pichon}, {Piersimoni}, {Pineau}, {Plachy}, {Plum}, {Poujoulet},
  {Pr{\v{s}}a}, {Pulone}, {Ragaini}, {Rago}, {Rambaux}, {Ramos-Lerate},
  {Ranalli}, {Rauw}, {Read}, {Regibo}, {Renk}, {Reyl{\'e}}, {Ribeiro},
  {Rimoldini}, {Ripepi}, {Riva}, {Rixon}, {Roelens}, {Romero-G{\'o}mez},
  {Rowell}, {Royer}, {Rudolph}, {Ruiz-Dern}, {Sadowski}, {Sagrist{\`a}
  Sell{\'e}s}, {Sahlmann}, {Salgado}, {Salguero}, {Sarasso}, {Savietto},
  {Schnorhk}, {Schultheis}, {Sciacca}, {Segol}, {Segovia}, {Segransan},
  {Serpell}, {Shih}, {Smareglia}, {Smart}, {Smith}, {Solano}, {Solitro},
  {Sordo}, {Soria Nieto}, {Souchay}, {Spagna}, {Spoto}, {Stampa}, {Steele},
  {Steidelm{\"u}ller}, {Stephenson}, {Stoev}, {Suess}, {S{\"u}veges}, {Surdej},
  {Szabados}, {Szegedi-Elek}, {Tapiador}, {Taris}, {Tauran}, {Taylor},
  {Teixeira}, {Terrett}, {Tingley}, {Trager}, {Turon}, {Ulla}, {Utrilla},
  {Valentini}, {van Elteren}, {Van Hemelryck}, {van Leeuwen}, {Varadi},
  {Vecchiato}, {Veljanoski}, {Via}, {Vicente}, {Vogt}, {Voss}, {Votruba},
  {Voutsinas}, {Walmsley}, {Weiler}, {Weingrill}, {Werner}, {Wevers},
  {Whitehead}, {Wyrzykowski}, {Yoldas}, {{\v{Z}}erjal}, {Zucker}, {Zurbach},
  {Zwitter}, {Alecu}, {Allen}, {Allende Prieto}, {Amorim},
  {Anglada-Escud{\'e}}, {Arsenijevic}, {Azaz}, {Balm}, {Beck}, {Bernstein},
  {Bigot}, {Bijaoui}, {Blasco}, {Bonfigli}, {Bono}, {Boudreault}, {Bressan},
  {Brown}, {Brunet}, {Bunclark}, {Buonanno}, {Butkevich}, {Carret}, {Carrion},
  {Chemin}, {Ch{\'e}reau}, {Corcione}, {Darmigny}, {de Boer}, {de Teodoro}, {de
  Zeeuw}, {Delle Luche}, {Domingues}, {Dubath}, {Fodor}, {Fr{\'e}zouls},
  {Fries}, {Fustes}, {Fyfe}, {Gallardo}, {Gallegos}, {Gardiol}, {Gebran},
  {Gomboc}, {G{\'o}mez}, {Grux}, {Gueguen}, {Heyrovsky}, {Hoar}, {Iannicola},
  {Isasi Parache}, {Janotto}, {Joliet}, {Jonckheere}, {Keil}, {Kim},
  {Klagyivik}, {Klar}, {Knude}, {Kochukhov}, {Kolka}, {Kos}, {Kutka}, {Lainey},
  {LeBouquin}, {Liu}, {Loreggia}, {Makarov}, {Marseille}, {Martayan},
  {Martinez-Rubi}, {Massart}, {Meynadier}, {Mignot}, {Munari}, {Nguyen},
  {Nordlander}, {Ocvirk}, {O'Flaherty}, {Olias Sanz}, {Ortiz}, {Osorio},
  {Oszkiewicz}, {Ouzounis}, {Palmer}, {Park}, {Pasquato}, {Peltzer}, {Peralta},
  {P{\'e}turaud}, {Pieniluoma}, {Pigozzi}, {Poels}, {Prat}, {Prod'homme},
  {Raison}, {Rebordao}, {Risquez}, {Rocca-Volmerange}, {Rosen}, {Ruiz-Fuertes},
  {Russo}, {Sembay}, {Serraller Vizcaino}, {Short}, {Siebert}, {Silva},
  {Sinachopoulos}, {Slezak}, {Soffel}, {Sosnowska}, {Strai{\v{z}}ys}, {ter
  Linden}, {Terrell}, {Theil}, {Tiede}, {Troisi}, {Tsalmantza}, {Tur},
  {Vaccari}, {Vachier}, {Valles}, {Van Hamme}, {Veltz}, {Virtanen}, {Wallut},
  {Wichmann}, {Wilkinson}, {Ziaeepour}, \& {Zschocke}}]{Gaia2016}
{Gaia Collaboration}, {Prusti}, T., {de Bruijne}, J.~H.~J., {et~al.} 2016,
  \aap, 595, A1

\bibitem[{{Garc{\'\i}a} \& {Ballot}(2019)}]{GarciaBallot2019}
{Garc{\'\i}a}, R.~A. \& {Ballot}, J. 2019, Living Reviews in Solar Physics, 16,
  4

\bibitem[{{Garc{\'\i}a} {et~al.}(2014){Garc{\'\i}a}, {Ceillier}, {Salabert},
  {Mathur}, {van Saders}, {Pinsonneault}, {Ballot}, {Beck}, {Bloemen},
  {Campante}, {Davies}, {do Nascimento}, {Mathis}, {Metcalfe}, {Nielsen},
  {Su{\'a}rez}, {Chaplin}, {Jim{\'e}nez}, \& {Karoff}}]{garcia2014}
{Garc{\'\i}a}, R.~A., {Ceillier}, T., {Salabert}, D., {et~al.} 2014, \aap, 572,
  A34

\bibitem[{Granitto {et~al.}(2006)Granitto, Furlanello, Biasioli, \&
  Gasperi}]{Granitto2006}
Granitto, P.~M., Furlanello, C., Biasioli, F., \& Gasperi, F. 2006, Chemom
  Intell Lab Syst, 83, 83

\bibitem[{{Guzik} {et~al.}(2000){Guzik}, {Kaye}, {Bradley}, {Cox}, \&
  {Neuforge}}]{Guzik2000}
{Guzik}, J.~A., {Kaye}, A.~B., {Bradley}, P.~A., {Cox}, A.~N., \& {Neuforge},
  C. 2000, \apjl, 542, L57

\bibitem[{Hall(1999)}]{Hall1999}
Hall, M.~A. 1999, PhD thesis, University of Waikato Hamilton

\bibitem[{{Handberg} {et~al.}(2021){Handberg}, {Lund}, {White}, {Hall},
  {Buzasi}, {Pope}, {Hansen}, {von Essen}, {Carboneau}, {Huber}, {Vanderspek},
  {Fausnaug}, {Tenenbaum}, {Jenkins}, \& {the T'DA
  Collaboration}}]{handberg2021}
{Handberg}, R., {Lund}, M.~N., {White}, T.~R., {et~al.} 2021, arXiv e-prints,
  arXiv:2106.08341

\bibitem[{Harris {et~al.}(2020)Harris, Millman, van~der Walt, Gommers,
  Virtanen, Cournapeau, Wieser, Taylor, Berg, Smith, Kern, Picus, Hoyer, van
  Kerkwijk, Brett, Haldane, del R{'{\i}}o, Wiebe, Peterson,
  G{'{e}}rard-Marchant, Sheppard, Reddy, Weckesser, Abbasi, Gohlke, \&
  Oliphant}]{numpy2020}
Harris, C.~R., Millman, K.~J., van~der Walt, S.~J., {et~al.} 2020, Nature, 585,
  357

\bibitem[{{Hekker} \& {Christensen-Dalsgaard}(2017)}]{HekkerCD2017}
{Hekker}, S. \& {Christensen-Dalsgaard}, J. 2017, \aapr, 25, 1

\bibitem[{Hon {et~al.}(2018)Hon, Stello, \& Zinn}]{Hon_2018}
Hon, M., Stello, D., \& Zinn, J.~C. 2018, The Astrophysical Journal, 859, 64

\bibitem[{{Howell} {et~al.}(2014){Howell}, {Sobeck}, {Haas}, {Still},
  {Barclay}, {Mullally}, {Troeltzsch}, {Aigrain}, {Bryson}, {Caldwell},
  {Chaplin}, {Cochran}, {Huber}, {Marcy}, {Miglio}, {Najita}, {Smith},
  {Twicken}, \& {Fortney}}]{howell2014}
{Howell}, S.~B., {Sobeck}, C., {Haas}, M., {et~al.} 2014, \pasp, 126, 398

\bibitem[{{H{\"u}mmerich} {et~al.}(2018){H{\"u}mmerich}, {Mikul{\'a}{\v{s}}ek},
  {Paunzen}, {Bernhard}, {Jan{\'\i}k}, {Yakunin}, {Pribulla}, {Va{\v{n}}ko}, \&
  {Mat{\v{e}}chov{\'a}}}]{Hummerich2018}
{H{\"u}mmerich}, S., {Mikul{\'a}{\v{s}}ek}, Z., {Paunzen}, E., {et~al.} 2018,
  \aap, 619, A98

\bibitem[{{Jamal} \& {Bloom}(2020)}]{jamal2020}
{Jamal}, S. \& {Bloom}, J.~S. 2020, arXiv e-prints, arXiv:2003.08618

\bibitem[{{Kallinger}(2019)}]{Kallinger2019}
{Kallinger}, T. 2019, arXiv e-prints, arXiv:1906.09428

\bibitem[{{Kedem} \& {Slud}(1981)}]{Kedem1981}
{Kedem}, B. \& {Slud}, E. 1981, Biometrika, 68, 551

\bibitem[{Kedem \& Slud(1982)}]{Kedem1982}
Kedem, B. \& Slud, E. 1982, The Annals of Statistics, 10, 786

\bibitem[{Kgoadi {et~al.}(2019)Kgoadi, Engelbrecht, Whittingham, \&
  Tkachenko}]{Kgoadi2019}
Kgoadi, R., Engelbrecht, C., Whittingham, I., \& Tkachenko, A. 2019, arXiv
  preprint arXiv:1906.06628

\bibitem[{Kim \& Bailer-Jones(2016)}]{Kim2016}
Kim, D.-W. \& Bailer-Jones, C. A.~L. 2016, Astron Astrophys, 587, A18

\bibitem[{{Kirk} {et~al.}(2016){Kirk}, {Conroy}, {Pr{\v{s}}a}, {Abdul-Masih},
  {Kochoska}, {Matijevi{\v{c}}}, {Hambleton}, {Barclay}, {Bloemen}, {Boyajian},
  {Doyle}, {Fulton}, {Hoekstra}, {Jek}, {Kane}, {Kostov}, {Latham}, {Mazeh},
  {Orosz}, {Pepper}, {Quarles}, {Ragozzine}, {Shporer}, {Southworth},
  {Stassun}, {Thompson}, {Welsh}, {Agol}, {Derekas}, {Devor}, {Fischer},
  {Green}, {Gropp}, {Jacobs}, {Johnston}, {LaCourse}, {Saetre}, {Schwengeler},
  {Toczyski}, {Werner}, {Garrett}, {Gore}, {Martinez}, {Spitzer}, {Stevick},
  {Thomadis}, {Vrijmoet}, {Yenawine}, {Batalha}, \& {Borucki}}]{Kirk2016}
{Kirk}, B., {Conroy}, K., {Pr{\v{s}}a}, A., {et~al.} 2016, \aj, 151, 68

\bibitem[{{Kiss} \& {B{\'o}di}(2017)}]{Kiss2017}
{Kiss}, L.~L. \& {B{\'o}di}, A. 2017, \aap, 608, A99

\bibitem[{{Kjeldsen} \& {Bedding}(1995)}]{Kjeldsen_Bedding1995}
{Kjeldsen}, H. \& {Bedding}, T.~R. 1995, \aap, 293, 87

\bibitem[{{Kjeldsen} {et~al.}(1995){Kjeldsen}, {Bedding}, {Viskum}, \&
  {Frandsen}}]{Kjeldsen1995}
{Kjeldsen}, H., {Bedding}, T.~R., {Viskum}, M., \& {Frandsen}, S. 1995, \aj,
  109, 1313

\bibitem[{Kohonen(1990)}]{Kohonen:1990fd}
Kohonen, T. 1990, Proceedings of the IEEE, 78, 1464

\bibitem[{Kozachenko \& Leonenko(1987)}]{Kozachenko1987}
Kozachenko, L.~F. \& Leonenko, N.~N. 1987, Probl. Peredachi Inf., 23, 95–101

\bibitem[{Kraskov {et~al.}(2004)Kraskov, St\"ogbauer, \&
  Grassberger}]{Kraskov2004}
Kraskov, A., St\"ogbauer, H., \& Grassberger, P. 2004, Phys. Rev. E, 69, 066138

\bibitem[{{Kuszlewicz} {et~al.}(2020){Kuszlewicz}, {Hekker}, \&
  {Bell}}]{Kuszlewicz2020}
{Kuszlewicz}, J.~S., {Hekker}, S., \& {Bell}, K.~J. 2020, \mnras, 497, 4843

\bibitem[{{Li} {et~al.}(2020){Li}, {Van Reeth}, {Bedding}, {Murphy}, {Antoci},
  {Ouazzani}, \& {Barbara}}]{Li2020}
{Li}, G., {Van Reeth}, T., {Bedding}, T.~R., {et~al.} 2020, \mnras, 491, 3586

\bibitem[{{Lightkurve Collaboration} {et~al.}(2018){Lightkurve Collaboration},
  {Cardoso}, {Hedges}, {Gully-Santiago}, {Saunders}, {Cody}, {Barclay}, {Hall},
  {Sagear}, {Turtelboom}, {Zhang}, {Tzanidakis}, {Mighell}, {Coughlin}, {Bell},
  {Berta-Thompson}, {Williams}, {Dotson}, \& {Barentsen}}]{lightkurve2018}
{Lightkurve Collaboration}, {Cardoso}, J.~V.~d.~M., {Hedges}, C., {et~al.}
  2018, {Lightkurve: Kepler and TESS time series analysis in Python},
  Astrophysics Source Code Library

\bibitem[{{Lomb}(1976)}]{lomb1976}
{Lomb}, N.~R. 1976, \apss, 39, 447

\bibitem[{Lundberg {et~al.}(2020)Lundberg, Erion, Chen, DeGrave, Prutkin, Nair,
  Katz, Himmelfarb, Bansal, \& Lee}]{lundberg2019explainable}
Lundberg, S.~M., Erion, G., Chen, H., {et~al.} 2020, Nature Machine
  Intelligence, 2, 2522

\bibitem[{Lundberg \& Lee(2017)}]{NIPS2017_7062}
Lundberg, S.~M. \& Lee, S.-I. 2017, in Advances in Neural Information
  Processing Systems 30, ed. I.~Guyon, U.~V. Luxburg, S.~Bengio, H.~Wallach,
  R.~Fergus, S.~Vishwanathan, \& R.~Garnett (Curran Associates, Inc.),
  4765--4774

\bibitem[{{Manick} {et~al.}(2019){Manick}, {Kamath}, {Van Winckel}, {Jorissen},
  {Sekaran}, {Bowman}, {Oomen}, {Kluska}, {Bollen}, \&
  {Waelkens}}]{Manick2019a}
{Manick}, R., {Kamath}, D., {Van Winckel}, H., {et~al.} 2019, \aap, 628, A40

\bibitem[{{Mathys} {et~al.}(2020){Mathys}, {Kurtz}, \&
  {Holdsworth}}]{Mathys2020}
{Mathys}, G., {Kurtz}, D.~W., \& {Holdsworth}, D.~L. 2020, \aap, 639, A31

\bibitem[{{Matijevi{\v{c}}} {et~al.}(2012){Matijevi{\v{c}}}, {Pr{\v{s}}a},
  {Orosz}, {Welsh}, {Bloemen}, \& {Barclay}}]{Matijevic2012}
{Matijevi{\v{c}}}, G., {Pr{\v{s}}a}, A., {Orosz}, J.~A., {et~al.} 2012, \aj,
  143, 123

\bibitem[{{McQuillan} {et~al.}(2014){McQuillan}, {Mazeh}, \&
  {Aigrain}}]{McQuillan2014}
{McQuillan}, A., {Mazeh}, T., \& {Aigrain}, S. 2014, \apjs, 211, 24

\bibitem[{{Modak} {et~al.}(2018){Modak}, {Chattopadhyay}, \&
  {Chattopadhyay}}]{modak2018}
{Modak}, S., {Chattopadhyay}, T., \& {Chattopadhyay}, A.~K. 2018, arXiv
  e-prints, arXiv:1801.09406

\bibitem[{{Montgomery} \& {Odonoghue}(1999)}]{Montgomery1999}
{Montgomery}, M.~H. \& {Odonoghue}, D. 1999, Delta Scuti Star Newsletter, 13,
  28

\bibitem[{{Namekata} {et~al.}(2019){Namekata}, {Maehara}, {Notsu}, {Toriumi},
  {Hayakawa}, {Ikuta}, {Notsu}, {Honda}, {Nogami}, \& {Shibata}}]{Namekata2019}
{Namekata}, K., {Maehara}, H., {Notsu}, Y., {et~al.} 2019, \apj, 871, 187

\bibitem[{{Naul} {et~al.}(2018){Naul}, {Bloom}, {P{\'e}rez}, \& {van der
  Walt}}]{naul2018}
{Naul}, B., {Bloom}, J.~S., {P{\'e}rez}, F., \& {van der Walt}, S. 2018, Nature
  Astronomy, 2, 151

\bibitem[{{Nielsen} {et~al.}(2013){Nielsen}, {Gizon}, {Schunker}, \&
  {Karoff}}]{Nielsen2013}
{Nielsen}, M.~B., {Gizon}, L., {Schunker}, H., \& {Karoff}, C. 2013, \aap, 557,
  L10

\bibitem[{pandas~development team(2020)}]{pandassoftware2020}
pandas~development team, T. 2020, pandas-dev/pandas: Pandas

\bibitem[{{P{\'a}pics} {et~al.}(2012){P{\'a}pics}, {Briquet}, {Baglin},
  {Poretti}, {Aerts}, {Degroote}, {Tkachenko}, {Morel}, {Zima}, {Niemczura},
  {Rainer}, {Hareter}, {Baudin}, {Catala}, {Michel}, {Samadi}, \&
  {Auvergne}}]{papics2012}
{P{\'a}pics}, P.~I., {Briquet}, M., {Baglin}, A., {et~al.} 2012, \aap, 542, A55

\bibitem[{{P{\'a}pics} {et~al.}(2017){P{\'a}pics}, {Tkachenko}, {Van Reeth},
  {Aerts}, {Moravveji}, {Van de Sande}, {De Smedt}, {Bloemen}, {Southworth},
  {Debosscher}, {Niemczura}, \& {Gameiro}}]{Papics2017}
{P{\'a}pics}, P.~I., {Tkachenko}, A., {Van Reeth}, T., {et~al.} 2017, \aap,
  598, A74

\bibitem[{{Pedersen} {et~al.}(2020){Pedersen}, {Escorza}, {P{\'a}pics}, \&
  {Aerts}}]{Pedersen2020}
{Pedersen}, M.~G., {Escorza}, A., {P{\'a}pics}, P.~I., \& {Aerts}, C. 2020,
  \mnras, 495, 2738

\bibitem[{Pedregosa {et~al.}(2011)Pedregosa, Varoquaux, Gramfort, Michel,
  Thirion, Grisel, Blondel, Prettenhofer, Weiss, Dubourg, Vanderplas, Passos,
  Cournapeau, Brucher, Perrot, \& Duchesnay}]{scikit-learn}
Pedregosa, F., Varoquaux, G., Gramfort, A., {et~al.} 2011, Journal of Machine
  Learning Research, 12, 2825

\bibitem[{{Pinsonneault} {et~al.}(2018){Pinsonneault}, {Elsworth}, {Tayar},
  {Serenelli}, {Stello}, {Zinn}, {Mathur}, {Garc{\'\i}a}, {Johnson}, {Hekker},
  {Huber}, {Kallinger}, {M{\'e}sz{\'a}ros}, {Mosser}, {Stassun}, {Girardi},
  {Rodrigues}, {Silva Aguirre}, {An}, {Basu}, {Chaplin}, {Corsaro}, {Cunha},
  {Garc{\'\i}a-Hern{\'a}ndez}, {Holtzman}, {J{\"o}nsson}, {Shetrone}, {Smith},
  {Sobeck}, {Stringfellow}, {Zamora}, {Beers}, {Fern{\'a}ndez-Trincado},
  {Frinchaboy}, {Hearty}, \& {Nitschelm}}]{Pinsonneault2018}
{Pinsonneault}, M.~H., {Elsworth}, Y.~P., {Tayar}, J., {et~al.} 2018, \apjs,
  239, 32

\bibitem[{{Plachy} {et~al.}(2018){Plachy}, {B{\'o}di}, \&
  {Koll{\'a}th}}]{Plachy2018}
{Plachy}, E., {B{\'o}di}, A., \& {Koll{\'a}th}, Z. 2018, \mnras, 481, 2986

\bibitem[{{Pojmanski}(2002)}]{Pojmanski2002}
{Pojmanski}, G. 2002, \actaa, 52, 397

\bibitem[{{Preston}(1974)}]{Preston1974}
{Preston}, G.~W. 1974, \araa, 12, 257

\bibitem[{Provost(2000)}]{provost2000}
Provost, F. 2000, in Proceedings of the AAAI’2000 workshop on imbalanced data
  sets (AAAI Press)

\bibitem[{{Pr{\v{s}}a} {et~al.}(2011){Pr{\v{s}}a}, {Batalha}, {Slawson},
  {Doyle}, {Welsh}, {Orosz}, {Seager}, {Rucker}, {Mjaseth}, {Engle}, {Conroy},
  {Jenkins}, {Caldwell}, {Koch}, \& {Borucki}}]{Prsa2011}
{Pr{\v{s}}a}, A., {Batalha}, N., {Slawson}, R.~W., {et~al.} 2011, \aj, 141, 83

\bibitem[{{Rauer} {et~al.}(2014){Rauer}, {Catala}, {Aerts}, {Appourchaux},
  {Benz}, {Brandeker}, {Christensen-Dalsgaard}, {Deleuil}, {Gizon}, {Goupil},
  {G{\"u}del}, {Janot-Pacheco}, {Mas-Hesse}, {Pagano}, {Piotto}, {Pollacco},
  {Santos}, {Smith}, {Su{\'a}rez}, {Szab{\'o}}, {Udry}, {Adibekyan}, {Alibert},
  {Almenara}, {Amaro-Seoane}, {Eiff}, {Asplund}, {Antonello}, {Barnes},
  {Baudin}, {Belkacem}, {Bergemann}, {Bihain}, {Birch}, {Bonfils}, {Boisse},
  {Bonomo}, {Borsa}, {Brand{\~a}o}, {Brocato}, {Brun}, {Burleigh}, {Burston},
  {Cabrera}, {Cassisi}, {Chaplin}, {Charpinet}, {Chiappini}, {Church},
  {Csizmadia}, {Cunha}, {Damasso}, {Davies}, {Deeg}, {D{\'\i}az}, {Dreizler},
  {Dreyer}, {Eggenberger}, {Ehrenreich}, {Eigm{\"u}ller}, {Erikson}, {Farmer},
  {Feltzing}, {de Oliveira Fialho}, {Figueira}, {Forveille}, {Fridlund},
  {Garc{\'\i}a}, {Giommi}, {Giuffrida}, {Godolt}, {Gomes da Silva}, {Granzer},
  {Grenfell}, {Grotsch-Noels}, {G{\"u}nther}, {Haswell}, {Hatzes},
  {H{\'e}brard}, {Hekker}, {Helled}, {Heng}, {Jenkins}, {Johansen},
  {Khodachenko}, {Kislyakova}, {Kley}, {Kolb}, {Krivova}, {Kupka}, {Lammer},
  {Lanza}, {Lebreton}, {Magrin}, {Marcos-Arenal}, {Marrese}, {Marques},
  {Martins}, {Mathis}, {Mathur}, {Messina}, {Miglio}, {Montalban}, {Montalto},
  {Monteiro}, {Moradi}, {Moravveji}, {Mordasini}, {Morel}, {Mortier},
  {Nascimbeni}, {Nelson}, {Nielsen}, {Noack}, {Norton}, {Ofir}, {Oshagh},
  {Ouazzani}, {P{\'a}pics}, {Parro}, {Petit}, {Plez}, {Poretti}, {Quirrenbach},
  {Ragazzoni}, {Raimondo}, {Rainer}, {Reese}, {Redmer}, {Reffert},
  {Rojas-Ayala}, {Roxburgh}, {Salmon}, {Santerne}, {Schneider}, {Schou},
  {Schuh}, {Schunker}, {Silva-Valio}, {Silvotti}, {Skillen}, {Snellen}, {Sohl},
  {Sousa}, {Sozzetti}, {Stello}, {Strassmeier}, {{\v{S}}vanda}, {Szab{\'o}},
  {Tkachenko}, {Valencia}, {Van Grootel}, {Vauclair}, {Ventura}, {Wagner},
  {Walton}, {Weingrill}, {Werner}, {Wheatley}, \& {Zwintz}}]{Rauer2014}
{Rauer}, H., {Catala}, C., {Aerts}, C., {et~al.} 2014, Experimental Astronomy,
  38, 249

\bibitem[{Richards {et~al.}(2011)Richards, Starr, Butler, Bloom, Brewer,
  Crellin-Quick, Higgins, Kennedy, \& Rischard}]{richards2011}
Richards, J.~W., Starr, D.~L., Butler, N.~R., {et~al.} 2011, Astrophys J, 733
  [\eprint[arXiv]{1101.1959}]

\bibitem[{Richman \& Moorman(2000)}]{Richman2000}
Richman, J.~S. \& Moorman, J.~R. 2000, American Journal of Physiology-Heart and
  Circulatory Physiology, 278, H2039, pMID: 10843903

\bibitem[{{Ricker} {et~al.}(2015){Ricker}, {Winn}, {Vanderspek}, {Latham},
  {Bakos}, {Bean}, {Berta-Thompson}, {Brown}, {Buchhave}, {Butler}, {Butler},
  {Chaplin}, {Charbonneau}, {Christensen-Dalsgaard}, {Clampin}, {Deming},
  {Doty}, {De Lee}, {Dressing}, {Dunham}, {Endl}, {Fressin}, {Ge}, {Henning},
  {Holman}, {Howard}, {Ida}, {Jenkins}, {Jernigan}, {Johnson}, {Kaltenegger},
  {Kawai}, {Kjeldsen}, {Laughlin}, {Levine}, {Lin}, {Lissauer}, {MacQueen},
  {Marcy}, {McCullough}, {Morton}, {Narita}, {Paegert}, {Palle}, {Pepe},
  {Pepper}, {Quirrenbach}, {Rinehart}, {Sasselov}, {Sato}, {Seager},
  {Sozzetti}, {Stassun}, {Sullivan}, {Szentgyorgyi}, {Torres}, {Udry}, \&
  {Villasenor}}]{ricker2015}
{Ricker}, G.~R., {Winn}, J.~N., {Vanderspek}, R., {et~al.} 2015, Journal of
  Astronomical Telescopes, Instruments, and Systems, 1, 014003

\bibitem[{{Roberts} {et~al.}(1987){Roberts}, {Lehar}, \&
  {Dreher}}]{Roberts1987}
{Roberts}, D.~H., {Lehar}, J., \& {Dreher}, J.~W. 1987, \aj, 93, 968

\bibitem[{{Santos} {et~al.}(2019){Santos}, {Garc{\'\i}a}, {Mathur}, {Bugnet},
  {van Saders}, {Metcalfe}, {Simonian}, \& {Pinsonneault}}]{santos2019}
{Santos}, A.~R.~G., {Garc{\'\i}a}, R.~A., {Mathur}, S., {et~al.} 2019, \apjs,
  244, 21

\bibitem[{{Sarro} {et~al.}(2009){Sarro}, {Debosscher}, {L{\'o}pez}, \&
  {Aerts}}]{sarro2009}
{Sarro}, L.~M., {Debosscher}, J., {L{\'o}pez}, M., \& {Aerts}, C. 2009, \aap,
  494, 739

\bibitem[{{Scargle}(1982)}]{scargle1982}
{Scargle}, J.~D. 1982, \apj, 263, 835

\bibitem[{Schapire(1990)}]{schapire1990}
Schapire, R.~E. 1990, Machine Learning, 5, 197

\bibitem[{Shannon(1948)}]{Shannon1948}
Shannon, C.~E. 1948, Bell System Technical Journal, 27, 623

\bibitem[{Shapiro \& Wilk(1965)}]{ShapiroWilk1965}
Shapiro, S.~S. \& Wilk, M.~B. 1965, Biometrika, 52, 591

\bibitem[{{Slawson} {et~al.}(2011){Slawson}, {Pr{\v{s}}a}, {Welsh}, {Orosz},
  {Rucker}, {Batalha}, {Doyle}, {Engle}, {Conroy}, {Coughlin}, {Gregg},
  {Fetherolf}, {Short}, {Windmiller}, {Fabrycky}, {Howell}, {Jenkins}, {Uddin},
  {Mullally}, {Seader}, {Thompson}, {Sand erfer}, {Borucki}, \&
  {Koch}}]{Slawson2011}
{Slawson}, R.~W., {Pr{\v{s}}a}, A., {Welsh}, W.~F., {et~al.} 2011, \aj, 142,
  160

\bibitem[{{Stassun} {et~al.}(2019){Stassun}, {Oelkers}, {Paegert}, {Torres},
  {Pepper}, {De Lee}, {Collins}, {Latham}, {Muirhead}, {Chittidi},
  {Rojas-Ayala}, {Fleming}, {Rose}, {Tenenbaum}, {Ting}, {Kane}, {Barclay},
  {Bean}, {Brassuer}, {Charbonneau}, {Ge}, {Lissauer}, {Mann}, {McLean},
  {Mullally}, {Narita}, {Plavchan}, {Ricker}, {Sasselov}, {Seager}, {Sharma},
  {Shiao}, {Sozzetti}, {Stello}, {Vanderspek}, {Wallace}, \&
  {Winn}}]{Stassun2019}
{Stassun}, K.~G., {Oelkers}, R.~J., {Paegert}, M., {et~al.} 2019, \aj, 158, 138

\bibitem[{{Szab{\'o}}(2018)}]{Szabo2018}
{Szab{\'o}}, R. 2018, in The RR Lyrae 2017 Conference. Revival of the Classical
  Pulsators: from Galactic Structure to Stellar Interior Diagnostics, Vol.~6,
  119--123

\bibitem[{{Szab{\'o}} {et~al.}(2011){Szab{\'o}}, {Szabados}, {Ngeow}, {Smolec},
  {Derekas}, {Moskalik}, {Nuspl}, {Lehmann}, {F{\.z}r{\'e}sz},
  {Molenda-{\.Z}akowicz}, {Bryson}, {Henden}, {Kurtz}, {Stello}, {Nemec},
  {Benk{\'{o}}}, {Berdnikov}, {Bruntt}, {Evans}, {Gorynya}, {Pastukhova},
  {Simcoe}, {Grindlay}, {Los}, {Doane}, {Laycock}, {Mink}, {Champine},
  {Sliski}, {Handler}, {Kiss}, {Koll{\'a}th}, {Kov{\'a}cs},
  {Christensen-Dalsgaard}, {Kjeldsen}, {Allen}, {Thompson}, \& {van
  Cleve}}]{Szabo2011}
{Szab{\'o}}, R., {Szabados}, L., {Ngeow}, C.~C., {et~al.} 2011, \mnras, 413,
  2709

\bibitem[{{Thompson} {et~al.}(2016){Thompson}, {Caldwell}, {Jenkins},
  {Barclay}, {Barentsen}, {Bryson}, {Burke}, {Campbell}, {Catanzarite},
  {Christiansen}, {Clarke}, {Colon}, {Coughlin}, {Girouard}, {Haas},
  {Harrison}, {Ibrahim}, {Klaus}, {Li}, {McCauliff}, {Morris}, {Mullally},
  {Rowe}, {Sabale}, {Seader}, {Smith}, {Tenenbaum}, {Twicken}, {Kamal Uddin},
  \& {Van Cleve}}]{Thompson2016}
{Thompson}, S.~E., {Caldwell}, D.~A., {Jenkins}, J.~M., {et~al.} 2016, {Kepler
  Data Release 25 Notes}, Kepler Science Document KSCI-19065-002

\bibitem[{{Tkachenko} {et~al.}(2013){Tkachenko}, {Aerts}, {Yakushechkin},
  {Debosscher}, {Degroote}, {Bloemen}, {P{\'a}pics}, {de Vries}, {Lombaert},
  {Hrudkova}, {Fr{\'e}mat}, {Raskin}, \& {Van Winckel}}]{tkachenko2013}
{Tkachenko}, A., {Aerts}, C., {Yakushechkin}, A., {et~al.} 2013, \aap, 556, A52

\bibitem[{Udalski {et~al.}(2008)Udalski, Szyma{\'{n}}ski, Soszy{\'{n}}ski, \&
  Poleski}]{Udalski2008}
Udalski, A., Szyma{\'{n}}ski, M.~K., Soszy{\'{n}}ski, I., \& Poleski, R. 2008,
  \actaa, 58, 69

\bibitem[{Udalski {et~al.}(2015)Udalski, Szymanski, \& Szymanski}]{Udalski2015}
Udalski, A., Szymanski, M.~K., \& Szymanski, G. 2015, \actaa, 65, 1

\bibitem[{{Valenzuela} \& {Pichara}(2018)}]{valenzuela2018}
{Valenzuela}, L. \& {Pichara}, K. 2018, \mnras, 474, 3259

\bibitem[{{Van Reeth} {et~al.}(2016){Van Reeth}, {Tkachenko}, \&
  {Aerts}}]{VanReeth2016}
{Van Reeth}, T., {Tkachenko}, A., \& {Aerts}, C. 2016, \aap, 593, A120

\bibitem[{{Van Reeth} {et~al.}(2015{\natexlab{a}}){Van Reeth}, {Tkachenko},
  {Aerts}, {P{\'a}pics}, {Degroote}, {Debosscher}, {Zwintz}, {Bloemen}, {De
  Smedt}, {Hrudkova}, {Raskin}, \& {Van Winckel}}]{VanReeth2015a}
{Van Reeth}, T., {Tkachenko}, A., {Aerts}, C., {et~al.} 2015{\natexlab{a}},
  \aap, 574, A17

\bibitem[{{Van Reeth} {et~al.}(2015{\natexlab{b}}){Van Reeth}, {Tkachenko},
  {Aerts}, {P{\'a}pics}, {Triana}, {Zwintz}, {Degroote}, {Debosscher},
  {Bloemen}, {Schmid}, {De Smedt}, {Fremat}, {Fuentes}, {Homan}, {Hrudkova},
  {Karjalainen}, {Lombaert}, {Nemeth}, {{\O}stensen}, {Van De Steene}, {Vos},
  {Raskin}, \& {Van Winckel}}]{VanReeth2015b}
{Van Reeth}, T., {Tkachenko}, A., {Aerts}, C., {et~al.} 2015{\natexlab{b}},
  \apjs, 218, 27

\bibitem[{{Vega} {et~al.}(2017){Vega}, {Stassun}, {Montez}, {Boyd}, \&
  {Somers}}]{Vega2017}
{Vega}, L.~D., {Stassun}, K.~G., {Montez}, Rodolfo, J., {Boyd}, P.~T., \&
  {Somers}, G. 2017, \apj, 839, 48

\bibitem[{Virtanen {et~al.}(2020)Virtanen, Gommers, Oliphant, Haberland, Reddy,
  Cournapeau, Burovski, Peterson, Weckesser, Bright, {van der Walt}, Brett,
  Wilson, Millman, Mayorov, Nelson, Jones, Kern, Larson, Carey, Polat, Feng,
  Moore, {VanderPlas}, Laxalde, Perktold, Cimrman, Henriksen, Quintero, Harris,
  Archibald, Ribeiro, Pedregosa, {van Mulbregt}, \& {SciPy 1.0
  Contributors}}]{scipy2020}
Virtanen, P., Gommers, R., Oliphant, T.~E., {et~al.} 2020, Nature Methods, 17,
  261

\bibitem[{{Waelkens} {et~al.}(1998){Waelkens}, {Aerts}, {Kestens}, {Grenon}, \&
  {Eyer}}]{Waelkens1998}
{Waelkens}, C., {Aerts}, C., {Kestens}, E., {Grenon}, M., \& {Eyer}, L. 1998,
  \aap, 330, 215

\bibitem[{{Walker} {et~al.}(2003){Walker}, {Matthews}, {Kuschnig}, {Johnson},
  {Rucinski}, {Pazder}, {Burley}, {Walker}, {Skaret}, {Zee}, {Grocott},
  {Carroll}, {Sinclair}, {Sturgeon}, \& {Harron}}]{Walker2003}
{Walker}, G., {Matthews}, J., {Kuschnig}, R., {et~al.} 2003, \pasp, 115, 1023

\bibitem[{{W}es {M}c{K}inney(2010)}]{pandaspaper2010}
{W}es {M}c{K}inney. 2010, in {P}roceedings of the 9th {P}ython in {S}cience
  {C}onference, ed. {S}t\'efan van~der {W}alt \& {J}arrod {M}illman, 56 -- 61

\bibitem[{Wolpert(1992)}]{Wolpert1992}
Wolpert, D.~H. 1992, Neural Networks, 5, 241

\bibitem[{{Wyrzykowski} \& {Belokurov}(2008)}]{Wyrzykowski2008}
{Wyrzykowski}, {\L}. \& {Belokurov}, V. 2008, in American Institute of Physics
  Conference Series, Vol. 1082, Classification and Discovery in Large
  Astronomical Surveys, ed. C.~A.~L. {Bailer-Jones}, 201--206

\bibitem[{Youden(1950)}]{youden1950}
Youden, W.~J. 1950, Cancer, 3, 32

\bibitem[{{Yu} {et~al.}(2020){Yu}, {Bedding}, {Stello}, {Huber}, {Compton},
  {Gizon}, \& {Hekker}}]{Yu2020}
{Yu}, J., {Bedding}, T.~R., {Stello}, D., {et~al.} 2020, \mnras, 493, 1388

\end{thebibliography}

\appendix

\section{Feature importance plots}
\label{Appendix:features}

\begin{figure}
   \centering
   \includegraphics[width=\columnwidth]{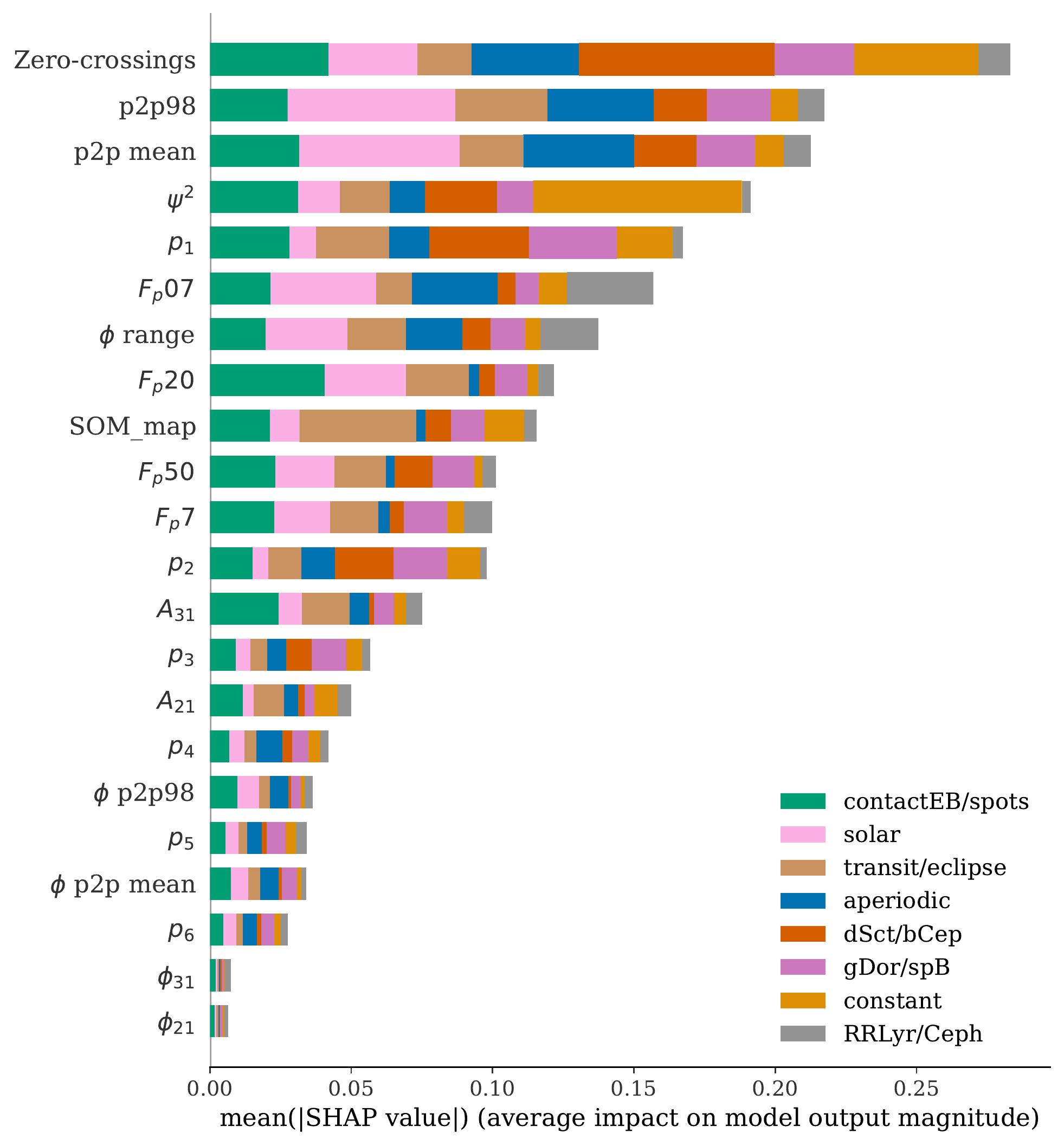}
      \caption{RFGC feature importances from SHAP. The hatched regions indicate the most important feature per class.}
         \label{Fig:RFGC_shap}
\end{figure}

\begin{figure}
   \centering
   \includegraphics[width=\columnwidth]{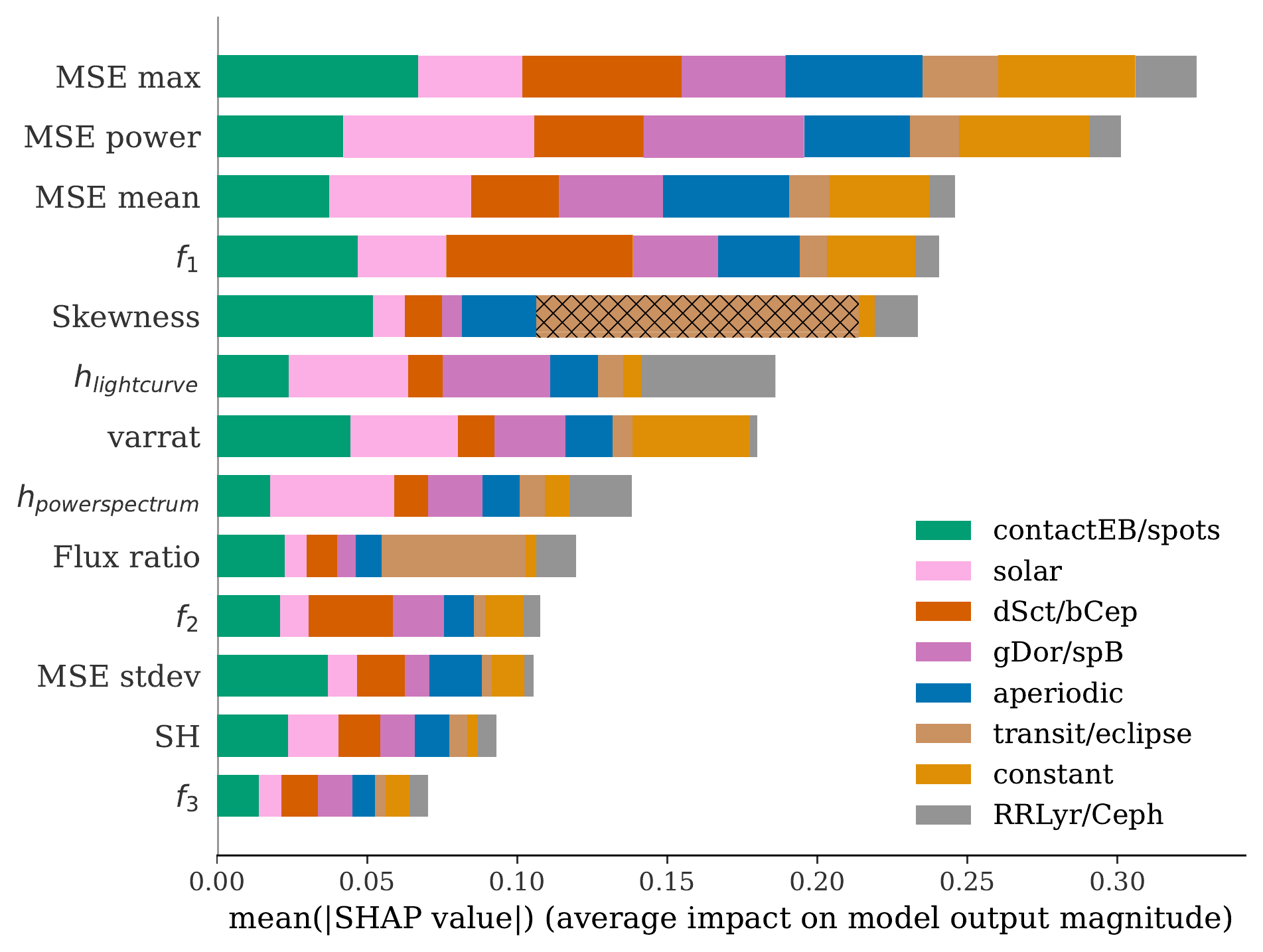}
      \caption{SORTING-HAT feature importances from SHAP. The hatched regions indicate the most important features per class.}
         \label{Fig:SORTINGHAT_shap}
\end{figure}

\begin{figure}
   \centering
   \includegraphics[width=\columnwidth]{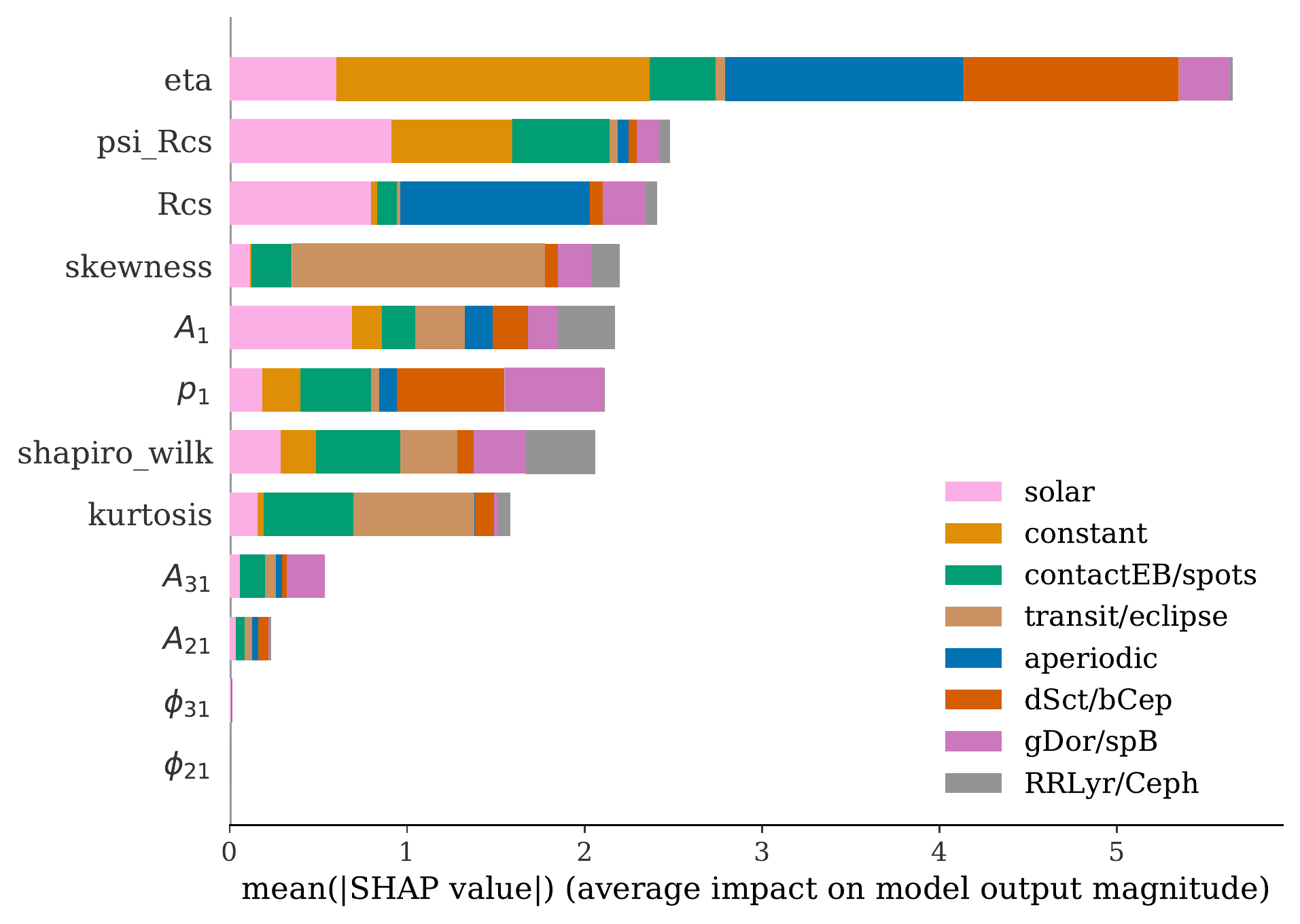}
      \caption{GBGC feature importances from SHAP. The hatched regions indicate the most important feature per class.} 
         \label{Fig:GBGC_shap}
\end{figure}

The SHAP\footnote{\url{https://github.com/slundberg/shap}} feature importance plots (see Sect \ref{Subsect:meta-testingval} for an explanation) allow us to evaluate the importance of the different attributes used by each classifier on a per class basis. The hatched regions in the plots indicate the most important feature per class. The plots for RFGC, SORTING-HAT and GBGC are respectively shown in Figs.~\ref{Fig:RFGC_shap}, \ref{Fig:SORTINGHAT_shap} and \ref{Fig:GBGC_shap}. Due to the fact that updated features have been used in the training of RFGC, the feature importances are different to those reported in \cite{Armstrong:2016br}.

For RFGC this reveals that the zero-crossings parameter is the most important for classifying contactEB/spots and dSct/bCep stars. The point-to-point differences are the primary features for solar-like oscillators and aperiodic stars, while respectively the coherency parameter, first fundamental period, FliPer value and the SOM are the most important for constant, gDor/SPB, RRLyr/Ceph and transit/eclipse stars.

In the case of SORTING-HAT we notice that the multiscale Entropy is by far the most important, as it is the primary feature for the contactEB/spots, aperiodic, constant, solar and gDor/SPB classes. In addition to that, the differential entropy is the primary feature for RRLyr/Ceph stars. For transit/eclipse stars the skewness is the most important followed by the flux ratio, which is logical given that these types of stars tend to have very skewed light curves. Lastly, for dSct/bCep stars the first fundamental frequency is most important.

For GBGC the variability index is the primary feature for constant, aperiodic and dSct/bCep stars. The range of the cumulative sum of the fluxes of the phase-folded light curve is the primary feature for solar-like oscillators and contactEB/spots stars. The skewness, first fundemental period and Shapiro-Wilk test for normality of the light curve are respectively the primary features for the transit/eclipse, gDor/SPB and RRLyr/Ceph classes.

\section{Effects of photon noise}
\label{Appendix:photon-noise}

\begin{figure*}
   \centering
   \includegraphics[width=18cm]{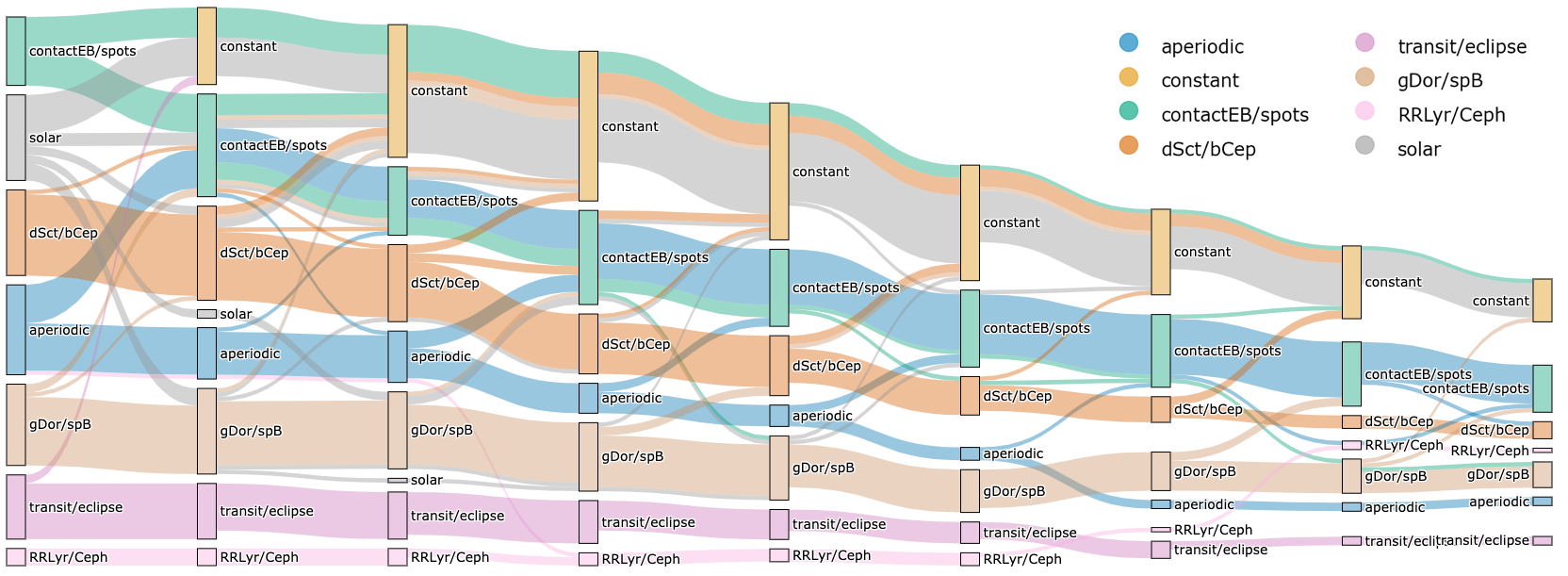}
      \caption{Sankey plot indicative of how stars move between classes when more noise is added to their light curves. The relatively brightest stars are shown on the left and relatively faintest on the right. The colors of the streams indicate the true label of the stars therein, and the bar labels and colors are representative of the predicted class. The height of the bars corresponds to the number of stars in that bin. The number of stars decreases from left to right because stars are eliminated once their newly calculated magnitude exceeds 15.5. Each step corresponds to a noise increase representative of 0.5 magnitude. Step 0 shows the original predictions by the metaclassifier.}
      \label{Fig:sankeyplot}
\end{figure*}

\begin{figure*}
   \centering
   \includegraphics[width=18cm]{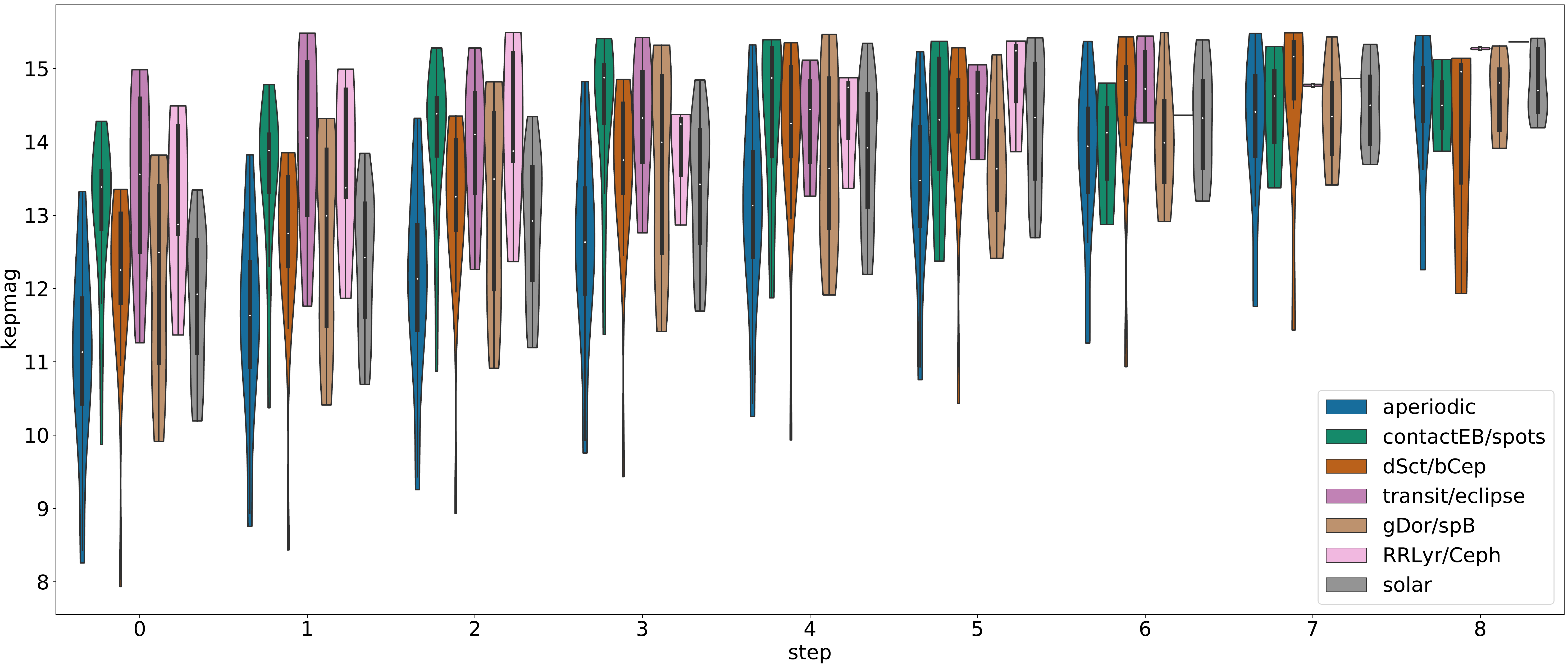}
      \caption{Violin plot illustrating the magnitude distribution at each step or bar of the Sankey plot in Fig.~\ref{Fig:sankeyplot}.}
         \label{Fig:violinplot}
\end{figure*}

\begin{figure}
   \centering
   \includegraphics[width=9cm]{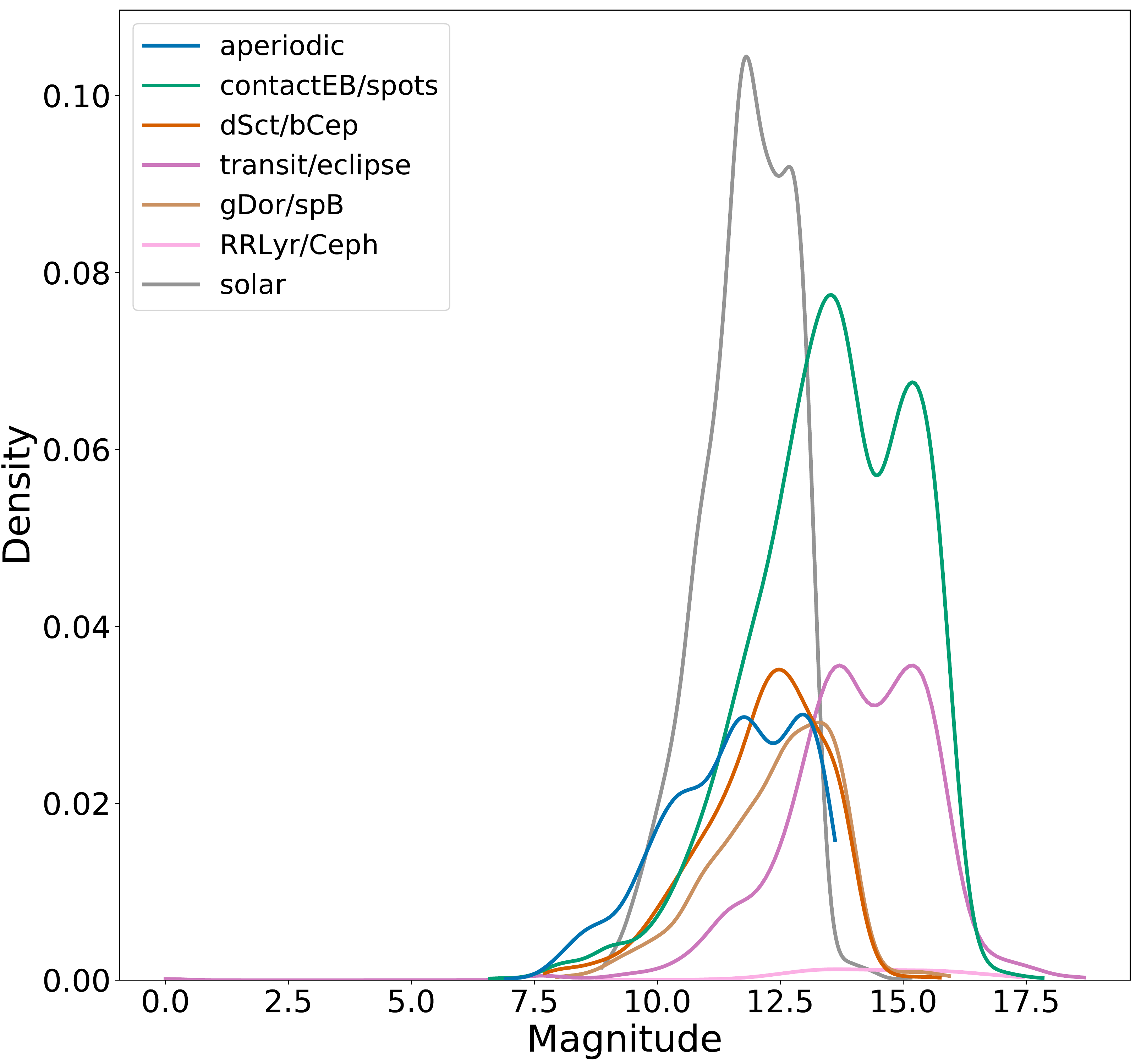}
      \caption{KDE plot illustrating the magnitude distribution of all stars in the training set described in Sect.~\ref{sec:training set}.}
         \label{Fig:kde-mags}
\end{figure}

\section{Class probabilities \textit{Kepler} Q9}
\label{Appendix:tables}

\begin{table*}
    \setlength{\tabcolsep}{0.75\tabcolsep}
	\centering
	\caption{First five rows of the electronic table with class probabilities and assigned labels.}
	\label{tab:classprob_table}
	\resizebox{0.9\width}{!}{
	\rotatebox{90}{
	\begin{tabular}{l l l l l l l l l l l}
	\hline
	kic & p(aperiodic) & p(constant) & p(contactEB/spots) & p(dSct/bCep) & p(eclipse/transit) & p(gDor/SPB) & p(RRLyr/Ceph) & p(solar) & Label & Label (Youden) \\
	\hline
	11513597 & 0.0641 & 0.0221 & 0.5526 & 0.0354 &  0.0327 & 0.1211 & 0.1078 & 0.0642 & contactEB/spots & contactEB/spots \\
	9266835 & 0.0299 & 0.0121 & 0.5285 & 0.0639 & 0.0819 & 0.1530 & 0.1001 & 0.0305 &  contactEB/spots & contactEB/spots \\
	11241837 & 0.02449 & 0.0012 & 0.7854 & 0.01607 & 0.0259 & 0.0477 & 0.0858 & 0.0134 &  contactEB/spots & contactEB/spots \\
	9591826 & 0.0362 &  0.0849 & 0.3630 & 0.1165 & 0.0542 & 0.1967 & 0.0971 & 0.0514 & contactEB/spots & contactEB/spots \\
	\hline
	\end{tabular}}}
\end{table*}

\end{document}